\documentstyle[12pt,graphicx,axodraw,amsmath]{article}
\begin{document}
\baselineskip 0.6cm

\def\simgt{\mathrel{\lower2.5pt\vbox{\lineskip=0pt\baselineskip=0pt
           \hbox{$>$}\hbox{$\sim$}}}}
\def\simlt{\mathrel{\lower2.5pt\vbox{\lineskip=0pt\baselineskip=0pt
           \hbox{$<$}\hbox{$\sim$}}}}

\begin{titlepage}

\begin{flushright}
SLAC-PUB-11674 \\
UCB-PTH-06/01 \\
LBNL-59605
\end{flushright}

\vskip 1.0cm

\begin{center}

{\Large \bf 
Supersymmetry, Naturalness, and Signatures \\ at the LHC
}

\vskip 1.0cm

{\large
Ryuichiro Kitano$^a$ and Yasunori Nomura$^{b,c}$}

\vskip 0.4cm

$^a$ {\it Stanford Linear Accelerator Center, 
                Stanford University, Stanford, CA 94309} \\
$^b$ {\it Department of Physics, University of California,
                Berkeley, CA 94720} \\
$^c$ {\it Theoretical Physics Group, Lawrence Berkeley National Laboratory,
                Berkeley, CA 94720} \\

\vskip 1.2cm

\abstract{Weak scale supersymmetry is often said to be fine-tuned, 
especially if the matter content is minimal.  This is not true if 
there is a large $A$ term for the top squarks.  We present a systematic 
study on fine-tuning in minimal supersymmetric theories and identify 
low energy spectra that do not lead to severe fine-tuning.  Characteristic 
features of these spectra are: a large $A$ term for the top squarks, 
small top squark masses, moderately large $\tan\beta$, and a small $\mu$ 
parameter.  There are classes of theories leading to these features, 
which are discussed.  In one class, which allows a complete elimination 
of fine-tuning, the Higgsinos are the lightest among all the superpartners 
of the standard model particles, leading to three nearly degenerate 
neutralino/chargino states.  This gives interesting signals at the LHC 
--- the dilepton invariant mass distribution has a very small endpoint 
and shows a particular shape determined by the Higgsino nature of the 
two lightest neutralinos.  We demonstrate that these signals are indeed 
useful in realistic analyses by performing Monte Carlo simulations, 
including detector simulations and background estimations.  We also 
present a method that allows the determination of all the relevant 
superparticle masses without using input from particular models, 
despite the limited kinematical information due to short cascades. 
This allows us to test various possible models, which is demonstrated 
in the case of a model with mixed moduli-anomaly mediation.  We also 
give a simple derivation of special renormalization group properties 
associated with moduli mediated supersymmetry breaking, which are 
relevant in a model without fine-tuning.}

\end{center}
\end{titlepage}

\section{Introduction}
\label{sec:intro}

What is the physics at the TeV scale and how can we test it?  These 
questions become more pressing as we approach the LHC era, which 
will start within two years.  It is extremely important now to 
consider what we expect to see at these energies, especially because 
the LHC is a hadron collider experiment, in which relations between 
experimental data and the underlying theory are not so simple. 
Knowing what we are looking for would certainly help to identify 
the physics at the TeV scale and may even be necessary, as the 
determination of the TeV physics at the LHC and other experiments 
will most likely take the form of a slow elimination process.

There are already several hints on possible physics at the TeV scale. 
They come from combining a theoretical criterion of naturalness and 
precision measurements of electroweak observables and rare flavor- 
and $CP$-violating processes.  Among these, the combination of 
naturalness and the electroweak data seems to give the most unambiguous 
hint, because these constraints cannot be satisfied simply by imposing 
approximate symmetries already present in the standard-model gauge 
and matter sector.  Interpreted naively, the precision electroweak 
data suggest the existence of a light Higgs boson, with the 
contributions to the electroweak observables from other physics 
highly suppressed~\cite{unknown:2005em}.  Naturalness then implies 
that there must be a new weakly-interacting physics at a TeV scale 
or below, which cuts off quadratically divergent contributions to 
the Higgs mass-squared parameter arising from loops of the standard 
model particles.

Weak scale supersymmetry is an ideal candidate for the new physics. 
Loops of superparticles cancel the quadratic divergences from those 
of the standard model particles.  The new interactions for the 
superparticles are necessarily weak, as they are related to the 
standard model interactions by supersymmetry.  Moreover, the mass 
of the lightest Higgs boson is predicted to be small, $M_{\rm Higgs} 
\simlt 200~{\rm GeV}$ in most (even extended) theories, which is 
very much consistent with the precision electroweak data.  With the 
introduction of $R$ parity and the assumption of flavor universality 
for the superparticle masses, weak scale supersymmetry can provide 
a fully consistent framework for physics of electroweak symmetry 
breaking.

Postulating weak scale supersymmetry alone, however, does not 
much narrow down signatures at the LHC.  Depending on the relative 
sizes for the soft supersymmetry breaking parameters, one can 
have drastically different signatures at the LHC.  The number of 
relatively model-independent signals is also small, making it 
difficult to discriminate supersymmetry from other TeV-scale physics. 
A generic signal of weak scale supersymmetry is large missing 
transverse energy in association with jets and/or isolated leptons. 
Such a signal, however, arises in almost any theory where 
the lightest TeV-scale particle is stable and neutral, which is 
suggested by the existence of the dark matter of the universe. 
It is then an important task to narrow down the parameter space 
of weak scale supersymmetry further and to perform a detailed study 
of the LHC signals there.  One of the important goals of such 
a study is to identify generic signals associated with a particular 
parameter region so that the non-observation of those signals 
will allow us to exclude the region.

How should we choose regions among the huge parameter space of 
weak scale supersymmetry?  An obvious way is to assume a particular 
supersymmetry breaking model, such as the minimal supergravity 
or gauge mediation model, based on the simplicity of the model. 
This selects a slice in the parameter space of soft supersymmetry 
breaking parameters, which depends only on a few free parameters. 
These studies have been performed by many authors, for example, 
in~[\ref{Baer:1995nq:X}~--~\ref{Weiglein:2004hn:X}].  In this paper 
we take a different criterion to choose the region.  We use the 
hint from naturalness to the maximal amount and consider what are 
generic implications of it on the spectrum of superparticles and 
on LHC signals.  Fortunately, or unfortunately, generic parameter 
regions of weak scale supersymmetry satisfying existing experimental 
constraints do not lead to the correct scale for electroweak 
symmetry breaking without significant fine-tuning.  This information, 
therefore, can be used to constrain the parameter space of soft 
supersymmetry breaking parameters and thus to narrow down possible 
signatures at the LHC.  Of course, the input from naturalness 
alone does not lead to unambiguous signatures in a wide variety 
of possible supersymmetric theories.  In this paper we focus 
our attention to the case where the matter content at the weak 
scale is minimal, i.e. given by that of the minimal supersymmetric 
standard model (MSSM).

What are generic implications of fine-tuning on the spectrum of 
superparticles in a theory with the minimal matter content?  As 
discussed in Ref.~\cite{Kitano:2005wc}, the fine-tuning problem of 
minimal supersymmetry can be solved without extending its matter 
content if the trilinear scalar term ($A$ term) for the top squarks 
is large and the holomorphic supersymmetry-breaking term ($\mu B$ 
term) for the Higgs doublets is small.  This allows us to evade 
the LEP~II bound on the Higgs boson mass with relatively small 
superparticle, specifically top squark, masses.  Then, if soft 
supersymmetry breaking parameters are generated (effectively) 
at low energies, the sensitivity of the electroweak scale to the 
fundamental parameters of the theory can be very mild.  One of 
the consequences of such a scenario is that the top squarks are 
relatively light and have a large mass splitting between the light 
and heavy ones.  Another important consequence is that the Higgsinos 
are rather light, with the masses smaller than about $190~{\rm 
GeV}$ ($270~{\rm GeV}$) for fine-tuning better than $\approx 20\%$ 
($10\%$), which is because naturalness requires any contribution 
to the Higgs boson squared mass to be small, including the 
supersymmetric contribution.  We argue that these features are 
robust and appear quite generically in a minimal supersymmetric 
theory without significant fine-tuning.

This argument provides a strong motivation to consider the case in 
which the lightest neutral Higgsino is the lightest supersymmetric 
particle (LSP).  While it is not a necessary consequence of solving 
the supersymmetric fine-tuning problem, the Higgsino LSP in fact 
arises in a large class of theories in which the pattern of soft 
supersymmetry breaking parameters described above is naturally 
obtained.  It is, therefore, quite important to perform an LHC 
study for the case of the Higgsino LSP and identify possible 
signatures.  An important consequence of the Higgsino LSP scenario 
is that there are three nearly degenerate neutralino/chargino states, 
$\tilde{\chi}^0_1$, $\tilde{\chi}^0_2$ and $\tilde{\chi}^+_1$, with 
$\tilde{\chi}^0_1$ being the LSP.  We show that this structure can 
give interesting signatures at the LHC in the dilepton invariant 
mass distribution arising from the decay $\tilde{\chi}^0_2 \rightarrow 
\tilde{\chi}^0_1\, l^+ l^-$.  We discuss in what sense these are 
characteristic signatures of the Higgsino LSP, and under what 
circumstances the signals can be used in realistic analyses.

To demonstrate the usefulness of the signatures, we need to choose 
specific parameter points and perform Monte Carlo simulations, 
including detector simulations and standard model background. 
We do this in the model discussed in Refs.~[\ref{Kitano:2005wc:X}~%
--~\ref{Kitano:2005ew:X}], where the desired pattern of the 
soft supersymmetry breaking masses, a large $A$ term and a small 
$\mu B$ term, are obtained while evading the existing experimental 
constraints such as the one from $b \rightarrow s \gamma$.  We show 
that the dilepton invariant mass distribution from $\tilde{\chi}^0_2 
\rightarrow \tilde{\chi}^0_1\, l^+ l^-$ is indeed useful to test 
the Higgsino nature of the LSP and to extract the information on a 
small mass difference between $\tilde{\chi}^0_1$ and $\tilde{\chi}^0_2$. 
We also show that important parameters of the model, the overall 
mass scale and the $\mu$ parameter, are determined by various other 
endpoint analyses.  In fact, we show that these parameters are 
overconstrained, so that we can test some of the model predictions. 
We perform these analyses for an integrated luminosity of 
$30~{\rm fb}^{-1}$, but essentially the same conclusion is 
obtained with $10~{\rm fb}^{-1}$.  The technique presented here 
can also be used in a larger class of theories having similar 
superparticle spectra.

The organization of the paper is as follows.  In 
section~\ref{sec:naturalness}, we present a systematic study on 
naturalness in general supersymmetric theories, especially focusing 
on the case where the matter content at the weak scale is minimal. 
We give general criteria that natural supersymmetric models with 
the minimal matter content must satisfy, and present characteristic 
patterns for the superparticle spectrum arising from these models. 
In section~\ref{sec:higgsino}, we discuss LHC signals of the Higgsino 
LSP scenario, which naturally arises in a class of models that do not 
suffer from fine-tuning.  We find that a combination of the endpoint 
and the shape of the dilepton invariant mass distribution provides 
a powerful tool to test the scenario.  In section~\ref{sec:model}, 
we perform Monte Carlo simulations to demonstrate that these signals 
are indeed useful in a realistic situation.  We also present an 
analysis that allows us to determine the masses of the gluino, squarks, 
and the two lightest neutralinos in the Higgsino LSP scenario, without 
relying on details of the underlying model.  We illustrate that these 
information can be used to test and/or discriminate between possible 
models.  Discussion and conclusions are given in section~\ref{sec:concl}. 
In the Appendix, we give a simple derivation of special renormalization 
group properties of moduli mediated supersymmetry breaking models, 
which are relevant in the model studied in section~\ref{sec:model}.

\section{Supersymmetry and Naturalness}
\label{sec:naturalness}

One of the principal motivations for weak scale supersymmetry is 
to provide a solution to the naturalness problem of the standard 
model.  This has, however, been put in a subtle position after 
non-discovery of both superparticles and a light Higgs boson 
at LEP~II.  In a generic parameter region motivated by simple 
supersymmetry breaking models, fine-tuning of order a few percent 
is required to reproduce the correct scale for electroweak 
symmetry breaking while evading the constraints from LEP~II. 
This problem, called the supersymmetric fine-tuning problem, has 
attracted much attention recently, and several solutions have been 
proposed, e.g., in~[\ref{Kitano:2005wc:X},~\ref{Choi:2005hd:X},%
~\ref{Bastero-Gil:2000bw:X}~--~\ref{Dermisek:2006ey:X}].
In this section we reconsider the problem and see what are generic 
implications of it, especially in the context of theories with the 
minimal matter content.  One of our emphases here is on the fact 
that the supersymmetric fine-tuning problem may simply be a problem 
of the supersymmetry breaking mechanism and not necessarily that 
of minimal supersymmetry itself.

\subsection{Large {\boldmath $A_t$} and small {\boldmath $\mu B$} 
in minimal supersymmetry}
\label{subsec:large-At}

We begin our discussion by considering fine-tuning in the Higgs 
potential in general weak scale supersymmetric theories.  Let 
$h$ be the Higgs field whose vacuum expectation value (VEV) 
breaks the electroweak symmetry.  In minimal supersymmetry, $h$ 
is a linear combination of the two Higgs doublets, $H_u$ and 
$H_d$.  The potential for $h$ is given by
\begin{equation}
  V = m_h^2\, |h|^2 + \frac{\lambda_h}{4} |h|^4,
\label{eq:higgs-pot}
\end{equation}
where $m_h^2$ is negative and $\lambda_h$ is positive.  By 
minimizing it, we obtain $v^2 \equiv \langle h \rangle^2 = 
-2 m_h^2/\lambda_h$ and $M_{\rm Higgs}^2 = \lambda_h v^2$, where 
$M_{\rm Higgs}$ is the mass of the physical Higgs boson, so that 
\begin{equation}
  \frac{M_{\rm Higgs}^2}{2} = - m_h^2.
\label{eq:higgs-mass}
\end{equation}
We thus find that $|m_h^2|$ cannot be large for a light Higgs 
boson: $|m_h^2|^{1/2} \simlt 140~{\rm GeV}$ ($90~{\rm GeV}$) 
for $M_{\rm Higgs} \simlt 200~{\rm GeV}$ ($130~{\rm GeV}$). 

What is $m_h^2$ in supersymmetric theories?  For moderately large 
$\tan\beta \equiv \langle H_u \rangle/\langle H_d \rangle$, e.g. 
$\tan\beta \simgt 2$, $m_h^2$ can be written as
\begin{equation}
  m_h^2 = |\mu|^2 + m_{H_u}^2|_{\rm tree} + m_{H_u}^2|_{\rm rad},
\label{eq:mh2}
\end{equation}
where $\mu$ is the supersymmetric mass for the Higgs doublets, 
and $m_{H_u}^2|_{\rm tree}$ and $m_{H_u}^2|_{\rm rad}$ 
represent the tree-level and radiative contributions to the soft 
supersymmetry-breaking mass squared for $H_u$.  The dominant 
contribution to $m_{H_u}^2|_{\rm rad}$ arises from top-stop loop:
\begin{equation}
  m_{H_u}^2|_{\rm rad} \simeq -\frac{3y_t^2}{8\pi^2} 
    \bigl( m_{Q_3}^2 + m_{U_3}^2 + |A_t|^2 \bigr) 
    \ln\Biggl( \frac{M_{\rm mess}}{m_{\tilde{t}}} \Biggr),
\label{eq:corr-Higgs}
\end{equation}
where $y_t$ is the top Yukawa coupling, $m_{Q_3}^2$ and $m_{U_3}^2$ 
soft supersymmetry breaking masses for the third-generation doublet 
quark, $Q_3$, and singlet up-type quark, $U_3$, and $A_t$ the trilinear 
scalar interaction parameter for the top squarks (our definition 
for the $A$ parameters is such that a scalar trilinear coupling 
is given by the product of the Yukawa coupling and the $A$ parameter, 
e.g., ${\cal L} = -y_t A_t \tilde{q}_3 \tilde{u}_3 H_u + {\rm h.c.}$). 
The quantity $M_{\rm mess}$ represents the scale at which squark and 
slepton masses are generated, and $m_{\tilde{t}}$ the scale of the 
top squark masses determined by $m_{Q_3}^2$, $m_{U_3}^2$ and $A_t$. 
Note that $M_{\rm mess}$ can be an effective scale different from 
the true scale of scalar mass generation in a case that the theory 
possesses special relations among various parameters.

For fine-tuning to be absent, each term in the right-hand-side 
of Eq.~(\ref{eq:mh2}) should not be much larger than the 
left-hand-side, which is related to the physical Higgs boson 
mass by Eq.~(\ref{eq:higgs-mass}).  Let us first consider 
$m_{H_u}^2|_{\rm rad}$.  The amount of fine-tuning from this term 
is given by $M_{\rm Higgs}^2/2 m_{H_u}^2|_{\rm rad}$, so that 
requiring the absence of fine-tuning worse than $\Delta^{-1}$ 
leads to the condition
\begin{equation}
  m_{\tilde{t}}^2
  \simlt \frac{2\pi^2}{3y_t^2}\frac{M_{\rm Higgs}^2}
    {\Bigl(1+\frac{x^2}{2}\Bigr)\Delta^{-1}
    \ln\frac{M_{\rm mess}}{m_{\tilde{t}}}}
  \approx (700~{\rm GeV})^2 \frac{1}{1+\frac{x^2}{2}}
    \Biggl( \frac{20\%}{\Delta^{-1}} \Biggr)
    \Biggl( \frac{3}{\ln\frac{M_{\rm mess}}{m_{\tilde{t}}}} \Biggr)
    \Biggl( \frac{M_{\rm Higgs}}{200~{\rm GeV}} \Biggr)^2,
\label{eq:stop-bound}
\end{equation}
where we have set $m_{Q_3}^2 \simeq m_{U_3}^2 \simeq m_{\tilde{t}}^2$ 
for simplicity, and $x \equiv |A_t|/m_{\tilde{t}}$.  This has the 
following implication on the properties of the supersymmetry breaking 
sector~\cite{Chacko:2005ra}.  Unless $M_{\rm mess}$ is extremely small, 
e.g. $M_{\rm mess} \simlt 10~{\rm TeV}$, the absence of fine-tuning, 
defined by $\Delta^{-1} \geq 20\%$, requires $m_{\tilde{t}} \simlt 
700~{\rm GeV}$, where we have used $M_{\rm Higgs} \simlt 200~{\rm GeV}$ 
as suggested by the precision electroweak data.  This implies that 
a naive low-scale mediation model, leading to the ``minimal gauge 
mediated mass relation'' $m_{\tilde{t}}^2/m_{\tilde{e}}^2 \approx 
g_3^4/g_1^4$, is unlikely to solve the fine-tuning problem because 
it gives too light right-handed sleptons.  For a lighter Higgs boson, 
we obtain severer bounds on $m_{\tilde{t}}$: for $M_{\rm Higgs} 
\simeq 140~{\rm GeV}$, for example, we find $m_{\tilde{t}} \simlt 
700~{\rm GeV}$ even for $\Delta^{-1} \simeq 10\%$.  Note that the 
condition of Eq.~(\ref{eq:stop-bound}) applies independently of 
any other considerations.%
\footnote{In the case that $\ln(M_{\rm mess}/m_{\tilde{t}})$ is 
large, for example in gravity mediated models, the expression in 
Eq.~(\ref{eq:corr-Higgs}) is not reliable and we should sum up the 
leading logarithms using renormalization group equations.  This 
case will be addressed in the next subsection.}

Light top squarks suggested by Eq.~(\ref{eq:stop-bound}) leads 
to a tension with the LEP~II bound on the Higgs boson mass, 
$M_{\rm Higgs} \simgt 114.4~{\rm GeV}$~\cite{Barate:2003sz}, 
since in the MSSM having $M_{\rm Higgs}$ larger than the $Z$ boson 
mass, $m_Z$, requires radiative corrections arising from top-stop 
loop, which grow with the top squark masses~\cite{Okada:1990vk}. 
A simple way to avoid the conflict is to introduce an additional 
contribution to the Higgs boson mass other than that in the MSSM. 
An example of such theories can be found in Ref.~\cite{Chacko:2005ra}, 
where the required properties for the supersymmetry breaking sector 
are realized by strong gauge dynamics breaking supersymmetry at 
a scale of $(10\!\sim\!100)~{\rm TeV}$.  What if we do not introduce 
any other contribution to $M_{\rm Higgs}$ than that in the MSSM? 
In this case it is unlikely that $M_{\rm Higgs}$ can be larger 
than $130~{\rm GeV}$, so Eq.~(\ref{eq:stop-bound}) leads to 
a severer bound
\begin{equation}
  m_{\tilde{t}}^2
  \simlt (450~{\rm GeV})^2 \frac{1}{1+\frac{x^2}{2}}
    \Biggl( \frac{20\%}{\Delta^{-1}} \Biggr)
    \Biggl( \frac{3}{\ln\frac{M_{\rm mess}}{m_{\tilde{t}}}} \Biggr).
\label{eq:stop-bound-2}
\end{equation}
While this bound is strong, it still leaves a room for evading 
the LEP~II constraint on $M_{\rm Higgs}$ without introducing 
severe fine-tuning.  As discussed in~\cite{Kitano:2005wc}, this 
happens if $A_t$ is large, $\tan\beta$ is (moderately) large, 
and $M_{\rm mess}$ is small.  In particular, it is crucial to 
have large $A_t$, compared with $m_{\tilde{t}}$, to evade the 
Higgs boson mass bound while keeping $\Delta^{-1}$ modest.

\begin{figure}[t]
\begin{center}
  \includegraphics[height=6.5cm]{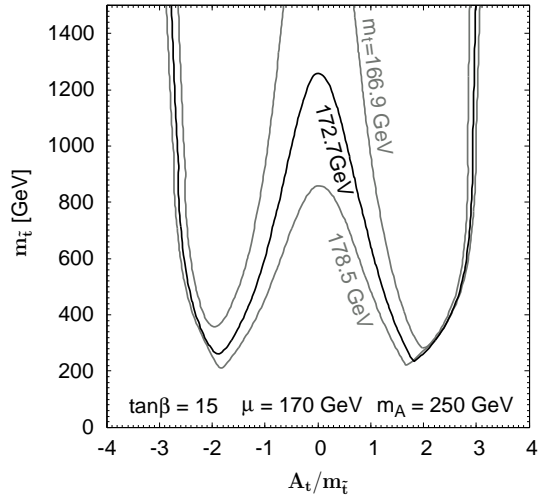}
\end{center}
\caption{Minimal values of $m_{\tilde{t}} \equiv (m_{Q_3}^2)^{1/2} 
 = (m_{U_3}^2)^{1/2}$ giving $M_{\rm Higgs} \simgt 114.4~{\rm GeV}$ 
 as a function of $A_t/m_{\tilde{t}}$.  The other parameters 
 are fixed to be $500~{\rm GeV}$ for the gaugino and sfermion 
 masses other than $m_{Q_3}^2$ and $m_{U_3}^2$, $(500~{\rm GeV})
 (A_t/m_{\tilde{t}})$ for the $A$ parameters other than $A_t$, 
 $\tan\beta = 15$, $\mu = +170~{\rm GeV}$, and $m_A = 250~{\rm GeV}$.}
\label{fig:minimal-mt}
\end{figure}

In Fig.~\ref{fig:minimal-mt}, we plot minimal values of $m_{\tilde{t}} 
\equiv (m_{Q_3}^2)^{1/2} = (m_{U_3}^2)^{1/2}$, $m_{\tilde{t}}|_{\rm min}$, 
that give $M_{\rm Higgs} \simgt 114.4~{\rm GeV}$ as a function 
of $A_t/m_{\tilde{t}}$.  The other parameters are fixed to be 
$500~{\rm GeV}$ for the gaugino and sfermion masses other than 
$m_{Q_3}^2$ and $m_{U_3}^2$, $(500~{\rm GeV})(A_t/m_{\tilde{t}})$ 
for the $A$ parameters other than $A_t$, $\tan\beta = 15$, 
$\mu = +170~{\rm GeV}$, and $m_A = 250~{\rm GeV}$, where $m_A$ 
is the mass of the pseudo-scalar Higgs boson. 
The dependence of the results on the fixed parameters is not 
significant, as long as $\tan\beta$ is sufficiently large, e.g., 
$\tan\beta \simgt 10$.  In the figure we plot $m_{\tilde{t}}|_{\rm 
min}$ for three different values of the top quark mass, $m_t = 166.9$, 
$172.7$ and $178.5~{\rm GeV}$, corresponding to the central value and 
the $2 \sigma$ range of the the latest experimental data: $m_t = 172.7 
\pm 2.9~{\rm GeV}$~\cite{Group:2005cc}.  The calculation here has 
been performed using {\it FeynHiggs\,2.2}~\cite{Heinemeyer:1998yj} 
(for earlier analyses for the Higgs boson mass in the MSSM, 
see e.g.~\cite{Carena:1995wu}).  If we instead use a code 
based on the pure $\overline{\rm DR}$ scheme, such as {\tt 
SuSpect\,2.3}~\cite{Djouadi:2002ze}, we obtain slightly different 
values of $m_{\tilde{t}}|_{\rm min}$: for $|A_t/m_{\tilde{t}}| 
\sim 2$ the differences are of order $50~{\rm GeV}$ but for 
$|A_t/m_{\tilde{t}}| \ll 1$ the $\overline{\rm DR}$ scheme 
calculation can give $m_{\tilde{t}}|_{\rm min}$ larger than 
that in the figure by about $200~{\rm GeV}$.  These differences 
give an estimate for the size of higher order corrections. 
Throughout the paper, our sign convention for $\mu$ and the 
soft supersymmetry breaking parameters follows that of SUSY 
Les Houches Accord~\cite{Skands:2003cj}.

The figure clearly shows that in order to have light top squarks 
suggested by Eq.~(\ref{eq:stop-bound-2}) the existence of 
a substantial $A_t$ term ($|A_t/m_{\tilde{t}}| \simgt 1$) is 
required.  For $m_t = 172.7~{\rm GeV}$, the existence of $A_t$ with 
$|A_t/m_{\tilde{t}}| \simgt 1$ can allow $m_{\tilde{t}}$ as small 
as $(200\!\sim\!400)~{\rm GeV}$ while $m_{\tilde{t}}$ should be 
larger than $\approx 1.2~{\rm TeV}$ for $|A_t/m_{\tilde{t}}| \ll 1$. 
Another important point is that for $|A_t/m_{\tilde{t}}| \sim 2$, 
which gives the minimal value of $m_{\tilde{t}}|_{\rm min}$, the 
sensitivity of $m_{\tilde{t}}|_{\rm min}$ to the value of $m_t$ 
is mild, while for $|A_t/m_{\tilde{t}}| \ll 1$ the sensitivity is 
huge.  This implies, for example, that if $m_t$ turns out to be 
smaller than $\simeq 170~{\rm GeV}$ theories with $|A_t/m_{\tilde{t}}| 
\ll 1$ at the weak scale will pretty much be ``excluded.''

We conclude that to have natural electroweak symmetry breaking 
in supersymmetric theories with the minimal, i.e. MSSM, matter 
content, the existence of a substantial $A_t$ term at the weak 
scale is crucial.  Another important ingredient is a moderately 
large $\tan\beta$, e.g. $\tan\beta \simgt 5$, to have a sufficiently 
large tree-level Higgs boson mass, which requires the holomorphic 
supersymmetry breaking mass squared for the Higgs doublets, 
the $\mu B$ term, to be (significantly) smaller than $2|\mu|^2 
+ m_{H_u}^2 + m_{H_d}^2$.  In fact, these ingredients dominantly 
control the amount of fine-tuning in almost any theory with the 
MSSM matter content.  To demonstrate this, we will now analyze 
the situations in the case of high scale supersymmetry breaking 
and in gauge mediation models from the viewpoint of the size 
of $A_t$ at the weak scale.  For earlier analyses of fine-tuning 
in these models, see e.g.~\cite{Giusti:1998gz}.

\subsection{High scale supersymmetry breaking and gauge mediation}
\label{subsec:mSUGRA-GMSB}

Let us first consider the minimal supergravity (mSUGRA) 
scenario~\cite{Chamseddine:1982jx}.  We start by considering 
the constrained mSUGRA, in which the soft supersymmetry breaking 
parameters are specified by the universal gaugino mass $M_{1/2}$, 
universal scalar mass squared $m_0^2$, universal $A$ term $A_0$, 
and the $\mu B$ term at the unification scale, $M_{\rm unif} 
\approx 10^{16}~{\rm GeV}$.  While this scenario is sometimes 
criticized due to a lack of a strong theoretical motivation, 
it is not so bad in term of fine-tuning, compared with other 
models such as gauge mediation models.  This is because we can 
obtain a reasonable size of $A_t/m_{\tilde{t}}$ at the weak 
scale, so that the top squark masses can be made smaller 
compared with the models giving smaller values of 
$A_t/m_{\tilde{t}}$ at the weak scale.

\begin{figure}[t]
\begin{center}
  \includegraphics[width=7.0cm]{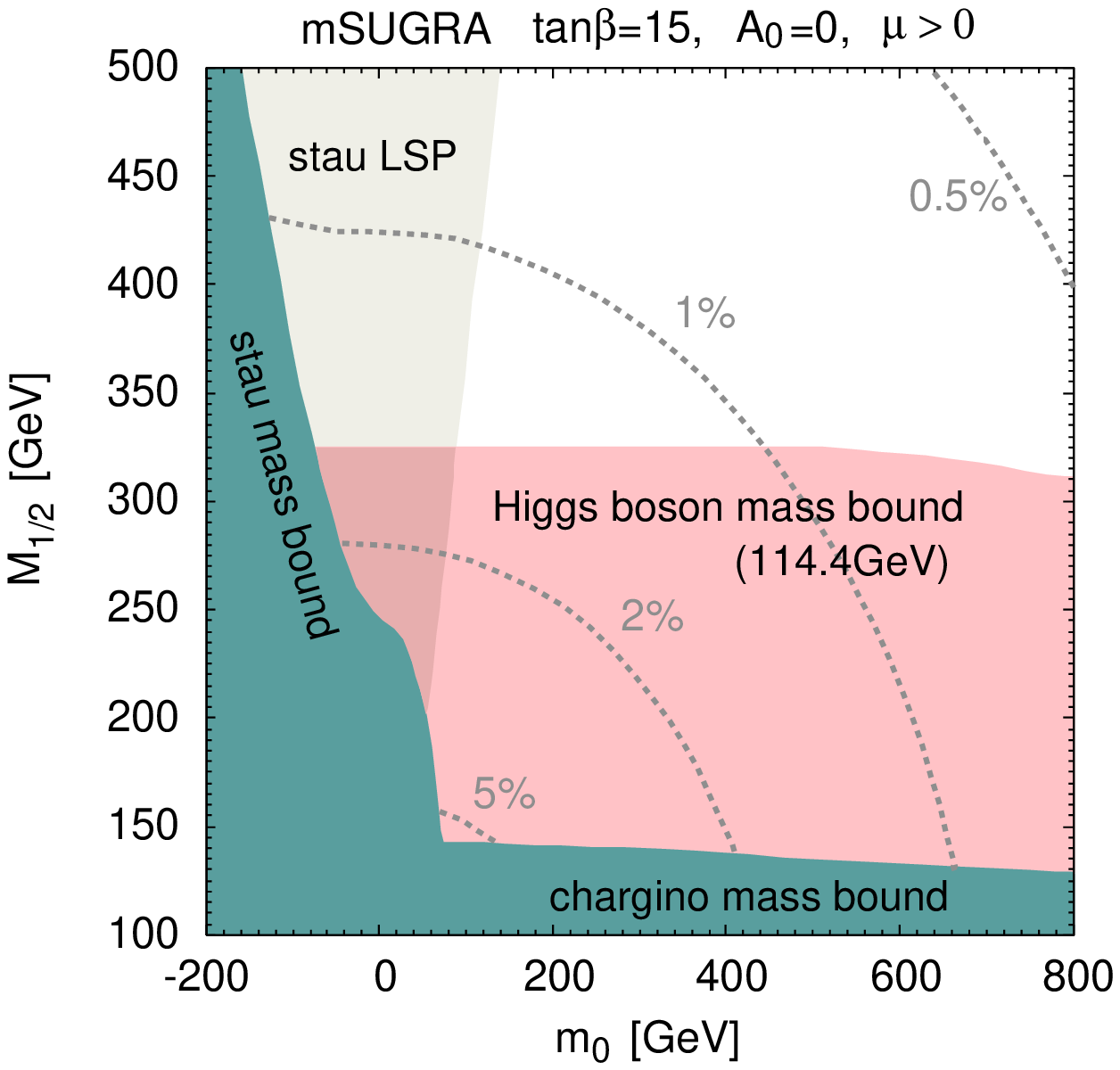}
\hspace{1cm}
  \includegraphics[width=7.0cm]{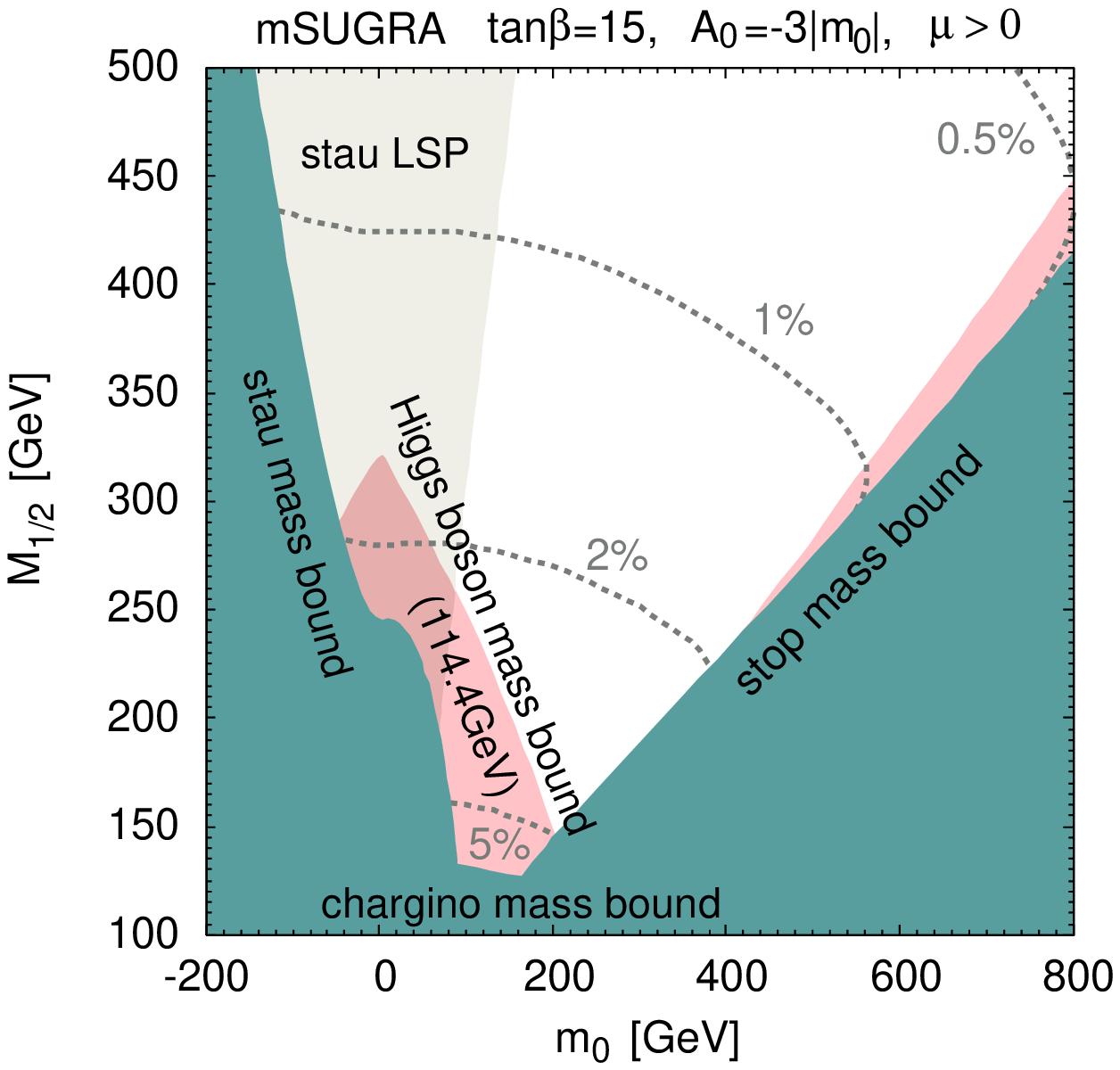}
\end{center}
\caption{Contours of $\Delta^{-1}$ on the $m_0$-$M_{1/2}$ plane for 
 the constrained mSUGRA with $A_0 = 0$ (left) and $A_0 = -3|m_0|$ 
 (right).  The sign of $\mu$ is chosen to be positive.  The constraints 
 from direct superparticle search, the Higgs boson mass bound, and 
 the stau LSP are also shown.}
\label{fig:cMSSM}
\end{figure}

In Fig.~\ref{fig:cMSSM} we plot values of the fine-tuning 
parameter $\Delta^{-1}$ in the constrained mSUGRA for two 
different choices of $A_0$: $A_0 = 0$ and $A_0 = -3|m_0|$. 
The parameter $\Delta^{-1}$ is defined by the fractional 
sensitivities of the electroweak VEV, $v \simeq 174~{\rm GeV}$, 
to the fundamental parameters of the theory, with generic 
sensitivities of $v$ to the parameters appropriately 
corrected~\cite{Barbieri:1987fn}.  We plot the contours of 
$\Delta^{-1}$ on the $m_0$-$M_{1/2}$ plane for $\mu > 0$, where 
$m_0 \equiv {\rm sgn}(m_0^2) |m_0^2|^{1/2}$.  The values of $\mu$ 
and $\mu B$ at $M_{\rm unif}$ are determined by $v$ and $\tan\beta$, 
and we take $\tan\beta = 15$.  We find that for $A_0 = 0$ the 
fine-tuning is worse than $2\%$, while for $A_0 = -3|m_0|$ it 
can be as mild as $5\%$ for $M_{1/2} \simeq 150~{\rm GeV}$ 
and $m_0^2 \simeq (200~{\rm GeV})^2$.  This can be understood 
as follows.  For $A_0 = 0$, renormalization group equations 
give low-energy values for $A_t$ and $m_{\tilde{t}}$ that 
satisfy $A_t/m_{\tilde{t}} \sim -1$.  While this value of 
$|A_t/m_{\tilde{t}}|$ is not totally negligible, it is still 
not large enough to give $M_{\rm Higgs} \simgt 114.4~{\rm GeV}$ 
with top squark masses smaller than about $600~{\rm GeV}$ (see 
Fig.~\ref{fig:minimal-mt}).  This gives a high sensitivity of $v$ 
to $y_t$ (the top-stop contribution to $m_{H_u}^2$), leading to 
$\Delta^{-1} \simlt 2\%$.  The situation can be made better by 
introducing non-vanishing $A_0$ at $M_{\rm unif}$.  While the 
sensitivity of low-energy $A_t$ to $A_0$ is rather weak, $A_0 
= -3|m_0|$ can give a low-energy value of $A_t/m_{\tilde{t}}$ about 
$-1.8$, which allows $m_{\tilde{t}}$ as small as $\simeq 250~{\rm 
GeV}$ to evade the Higgs boson mass bound, and thus $\Delta^{-1}$ 
as large as $5\%$.  Here $m_{\tilde{t}}$ is defined by $m_{\tilde{t}} 
\equiv (m_{Q_3}^2 m_{U_3}^2)^{1/4}$.  In fact, larger values of 
$A_0$ do not help in reducing fine-tuning because of a shrinking 
of the phenomenologically acceptable parameter region, and we obtain 
$\Delta^{-1}|_{\rm max} \approx 5\%$ in the constrained mSUGRA.

\begin{figure}[t]
\begin{center}
  \includegraphics[width=7.0cm]{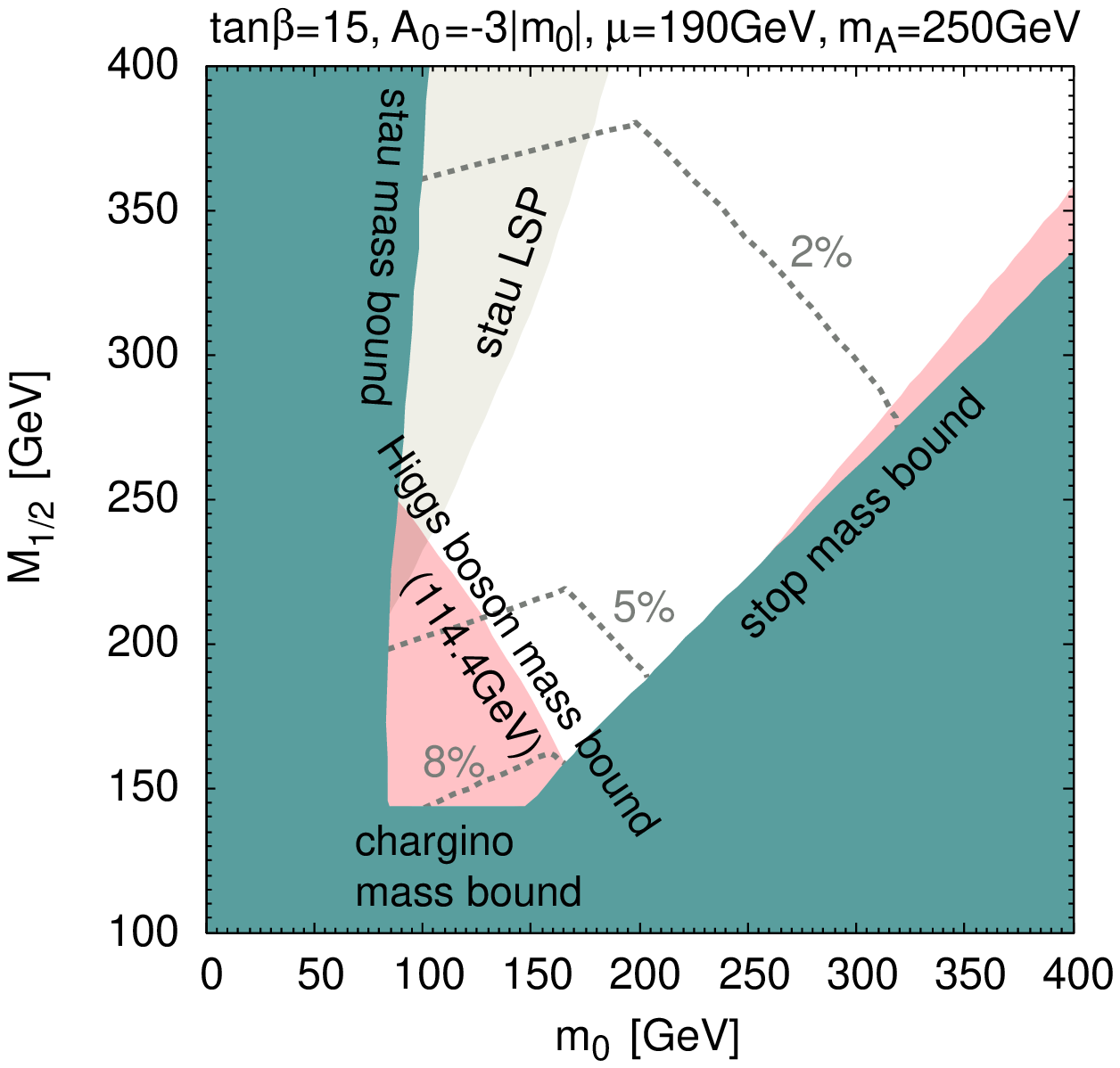}
\hspace{1cm}
  \includegraphics[width=7.0cm]{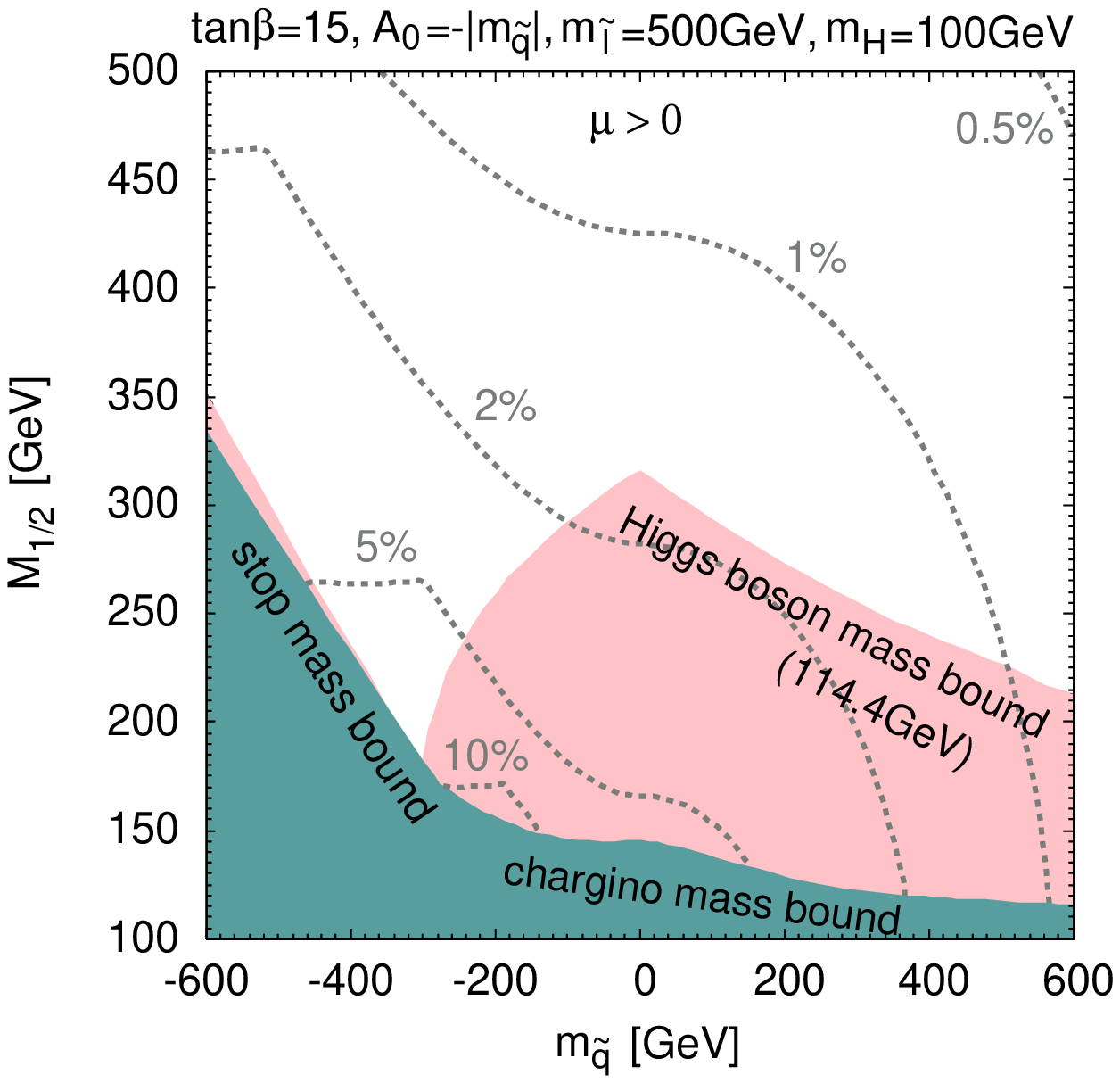}
\end{center}
\caption{Contours of $\Delta^{-1}$ on the $m_0$-$M_{1/2}$ plane in 
 the mSUGRA model with $\mu = 190~{\rm GeV}$ and $m_A = 250~{\rm GeV}$ 
 fixed at the weak scale and $A_0 = -3|m_0|$ at $M_{\rm unif}$ (left). 
 Contours of $\Delta^{-1}$ on the $m_{\tilde{q}}$-$M_{1/2}$ 
 plane with $\mu > 0$ and $m_{\tilde{l}}^2 = (500~{\rm GeV})^2$, 
 $A_0 = -|m_{\tilde{q}}|$ and $m_H^2 \equiv m_{H_u}^2 = m_{H_d}^2 
 = (100~{\rm GeV})^2$ at $M_{\rm unif}$, where $m_{\tilde{q}}^2$ 
 and $m_{\tilde{l}}^2$ are the squark and slepton masses, respectively 
 (right).  In both cases $\tan\beta = 15$.}
\label{fig:Mmess=Munif}
\end{figure}

In the case of the constrained mSUGRA described above, $\Delta^{-1}$ 
is determined by the sensitivity of $v$ to $y_t$ and $\mu$, which 
implies that the dominant source of fine-tuning comes from the 
sensitivity of $m_{H_u}^2$ to the top-stop loop contribution. 
We can make this sensitivity weaker by deviating from the constrained 
mSUGRA.  A simple way of doing this is to make $m_{H_u}^2$ and 
$m_{H_d}^2$ differ from $m_0^2$ at $M_{\rm unif}$.  Practically, 
this implies that we can take low-energy values of $\mu$ and $m_A$ 
as free parameters.  Then, for certain values of $\mu$ and $m_A$, 
which corresponds to choosing certain values of $m_{H_u}^2$ and 
$m_{H_d}^2$ at $M_{\rm unif}$, we find that the sensitivity of $v$ 
to $y_t$ can be made weaker due to renormalization group properties 
of the soft supersymmetry breaking parameters.  This is illustrated 
in the left panel of Fig.~\ref{fig:Mmess=Munif}, where we plot 
the contours of $\Delta^{-1}$ on the $m_0$-$M_{1/2}$ plane, with 
$\mu = 190~{\rm GeV}$, $m_A = 250~{\rm GeV}$ and $A_0 = -3|m_0|$. 
We find that fine-tuning can be made as mild as $\Delta^{-1} \simeq 
8\%$.  A similar reduction of tuning can also occur in the region 
with $m_{\tilde{t}}^2 < 0$ at $M_{\rm unif}$, which is allowed if 
we violate the universality of $M_{1/2}$ and/or $m_0^2$ to avoid 
the slepton mass bound.  This is illustrated in the right panel 
of Fig.~\ref{fig:Mmess=Munif} for $\mu > 0$, where we plot 
$\Delta^{-1}$ as a function of the gaugino mass, $M_{1/2}$, and 
the squark mass, $m_{\tilde{q}} \equiv {\rm sgn}(m_{\tilde{q}}^2)
|m_{\tilde{q}}^2|^{1/2}$, at the unification scale $M_{\rm unif}$. 
The other parameters are chosen as $m_{\tilde{l}}^2 = (500~{\rm GeV})^2$ 
for the sleptons, $A_0 = -|m_{\tilde{q}}|$, and $m_{H_u}^2 = m_{H_d}^2 
= (100~{\rm GeV})^2$ at $M_{\rm unif}$.  We find that fine-tuning 
can be as mild as $\Delta^{-1} \simeq 8\%$.%
\footnote{The region with $m_{\tilde{t}}^2 < 0$ has been discussed 
recently in~\cite{Dermisek:2006ey} in the context of finding 
a relation among soft supersymmetry breaking parameters which 
reduces fine-tuning.  Our approach here is different: we do not 
assume any special relations among the supersymmetry breaking 
masses, e.g., between the gaugino and squark masses.  We then do 
not find a region with $\Delta^{-1}$ better than $\approx 10\%$.}

How much can we reduce fine-tuning in a theory with high scale 
supersymmetry breaking?  It is possible, after all, that the soft 
supersymmetry breaking parameters are not calculable (easily) 
if they are generated through physics at the gravitational scale. 
Suppose, for example, that grand unification is realized in 
five dimensions and supersymmetry is broken on a brane on which 
the active gauge group is only $SU(3)_C \times SU(2)_L \times 
U(1)_Y$~\cite{Hall:2001pg}.  Suppose also that all the gauge, 
matter and Higgs fields propagate in the bulk, so that they 
all feel supersymmetry breaking through the operators $[Z 
{\cal W}^\alpha {\cal W}_\alpha]_{\theta^2}$, $[(Z+Z^\dagger) 
\Phi^\dagger \Phi]_{\theta^4}$ and $[Z^\dagger Z \Phi^\dagger 
\Phi]_{\theta^4}$, where $Z$ is the supersymmetry breaking field, 
${\cal W}_\alpha$ the gauge field-strength superfields, and 
$\Phi$ the matter and Higgs chiral superfields.  (The $\mu$ 
and $\mu B$ terms can also be generated through $[Z^\dagger 
H_u H_d + {\rm h.c.}]_{\theta^4}$ and $[Z^\dagger Z H_u H_d 
+ {\rm h.c.}]_{\theta^4}$.)  Then, if there is a flavor symmetry 
in the bulk and on the supersymmetry breaking brane, the generated 
supersymmetry breaking masses can be flavor universal.  The flavor 
symmetry is broken only on the other brane on which the Yukawa 
couplings are located.  This setup leads to soft supersymmetry 
breaking masses that are completely general other than the fact 
that they are flavor universal.  In particular, there is no 
imprint on the superparticle masses from the underlying gauge 
unification.

It is, therefore, important to figure out the maximum 
value of $\Delta^{-1}$ one can obtain in generic high scale 
supersymmetry breaking scenarios.  We first note that there 
is a ``model-independent'' source of the sensitivity of $v$ to 
the fundamental parameters --- the sensitivity of $v$ to the 
gluino mass, $M_3$, at $M_{\rm unif}$.  This is because $M_3$ 
always gives contributions that make $m_{\tilde{t}}$ grow at 
the infrared, which always pushes down the value of $m_{H_u}^2$ 
at the weak scale.  Since the sign of the contributions is definite, 
we cannot weaken this sensitivity by complicating renormalization 
group evolutions for a fixed value of $M_3$.  Now, there is a lower 
bound on $M_3$ at $M_{\rm unif}$ arising from the requirement that 
a sufficiently large $A_t/m_{\tilde{t}}$ is obtained at the weak 
scale, without making $A_t$ at $M_{\rm unif}$ extremely large and 
thus introducing a large sensitivity of $v$ to $A_t$.  We find 
$M_3(M_{\rm mess}) \simgt 150~{\rm GeV}$, which leads to a factor 
of $\approx 10$ stronger fractional variation of $v$ when we vary 
$M_3$ at $M_{\rm unif}$: $\partial \ln v/\partial \ln M_3 \approx 10$. 
We conclude that, without a special relation(s) among various soft 
supersymmetry breaking parameters, the maximum value of $\Delta^{-1}$ 
in high scale supersymmetry breaking is
\begin{equation}
  \Delta^{-1}_{\rm max} \Bigr|_{M_{\rm mess} \sim M_{\rm unif}}
  \approx 10\%.
\label{eq:Delta-max}
\end{equation}
This occurs in parameter regions in which $|A_t/m_{\tilde{t}}| 
\approx O(1.5\!\sim\!2.5)$, $|\mu| \simlt 250~{\rm GeV}$ and 
$M_{\tilde{g}} \simgt 450~{\rm GeV}$, where $M_{\tilde{g}}$ is 
the gluino mass at the weak scale.  In particular, the best points 
in Fig.~\ref{fig:Mmess=Munif} both occur at $A_t/m_{\tilde{t}} 
\simeq -1.8$ and $m_{\tilde{t}} = (m_{Q_3}^2 m_{U_3}^2)^{1/4} \simeq 
250~{\rm GeV}$.  The electroweak VEVs satisfy $\tan\beta \simgt 5$, 
and the Higgs boson mass is bounded by $M_{\rm Higgs} \simlt 
120~{\rm GeV}$.

\begin{figure}[t]
\begin{center}
  \includegraphics[width=7.0cm]{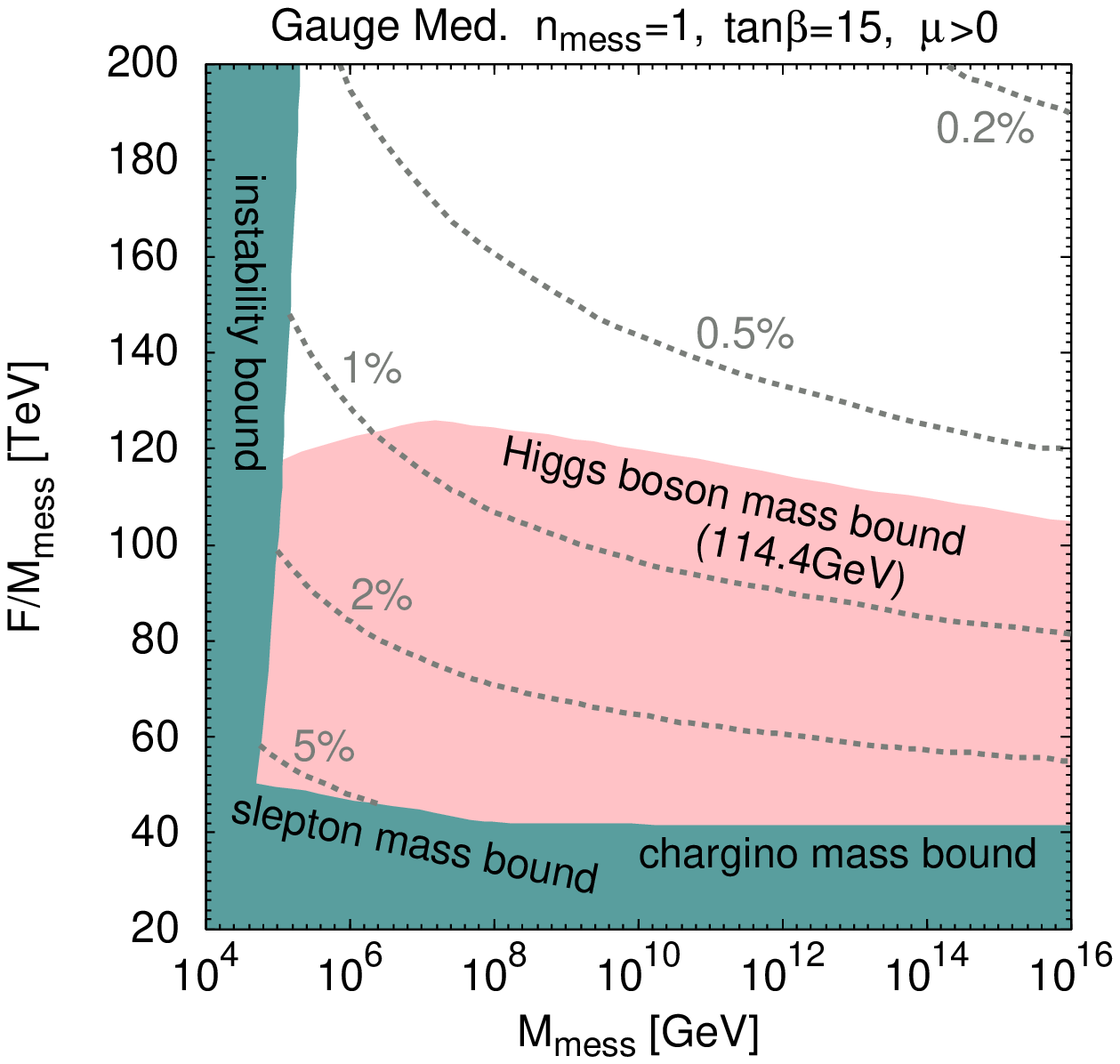}
\hspace{1cm}
  \includegraphics[width=7.0cm]{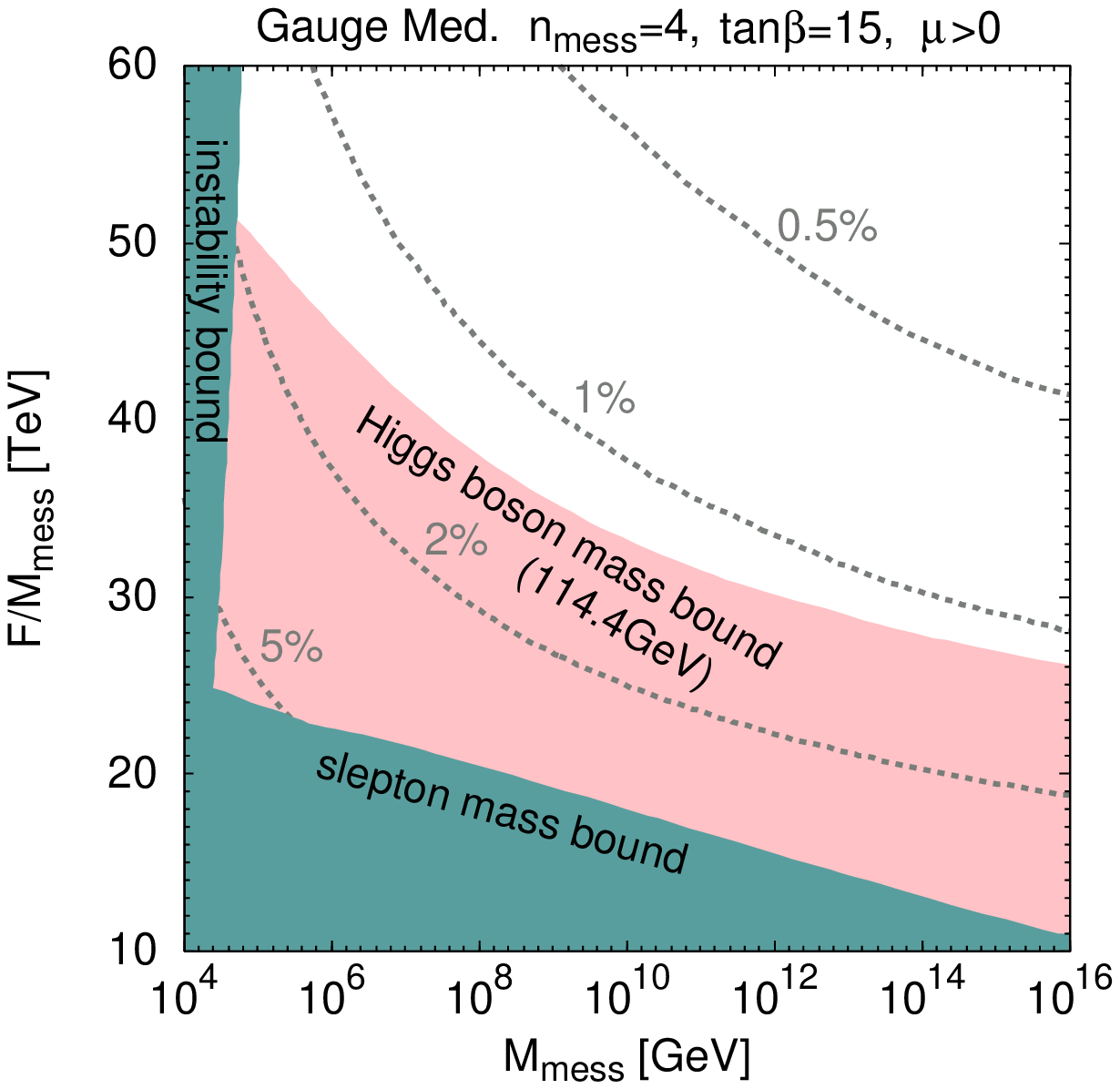}
\end{center}
\caption{Contours of $\Delta^{-1}$ on the $M_{\rm mess}$-$F/M_{\rm mess}$ 
 plane for the minimal gauge mediation models with $n_{\rm mess} = 1$ 
 (left) and $n_{\rm mess} = 4$ (right) pairs of messenger fields in 
 the ${\bf 5} + {\bf 5}^*$ representation.  The instability bound 
 implies the region in which the messenger fields are tachyonic, 
 $F/|M_{\rm mess}|^2 > 1$.  The Higgs sector parameters are fixed 
 as $\tan\beta = 15$ and $\mu > 0$.}
\label{fig:gauge-med}
\end{figure}

If we want to improve fine-tuning further, we must consider 
a theory that gives smaller values of $M_{\rm mess}$, at least 
effectively, because an ultimate reason for the $10\%$ tuning 
in Eq.~(\ref{eq:Delta-max}) is the large logarithm $\ln(M_{\rm 
mess}/m_{\tilde{t}}) \simeq \ln(M_{\rm unif}/m_{\tilde{t}})$. 
An important class of theories giving small $M_{\rm mess}$ is 
gauge mediation models~\cite{Dine:1981gu,Dine:1994vc}.  In these 
models, however, the size of $A$ terms at $M_{\rm mess}$ is 
small, so that $|A_t/m_{\tilde{t}}| \simlt 1$ at the weak 
scale.  This requires large top squark masses to evade the 
Higgs boson mass bound, and thus leads to severe fine-tuning. 
In Fig.~\ref{fig:gauge-med} we plot the contours of $\Delta^{-1}$ 
as a function of $F/M_{\rm mess}$ and $M_{\rm mess}$ for $\mu > 0$, 
in the minimal gauge mediation models with $n_{\rm mess} = 1$ 
and $4$ pairs of messenger fields in the ${\bf 5} + {\bf 5}^*$ 
representation of $SU(5) \supset SU(3)_C \times SU(2)_L \times 
U(1)_Y$.  Here, $M_{\rm mess}$ and $F$ are the supersymmetric 
and supersymmetry breaking masses for the messenger fields, 
respectively.
We find that fine-tuning cannot be better than $2\%$ over the 
entire parameter region.

\subsection{Large {\boldmath $A_t$} and small {\boldmath $\mu B$} 
with low scale {\boldmath $M_{\rm mess}$}}
\label{subsec:low-med}

Discussions in the previous subsections show that the fine-tuning 
problem in minimal supersymmetry is solved if soft supersymmetry 
breaking parameters are generated at a low scale with $|A_t|$ 
substantially (a factor of $\sim 1.5$ or so) larger than 
$m_{\tilde{t}}$.  It is, however, not so easy to achieve this in 
a simple manner.  Suppose, for example, that the superparticles 
obtain masses through direct couplings to the sector that dynamically 
breaks supersymmetry at a scale $\Lambda = O(10~{\rm TeV})$.  This 
will generate both $A$ terms and non-holomorphic supersymmetry 
breaking squared masses, $\tilde{m}^2$, through operators of the 
form $[(Z+Z^\dagger) \Phi^\dagger \Phi]_{\theta^4}$ and $[Z^\dagger 
Z \Phi^\dagger \Phi]_{\theta^4}$, where $Z$ is the supersymmetry 
breaking (composite) chiral superfield and $\Phi$ the quark, 
lepton and Higgs superfields.  Flavor universality of the soft 
masses can be ensured by imposing a flavor symmetry on these 
couplings.  Now, the strengths of these operators can be estimated 
using naive dimensional analysis~\cite{Luty:1997fk}.  We then find 
that the non-holomorphic masses are much larger than the $A$ terms, 
$\tilde{m}^2 \approx (\Lambda/4\pi)^2$ and $A \approx \Lambda/16\pi^2$, 
leading to an unwanted result of $|A_t/m_{\tilde{t}}| \approx 1/4\pi 
\ll 1$.  A possible way of avoiding this is to generate the operators 
given above in a perturbative regime at one and two loops, respectively, 
for example by using messenger-matter mixing in gauge mediation. 
This leads to $M_{\rm mess}$ in the $O(10\!\sim\!100~{\rm TeV})$ 
region, allowing fine-tuning to be relaxed to the $(10\!\sim\!20)\%$ 
level with a modest logarithm of $\ln(M_{\rm mess}/m_{\tilde{t}}) 
\simeq (3\!\sim\!6)$.

\ From the viewpoint of obtaining a large $A$ term at low energies, 
the simplest possibility is to have the operators $[(Z+Z^\dagger) 
\Phi^\dagger \Phi]_{\theta^4}$ (and $[Z^\dagger Z \Phi^\dagger 
\Phi]_{\theta^4}$) at tree level with $O(1)$ coefficients in units 
of the ``cutoff'' scale.  Such a situation arises naturally if 
supersymmetry is broken associated with an extra dimension(s) with 
the size of order $(10~{\rm TeV})^{-1}$~[\ref{Antoniadis:1990ew:X}%
~--~\ref{Barbieri:2002sw:X}] and if matter and/or Higgs fields 
propagate in the extra dimension(s).  Consider that the MSSM 
gauge, quark and lepton superfields propagate in an extra dimension 
compactified on $S^1/Z_2$ with the length $\pi R$.  The two Higgs 
doublets are located on a brane, and the Yukawa couplings and the 
$\mu$ term are introduced there.  Then, if the boundary conditions 
for these fields are twisted by the $SU(2)_R$ symmetry with an angle 
$\alpha$, the theory just below $1/R$ is the MSSM with the soft 
supersymmetry breaking parameters given by~\cite{Barbieri:2001dm}
\begin{equation}
  M_{1,2,3} = \frac{\alpha}{R},
\qquad
  m_{Q,U,D,L,E}^2 = \left( \frac{\alpha}{R} \right)^2,
\qquad
  A_{u,d,e} = -2 \frac{\alpha}{R},
\\
\label{eq:SS-gauge-matter}
\end{equation}
\begin{equation}
  m_{H_u,H_d}^2 = 0,
\qquad
  \mu B = 0,
\label{eq:SS-Higgs}
\end{equation}
where $M_{1,2,3}$ are the gaugino masses, and $m_{Q,U,D,L,E}^2$ 
and $A_{u,d,e}$ the flavor universal squark and slepton masses 
and $A$ terms, respectively.  Taking $\alpha/R$ to be a few hundred 
GeV and $1/R = O(10~{\rm TeV})$, this gives a perfect spectrum 
for electroweak symmetry breaking: the messenger scale is low, 
$M_{\rm mess} \simeq 1/R$, $|A_t/m_{\tilde{t}}|$ is slightly smaller 
than 2 at the weak scale, and there is no strong hierarchy between 
the colored and non-colored superparticles.  The origin of the small 
twist, $\alpha \approx (0.01\!\sim\!0.1)$, will lie in the dynamics 
of radius stabilization, as the $SU(2)_R$ twist in boundary conditions 
is equivalent to the supersymmetry breaking VEV in the radion 
supermultiplet~\cite{Marti:2001iw}.  A trade-off of this theory 
is that we lose a conventional picture of the supersymmetric desert, 
and thus a simple understanding of successful supersymmetric gauge 
coupling unification~\cite{Dimopoulos:1981zb}, although it might 
arise, for example, through some conformal property above $1/R$ 
with the conformality violation effect associated somehow with 
the zero-mode representations.%
\footnote{In such a scenario, the observed differences of the three 
low-energy gauge couplings should arise mainly from the differences 
of the gauge couplings in the bulk, and not from operators localized 
on a brane(s).  Otherwise, the soft supersymmetry breaking parameters 
in Eq.~(\ref{eq:SS-gauge-matter}) would receive large corrections 
from brane operators, and the colored superparticles would become 
much heavier than the non-colored ones, leading to the pattern which 
is not desirable in terms of electroweak symmetry breaking.}
In our view, the virtue of supersymmetric theories with a TeV extra 
dimension(s) lies in the fact that we can easily obtain large $A$ 
terms at low scales.  While the accommodation of the desert is 
nontrivial, we think that these theories provide a much more natural 
solution to the so-called little hierarchy problem~\cite{Barbieri:2000gf} 
compared with any other non-supersymmetric theories.

It is possible to obtain a picture similar to the one described 
above without losing the conventional supersymmetric desert. 
In a theory where the moduli~\cite{Ibanez:1992hc} and anomaly 
mediated~\cite{Randall:1998uk} contributions to supersymmetry breaking 
are comparable~[\ref{Choi:2004sx:X}~--~\ref{Falkowski:2005ck:X}], 
the mediation scale of supersymmetry breaking, $M_{\rm mess}$, 
can be effectively lowered without having a real physical threshold 
at $M_{\rm mess}$~\cite{Choi:2005uz}.  The soft supersymmetry 
breaking parameters at $M_{\rm mess}$ are then given essentially 
by those of boundary condition supersymmetry breaking or (equivalently) 
moduli mediated supersymmetry breaking.  (For a simple understanding 
of this property, see the Appendix.)  One of the challenges to make 
a natural theory using this property is to have a sufficiently small 
$\mu B$ term to obtain a sufficiently large $\tan\beta$.  Because 
of a large gravitino mass required to employ anomaly mediation, 
it is rather difficult to achieve the desired level of a (very) 
small $\mu B$ term.

A model having all the desired features to have natural electroweak 
symmetry breaking keeping the supersymmetric desert has been given 
in Refs.~\cite{Kitano:2005wc,Choi:2005hd}.  The effective messenger 
scale, $M_{\rm mess}$, is lowered to the TeV region, a large $A_t$ 
term with $A_t/m_{\tilde{t}} \sim -1.4$ is obtained, and a large 
enough $\tan\beta$, $\tan\beta \simgt 5$, is accommodated by making 
$\mu B$ small due to the renormalization group focusing effect 
and the elimination of the classical contribution.  In the minimal 
setup, the soft supersymmetry breaking parameters similar to those 
of Eqs.~(\ref{eq:SS-gauge-matter},~\ref{eq:SS-Higgs}) are obtained 
at $M_{\rm mess} = O({\rm TeV})$:
\begin{equation}
  M_{1,2,3} = M_0,
\qquad
  m_{Q,U,D,L,E}^2 = \frac{M_0^2}{2},
\qquad
  A_{u,d,e} = -M_0,
\\
\label{eq:MA-gauge-matter}
\end{equation}
\begin{equation}
  m_{H_u,H_d}^2 = 0,
\qquad
  \mu B = 0,
\label{eq:MA-Higgs}
\end{equation}
where $M_0$ is a parameter of order a few hundred GeV.%
\footnote{Note that the sign convention for the supersymmetry 
breaking masses adopted here, i.e. that of Ref.~\cite{Skands:2003cj}, 
is different from those in Refs.~[\ref{Kitano:2005wc:X}~--%
~\ref{Kitano:2005ew:X}].  In particular, the sign of the $A$ 
terms is opposite.}
Depending on the mechanism of $\mu$ and $\mu B$ term generation, 
the parameter $B \equiv \mu B/\mu$ may have a non-zero value 
of order $M_0/4\pi$.  As shown in Ref.~\cite{Kitano:2005wc}, 
this theory does not suffer from fine-tuning, i.e. $\Delta^{-1} 
\geq 20\%$, and the region giving $\Delta^{-1} \geq 20\%$ is 
consistent with the experimental constraints, such as the one 
from $b \rightarrow s \gamma$, as long as the sign of $\mu$ is 
positive~\cite{Kitano:2005ew}.%
\footnote{It is interesting to point out that the sign of $\mu$ 
is determined to be positive in the minimal setup, $\mu B = 0$ 
at $M_{\rm mess}$, as long as $M_{\rm mess}$ is larger than the 
weak scale.}
The essential point, again, is to generate a large $|A_t/m_{\tilde{t}}|$ 
and moderately large $\tan\beta$ with a small (effective) messenger 
scale, $M_{\rm mess} \sim {\rm TeV}$.  Note that small effective 
$M_{\rm mess}$ is achieved here by having special relations among 
various supersymmetry breaking parameters at $M_{\rm unif}$, which 
allows us to evade the general result of Eq.~(\ref{eq:Delta-max}). 
The top squarks should be light to eliminate fine-tuning, although 
they do not have to be as light as in the case of high scale 
supersymmetry breaking, because of rather small $M_{\rm mess}$. 
A detailed study of the LHC signatures in this particular model 
will be given in section~\ref{sec:model}.

\subsection{Characteristic spectra: light top squarks and 
light Higgsinos}
\label{subsec:spectra}

Considerations so far have highlighted certain generic features 
for a superparticle spectrum that leads to natural electroweak 
symmetry breaking in minimal supersymmetric theories.  These 
have been obtained by considering mainly the tension between the 
$m_{H_u}^2|_{\rm rad}$ term in Eq.~(\ref{eq:mh2}) and the Higgs 
boson mass bound from LEP~II.  Another important constraint on the 
spectrum comes from the $|\mu|^2$ term in Eq.~(\ref{eq:mh2}).  The 
fine-tuning arising from this term is about $M_{\rm Higgs}^2/2 |\mu|^2$, 
so that requiring $\Delta^{-1} \geq 20\%$ ($10\%$) leads to the bound 
$|\mu| \simlt 190~{\rm GeV}$ ($270~{\rm GeV}$).  This implies that 
the Higgsinos should be much lighter than in typical mSUGRA or gauge 
mediation models, where relatively large top squark masses lead to 
a large $\mu$ parameter.

Taking all these together, we find that a supersymmetric theory 
that has the minimal matter content and reproduces naturally 
the correct scale for electroweak symmetry breaking should have 
the following properties for the superparticle spectrum:
\begin{itemize}
\item
The $A$ term for the top squarks is large, $|A_t/m_{\tilde{t}}| 
\approx (1.5\!\sim\!2.5)$ at the weak scale.  This leads to a large 
mass splitting between the two top squarks:
\begin{equation}
  m_{\tilde{t}_2} - m_{\tilde{t}_1} \approx (1.5\!\sim\!2.5)\, m_t.
\label{eq:stop-split}
\end{equation}
Here, we have assumed $m_{Q_3}^2 \approx m_{U_3}^2$.  The splitting 
as small as $\simeq m_t$, however, may be allowed if $M_{\rm mess}$ 
is small, $M_{\rm mess} = O({\rm TeV})$.
\item
The top squarks should be light, i.e. $m_{\tilde{t}} \equiv 
(m_{Q_3}^2 m_{U_3}^2)^{1/4}$ should be small, to reduce the sensitivity 
of $v$ to $y_t$.  How small $m_{\tilde{t}}$ should be depends on the 
value of $M_{\rm mess}$ and the amount of $\Delta^{-1}$ required. 
For the case of high scale supersymmetry breaking with $\Delta^{-1} 
\approx 10\%$, the bound is very strong, $m_{\tilde{t}} \simlt 
300~{\rm GeV}$, leading to $m_{\tilde{t}_1} \sim 100~{\rm GeV}$. 
\item
Small values for the top squark masses imply that we cannot push 
up the Higgs boson mass much larger than the tree-level value. 
Typically, we find
\begin{equation}
  M_{\rm Higgs} \simlt 120~{\rm GeV}.
\label{eq:Higgs}
\end{equation}
\item
The ratio of the electroweak VEVs should also be moderately large
\begin{equation}
  \tan\beta \simgt 5,
\label{eq:tan-beta}
\end{equation}
to have a sufficiently large Higgs boson mass at tree level.  This 
implies that the $\mu B$ term should be (significantly) smaller than 
$2|\mu|^2 + m_{H_u}^2 + m_{H_d}^2$ at the weak scale.
\item
The $\mu$ parameter should be small
\begin{equation}
  |\mu| \simlt 190~{\rm GeV}\,\, (270~{\rm GeV}),
\label{eq:mu}
\end{equation}
for $\Delta^{-1} \geq 20\%$ ($10\%$).  This leads to light Higgsinos, 
which may be the LSP, or if not, may significantly mix with the LSP. 
\end{itemize}

\begin{figure}[t]
\begin{center} 
\begin{picture}(450,195)(-12,-38)
%
%
  \Text(100,-20)[t]{\Large (a)}
  \Line(5,0)(200,0)
  \LongArrow(10,-10)(10,150)
  \Text(-12,157)[b]{mass} \Text(-12,153)[t]{[GeV]}
  \Text(0,0)[r]{$0$}
  \Line(8,20)(12,20)   
  \Line(8,40)(12,40)   \Text(4,40)[r]{$200$}
  \Line(8,60)(12,60)   
  \Line(8,80)(12,80)   \Text(4,80)[r]{$400$}
  \Line(8,100)(12,100) 
  \Line(8,120)(12,120) \Text(4,120)[r]{$600$}
  \Line(8,140)(12,140) 
  \DashLine(10,90)(56,90){3} \LongArrow(33,90)(33,83)
  \Line(20,78)(46,78) \Text(49,78)[l]{$\tilde{t}_2$}
  \Line(20,24)(46,24) \Text(49,24)[l]{$\tilde{t}_1$}
  \DashLine(10,54)(101,54){3} \LongArrow(78,54)(78,47)
  \Line(65,40)(91,40)
  \Line(65,37)(91,37) \Text(93,37)[l]{$\tilde{h}^{0,+}$}
  \Line(65,31)(91,31)
  \Line(110,88)(136,88) \Text(139,88)[l]{$\tilde{g}$}
  \DashLine(121,90)(121,99){1} \LongArrow(121,98)(121,99)
  \DashLine(121,86)(121,77){1} \LongArrow(121,78)(121,77)
  \Line(110,31)(136,31) \Line(110,28)(136,28) \Text(139,31)[l]{$\tilde{W}$}
  \Line(110,16)(136,16) \Text(139,16)[l]{$\tilde{B}$}
  \DashLine(125,33)(125,42){1} \LongArrow(125,41)(125,42)
  \DashLine(125,14)(125,5){1} \LongArrow(125,6)(125,5)
  \Line(155,86)(181,86) \Line(155,82)(181,82)
  \Line(155,78)(181,78) \Text(185,83)[l]{$\tilde{q}$}
  \Line(155,65)(181,65) \Text(184,65)[l]{$\tilde{b}_L$}
  \DashLine(166,88)(166,97){1} \LongArrow(166,96)(166,97)
  \DashLine(166,63)(166,54){1} \LongArrow(166,55)(166,54)
  \Line(155,40)(181,40)
  \Line(155,30)(181,30) \Text(185,36)[l]{$\tilde{l}$}
  \DashLine(170,42)(170,51){1} \LongArrow(170,50)(170,51)
  \DashLine(170,28)(170,19){1} \LongArrow(170,20)(170,19)
%
%
  \Text(350,-20)[t]{\Large (b)}
  \Line(255,0)(450,0)
  \LongArrow(260,-10)(260,150)
  \Text(238,157)[b]{mass} \Text(238,153)[t]{[GeV]}
  \Text(250,0)[r]{$0$}
  \Line(258,20)(262,20)   
  \Line(258,40)(262,40)   \Text(254,40)[r]{$200$}
  \Line(258,60)(262,60)   
  \Line(258,80)(262,80)   \Text(254,80)[r]{$400$}
  \Line(258,100)(262,100) 
  \Line(258,120)(262,120) \Text(254,120)[r]{$600$}
  \Line(258,140)(262,140) 
  \DashLine(260,146)(306,146){3} \LongArrow(283,146)(283,126)
  \Line(270,107)(296,107) \Text(299,107)[l]{$\tilde{t}_2$}
  \Line(270,57)(296,57) \Text(299,57)[l]{$\tilde{t}_1$}
  \DashLine(260,38)(351,38){3} \LongArrow(328,38)(328,33)
  \Line(315,32)(341,32)
  \Line(315,31)(341,31) \Text(343,31)[l]{$\tilde{h}^{0,+}$}
  \Line(315,30)(341,30)
  \Line(360,118)(386,118)
  \Line(360,116)(386,116) \Text(389,116)[l]{$\tilde{g},\tilde{W},\tilde{B}$}
  \Line(360,114)(386,114)
  \DashLine(375,120)(375,140){1} \LongArrow(375,139)(375,140)
  \DashLine(375,112)(375,92){1} \LongArrow(375,93)(375,92)
  \Line(405,84)(431,84) \Line(405,83)(431,83)
  \Line(405,82)(431,82) \Text(435,83)[l]{$\tilde{q},\tilde{l}$}
  \Line(405,81)(431,81) \Line(405,80)(431,80)
  \DashLine(420,86)(420,106){1} \LongArrow(420,105)(420,106)
  \DashLine(420,78)(420,58){1} \LongArrow(420,59)(420,58)
\end{picture}
\caption{Characteristic spectra for the superparticles which give the 
 correct scale for electroweak symmetry breaking without significant 
 fine-tuning.  The spectrum in (a) arises typically in a high scale 
 supersymmetry breaking scenario with $\Delta^{-1} \approx 10\%$, 
 while that in (b) arises in a theory where supersymmetry is broken 
 by boundary conditions, or moduli contributions, with small 
 (effective) $M_{\rm mess}$.}
\label{fig:spectra}
\end{center}
\end{figure}
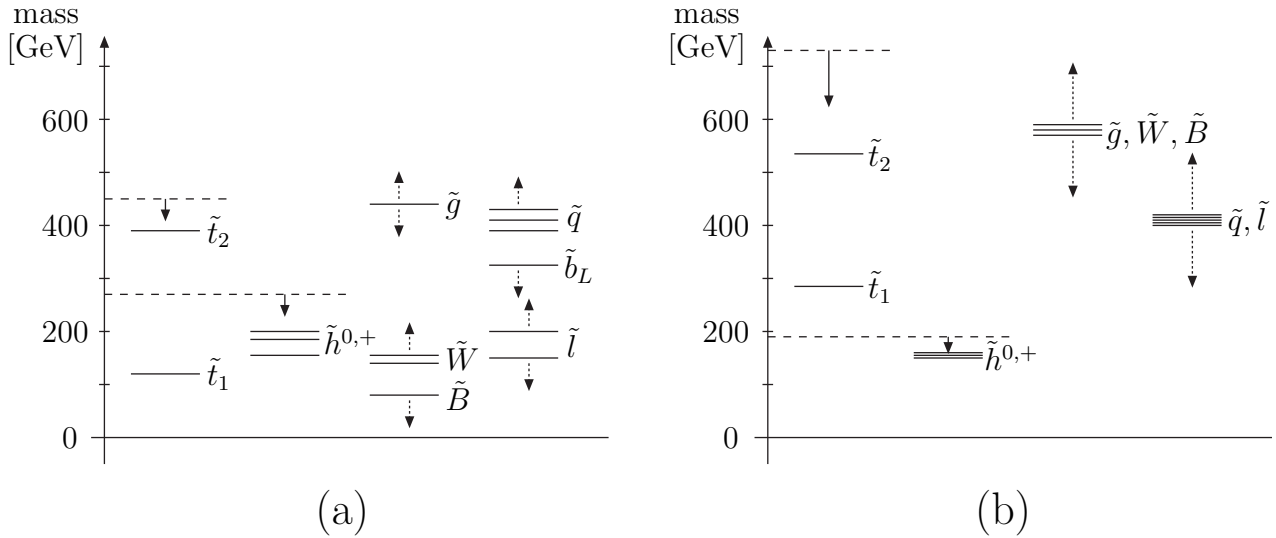

The features described above still leave several possible patterns 
for the superparticle spectrum, which can lead to somewhat different 
situations.  In Fig.~\ref{fig:spectra} we depict possible patterns 
which are representative for generic cases.
In Fig.~\ref{fig:spectra}(a), we depict a pattern that typically arises 
in a high scale supersymmetry breaking scenario with $\Delta^{-1} 
\approx 10\%$, for example in the mSUGRA model with $m_{H_u}^2, m_{H_d}^2 
\neq m_0^2$ (see Fig.~\ref{fig:Mmess=Munif}).  In this situation the top 
squarks are rather light, $m_{\tilde{t}} = (m_{Q_3}^2 m_{U_3}^2)^{1/4} 
\simlt 300~{\rm GeV}$, with the light top squark close to its experimental 
bound, $m_{\tilde{t}_1} \simeq 100~{\rm GeV}$.  The Higgsinos, $\tilde{h}$, 
are also light, although $\Delta^{-1} \approx 10\%$ allows $m_{\tilde{h}}$ 
as large as $\approx 270~{\rm GeV}$.  In principle there are little 
constraints on the other gaugino and sfermion masses, except that 
the gluino should be heavier than about $450~{\rm GeV}$ and that the 
squarks cannot be much lighter than the gluino.  The universal gaugino 
mass relation, $M_3/g_3^2 = M_2/g_2^2 = M_1/g_1^2$, may or may not be 
satisfied.  The LSP will be either bino-like, wino-like, Higgsino-like, 
or a mixture of these states.  If the LSP consists mainly of the bino, 
its thermal relic abundance can give the correct dark matter density 
either through a large mixture with the neutral Higgsinos and/or wino, 
coannihilation with the stau or stop, or resonant annihilation through 
$s$-channel Higgs boson exchange.  Note that some of these options are 
not available in typical mSUGRA points, in which top squarks are heavy 
and thus the $\mu$ parameter is large.  For other cases of wino-like 
and Higgsino-like LSPs, the production should be nonthermal.  It is 
interesting that a large $A_t$ term also makes it easier to satisfy 
the constraints from the precision electroweak data with small top 
squark masses~\cite{Cho:1999km}.  It is important to perform detailed 
LHC studies of this class of spectra, although a large ambiguity for 
the gaugino and sfermion masses will make a thorough study of the 
parameter space somewhat complicated.  These spectra will also be 
interesting for a future $e^+ e^-$ linear collider~\cite{Kitano:2002ss}.

If we want to reduce fine-tuning further, for example to eliminate 
it altogether ($\Delta^{-1} \geq 20\%$), we must generate the soft 
supersymmetry breaking parameters at low energies.  Because of a small 
logarithm, $\ln(M_{\rm mess}/m_{\tilde{t}})$, constraints on the top 
squark sector is slightly weaker in this case: $|A_t/m_{\tilde{t}}| 
\simgt O(1)$ and $m_{\tilde{t}} \simlt O(700~{\rm GeV})$.  On the other 
hand, the constraint on the $\mu$ parameter is stronger because of 
a stronger requirement on $\Delta^{-1}$: for $\Delta^{-1} \geq 20\%$, 
we obtain the bound on the Higgsino masses $m_{\tilde{h}} \simlt 
190~{\rm GeV}$.  In fact, generating these spectra with small $M_{\rm 
mess}$ is nontrivial, and one of the ways is to adopt (effectively) 
the scheme of boundary condition, or moduli, supersymmetry breaking 
at low energies.  This generically leads to a rather ordered spectrum 
at low energy, e.g. universal gaugino and sfermion masses at the weak 
scale.  This situation is depicted in Fig.~\ref{fig:spectra}(b), 
where almost degenerate gaugino and sfermion masses are assumed.  The 
relation between the gaugino and sfermion masses is model dependent. 
An important implication of this class of spectra is that the LSP 
is one of the neutral Higgsinos, unless the gravitino is lighter. 
To identify this LSP to be the dark matter, it must be produced 
nonthermally, for example, as in~\cite{Moroi:1999zb}.  As discussed 
in~\cite{Kitano:2005ew}, such dark matter can have an interesting 
implication on direct dark matter detection experiments such 
as CDMS~II. 

In the rest of the paper, we focus on studying LHC signatures for 
the latter class of spectra, given in Fig.~\ref{fig:spectra}(b). 
We do so partly because it gives milder (or no) fine-tuning, and 
partly because there is little LHC study directly related to this case. 
In particular, we first focus on signals expected from the Higgsino 
LSP at the LHC, and discuss under what conditions the signals are 
most useful.  We then demonstrate in section~\ref{sec:model} that 
the signals can indeed be used in realistic analyses, using the 
explicit model discussed at the end of subsection~\ref{subsec:low-med}. 
The determination of model parameters are also discussed there, which 
may be useful to discriminate between various possible models.

\section{Higgsino LSP at the LHC}
\label{sec:higgsino}

We have seen that in a large class of theories where the supersymmetric 
fine-tuning is solved, the Higgsinos are the lightest among the 
superpartners of the standard model particles.  A characteristic 
pattern for the superparticle spectrum in these theories is depicted 
in Fig.~\ref{fig:spectra}(b).  The three gauginos are almost degenerate 
at the weak scale, as well as the squarks and sleptons.  The ratio 
of the gaugino and the sfermion masses is model dependent.  In this 
section, we identify characteristic signals of these spectra at 
the LHC, and discuss under what conditions the signals are useful 
in realistic analyses.  We assume throughout that the gravitino 
is not lighter than the Higgsinos, so that the LSP is the lightest 
neutral Higgsino.%
\footnote{In fact, none of our analyses changes unless the gravitino 
mass, $m_{3/2}$, is smaller than $O(1\!\sim\!10~{\rm keV})$ because 
the lightest Higgsino then lives long enough so that it can be treated 
as a stable particle for collider purposes.  For a smaller gravitino 
mass, the lightest Higgsino decays into a Higgs boson and a gravitino, 
followed by the Higgs boson decay $h \rightarrow b\bar{b}$.  This 
can be used to measure the Higgs boson mass, e.g., by selecting 
four $b$-jet events and plotting $M_{bb}$ invariant masses.  For 
$m_{3/2} = O(0.01\!\sim\!1~{\rm keV})$, we may also have displaced 
$b\bar{b}$ vertex signals.}

\subsection{Dilepton invariant mass distribution from {\boldmath 
$\tilde{\chi}^0_2 \rightarrow \tilde{\chi}^0_1\, l^+ l^-$} in the 
Higgsino LSP scenario}
\label{subsec:dilepton}

An important feature of the spectra depicted in Fig.~\ref{fig:spectra}(b) 
is that there are three almost degenerate neutralino/chargino states, 
$\tilde{\chi}^0_1$, $\tilde{\chi}^0_2$ and $\tilde{\chi}^+_1$, with 
the masses $\approx |\mu| \simlt 190~{\rm GeV}$.  If the gaugino masses 
are sufficiently larger than $|\mu|$, which we assume to be the case, 
these states are almost purely the Higgsinos, with the mass splittings 
given by
\begin{eqnarray}
  m_{\tilde{\chi}^0_2} - m_{\tilde{\chi}^0_1} &\simeq& 
    m_Z^2 \biggl( \frac{\cos^2\!\theta_W}{M_2} 
    + \frac{\sin^2\!\theta_W}{M_1} \biggr),
\label{eq:chi0_2-chi0_1} \\
  m_{\tilde{\chi}^+_1} - m_{\tilde{\chi}^0_1} &\simeq& 
    \frac{m_Z^2}{2} \biggl( \frac{\cos^2\!\theta_W}{M_2} 
    + \frac{\sin^2\!\theta_W}{M_1} \biggr),
\label{eq:chi+_1-chi0_1}
\end{eqnarray}
where we have assumed a moderately large $\tan\beta$, e.g. $\tan\beta 
\simgt 5$, $\theta_W$ is the Weinberg angle, and $M_1$ and $M_2$ 
are the $U(1)_Y$ and $SU(2)_L$ gaugino mass parameters, respectively. 
This implies that $\tilde{\chi}^0_2$ and $\tilde{\chi}^+_1$ 
undergo three-body decays to $\tilde{\chi}^0_1$.  In particular, 
$\tilde{\chi}^0_2$ has the leptonic decay mode
\begin{equation}
  \tilde{\chi}^0_2 \rightarrow \tilde{\chi}^0_1\, l^+ l^-.
\label{eq:chi0_2-decay}
\end{equation}
At hadron colliders, this decay mode can give important 
information on the properties of the initial- and final-state 
neutralinos~[\ref{Hinchliffe:1996iu:X},~\ref{Baer:1986vf:X}~--~%
\ref{Nojiri:1999ki:X}].  Below we show that the dilepton arising 
from the decay of Eq.~(\ref{eq:chi0_2-decay}) can provide an 
important test for the Higgsino nature of the lightest two 
neutralinos $\tilde{\chi}^0_1$ and $\tilde{\chi}^0_2$.

\begin{figure}[t]
\begin{center}
\begin{picture}(450,65)(0,0)
  \Line(13,50)(60,50)  \Text(10,50)[r]{$\tilde{\chi}^0_2$}
  \Line(60,50)(115,65) \Text(118,65)[l]{$\tilde{\chi}^0_1$}
  \Photon(60,50)(70,23){3}{4} \Text(58,35)[tr]{$Z$}
  \Line(70,23)(115,35) \Text(119,35)[l]{$l^+$}
  \Line(70,23)(115,10) \Text(119,10)[l]{$l^-$}
  \Line(173,50)(220,50) \Text(170,50)[r]{$\tilde{\chi}^0_2$}
  \Line(220,50)(275,65) \Text(279,65)[l]{$l^+$}
  \DashLine(220,50)(230,23){2} \Text(223,39)[tr]{$\tilde{l}_{L,R}$}
  \Line(230,23)(275,35) \Text(279,35)[l]{$l^-$}
  \Line(230,23)(275,10) \Text(278,10)[l]{$\tilde{\chi}^0_1$}
  \Line(333,50)(380,50) \Text(330,50)[r]{$\tilde{\chi}^0_2$}
  \Line(380,50)(435,65) \Text(439,65)[l]{$l^-$}
  \DashLine(380,50)(390,23){2} \Text(383,39)[tr]{$\tilde{l}_{L,R}$}
  \Line(390,23)(435,35) \Text(439,35)[l]{$l^+$}
  \Line(390,23)(435,10) \Text(438,10)[l]{$\tilde{\chi}^0_1$}
\end{picture}
\caption{The diagrams contributing to the $\tilde{\chi}^0_2 \rightarrow 
 \tilde{\chi}^0_1\, l^+ l^-$ decay.} 
\label{fig:3-body}
\end{center}
\end{figure}
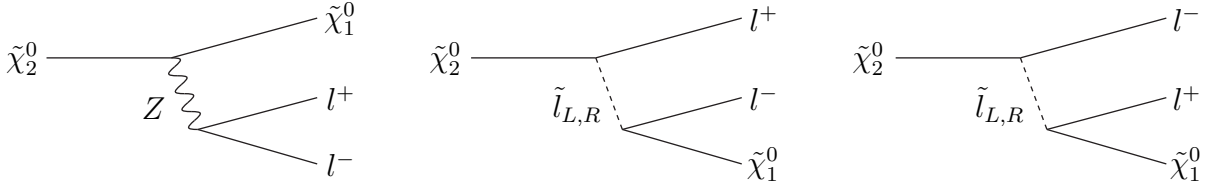

The three-body decay $\tilde{\chi}^0_2 \rightarrow \tilde{\chi}^0_1\, 
l^+ l^-$ occurs through the diagrams shown in Fig.~\ref{fig:3-body}. 
In the limit where the mass difference between the two neutralinos 
is much smaller than the $Z$ boson mass and where the slepton masses 
are much larger than the decaying neutralino, $m_{\tilde{\chi}^0_2} 
- m_{\tilde{\chi}^0_1} \ll m_Z$ and $m_{\tilde{l}_L}, m_{\tilde{l}_R} 
\gg m_{\tilde{\chi}^0_2}$, the effects of these diagrams are described 
by a single low-energy $\tilde{\chi}^0_1 \tilde{\chi}^0_2 l^+ l^-$ 
four-Fermi operator for each chirality of leptons.  This implies 
that the distribution shape of the dilepton invariant mass, 
$M_{ll} \equiv \sqrt{(p_{l^+} + p_{l^-})^2}$, is completely 
determined by the masses of the two neutralinos, $\tilde{\chi}^0_1$ 
and $\tilde{\chi}^0_2$, where $p_{l^+}$ and $p_{l^-}$ are the 
four-momenta of $l^+$ and $l^-$.  Let us now adopt the phase 
convention in which all the mass eigenvalues and the mixing matrix 
elements for the neutralinos are taken to be real.  This basis can 
always be taken as long as there is no $CP$ violating effect in 
the neutralino mass matrix, which we assume throughout.  We then 
obtain the following dilepton invariant mass distribution after 
performing appropriate phase space integrals:
\begin{eqnarray}
  \frac{d\Gamma(\tilde{\chi}^0_2 \rightarrow 
    \tilde{\chi}^0_1\, l^+ l^-)}{d M_{ll}} &\propto& 
    M_{ll}\, \sqrt{(m_{\tilde{\chi}^0_2}^2-m_{\tilde{\chi}^0_1}^2)^2
    -2(m_{\tilde{\chi}^0_1}^2+m_{\tilde{\chi}^0_2}^2)M_{ll}^2+M_{ll}^4}
\nonumber\\
  && \times \Bigl\{ (m_{\tilde{\chi}^0_2}^2-m_{\tilde{\chi}^0_1}^2)^2
    +(m_{\tilde{\chi}^0_1}^2+m_{\tilde{\chi}^0_2}^2)M_{ll}^2-2M_{ll}^4 
    +6\, \eta_\chi\, m_{\tilde{\chi}^0_1} m_{\tilde{\chi}^0_2} M_{ll}^2 \Bigr\},
\label{eq:Mll-distr-1}
\end{eqnarray}
for $0 \leq M_{ll} \leq m_{\tilde{\chi}^0_2}-m_{\tilde{\chi}^0_1}$ and 
$d\Gamma(\tilde{\chi}^0_2 \rightarrow \tilde{\chi}^0_1\, l^+ l^-)/d M_{ll} 
= 0$ for $M_{ll} > m_{\tilde{\chi}^0_2}-m_{\tilde{\chi}^0_1}$.  Here, 
$m_{\tilde{\chi}^0_1} = |M_{\tilde{\chi}^0_1}|$ and $m_{\tilde{\chi}^0_2} 
= |M_{\tilde{\chi}^0_2}|$ are the absolute values for the two 
smallest neutralino mass eigenvalues $M_{\tilde{\chi}^0_1}$ and 
$M_{\tilde{\chi}^0_2}$ with $m_{\tilde{\chi}^0_1} < m_{\tilde{\chi}^0_2}$, 
and $\eta_\chi \equiv {\rm sgn}(M_{\tilde{\chi}^0_1})\, 
{\rm sgn}(M_{\tilde{\chi}^0_2})$ is the relative sign between them. 
In fact, with the LEP~II bound on the Higgsino masses, the assumption 
of $m_{\tilde{\chi}^0_2} - m_{\tilde{\chi}^0_1} \ll m_Z$ implies 
that the two neutralinos are nearly degenerate: $\varDelta m \equiv 
m_{\tilde{\chi}^0_2} - m_{\tilde{\chi}^0_1} \ll m_{\tilde{\chi}^0_1}$. 
The $M_{ll}$ distribution is then further simplifies to
\begin{equation}
  \frac{d\Gamma(\tilde{\chi}^0_2 \rightarrow 
    \tilde{\chi}^0_1\, l^+ l^-)}{d M_{ll}}
  \propto M_{ll} \sqrt{\varDelta m^2-M_{ll}^2}\,
    \Bigl\{ 2 \varDelta m^2 + (1+3\eta_\chi)M_{ll}^2 \Bigr\},
\label{eq:Mll-distr-2}
\end{equation}
for $0 \leq M_{ll} \leq \varDelta m$ and $d\Gamma(\tilde{\chi}^0_2 
\rightarrow \tilde{\chi}^0_1\, l^+ l^-)/d M_{ll} = 0$ for 
$M_{ll} > \varDelta m$.  

There are two important features for the $M_{ll}$ distribution 
in Eq.~(\ref{eq:Mll-distr-2}) which can be used to test the 
Higgsino LSP scenario.  First, the endpoint of the distribution, 
$m_{\tilde{\chi}^0_2}-m_{\tilde{\chi}^0_1}$, is expected to be 
very small:
\begin{equation}
  M_{ll}^{\rm max} = m_{\tilde{\chi}^0_2} - m_{\tilde{\chi}^0_1}
  \simeq \frac{m_Z^2}{M_0} = O(10~{\rm GeV}),
\label{eq:endpoint}
\end{equation}
where we have set $M_0 \equiv M_1 \simeq M_2$.  Given the LEP~II 
bound on the chargino mass, such a small mass splitting between 
$\tilde{\chi}^0_1$ and $\tilde{\chi}^0_2$ cannot arise in a theory 
where the LSP is gaugino-like and the three gauginos respect the 
universal mass relation, $M_3/g_3^2 = M_2/g_2^2 = M_1/g_1^2$. 
Second, the Higgsino LSP necessarily leads to the opposite signs 
between $M_{\tilde{\chi}^0_1}$ and $M_{\tilde{\chi}^0_2}$, so that 
$\eta_\chi = -1$ in Eqs.~(\ref{eq:Mll-distr-1},~\ref{eq:Mll-distr-2}). 
This is because in the gauge eigenbasis the $2 \times 2$ neutral 
Higgsino mass matrix takes a purely off-diagonal form $((0,-\mu),
(-\mu,0))$, which gives one positive and one negative eigenvalues 
after diagonalization including the effects of mixing with the 
gaugino states.  The resulting distribution is quite different 
from the one with $\eta_\chi = +1$, which arises in the case of the 
gaugino-like LSP with ${\rm sgn}(M_1) = {\rm sgn}(M_2)$.

\begin{figure}[t]
\begin{center}
  \includegraphics[width=5.5cm]{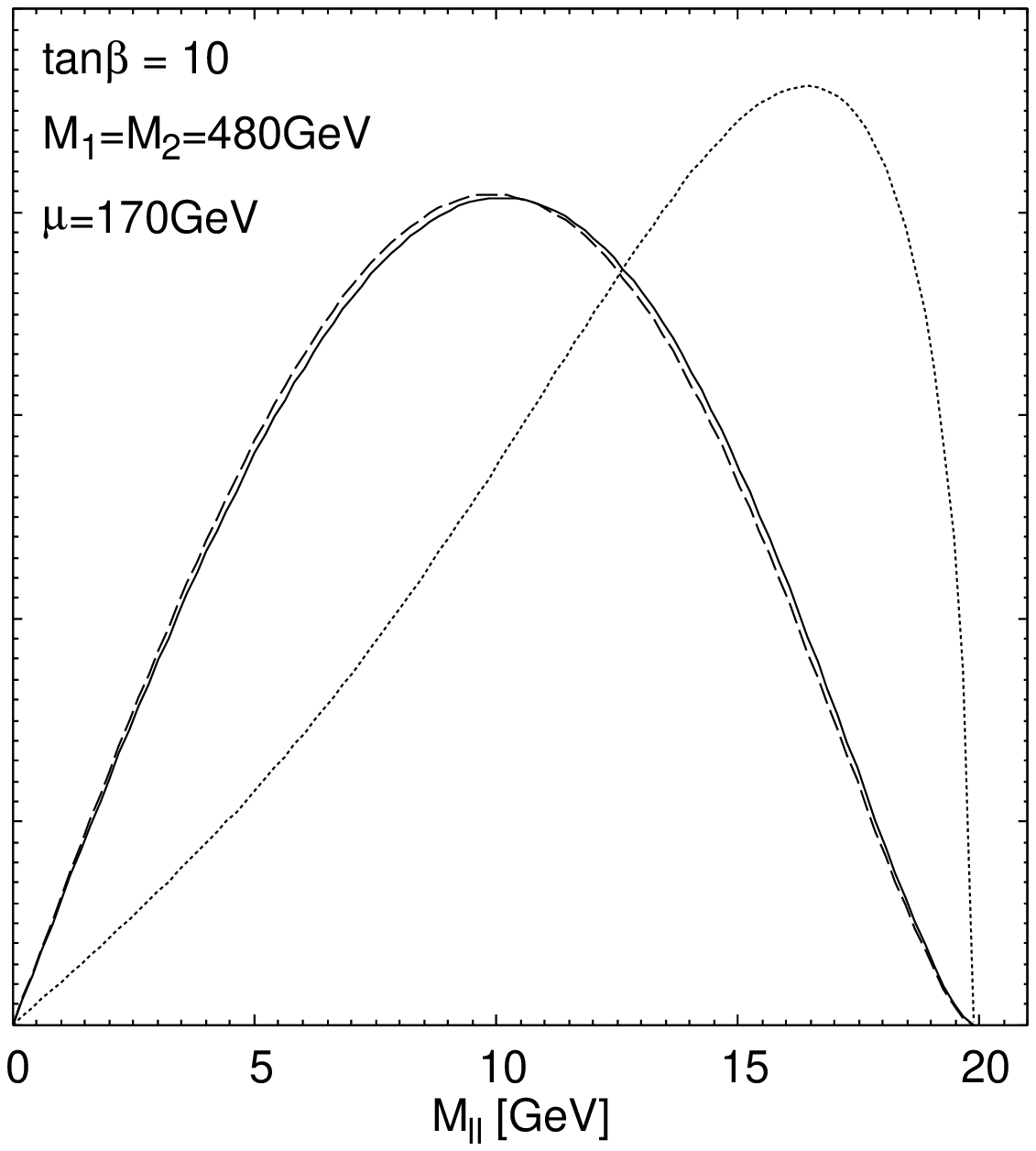}
\hspace{2cm}
  \includegraphics[width=5.5cm]{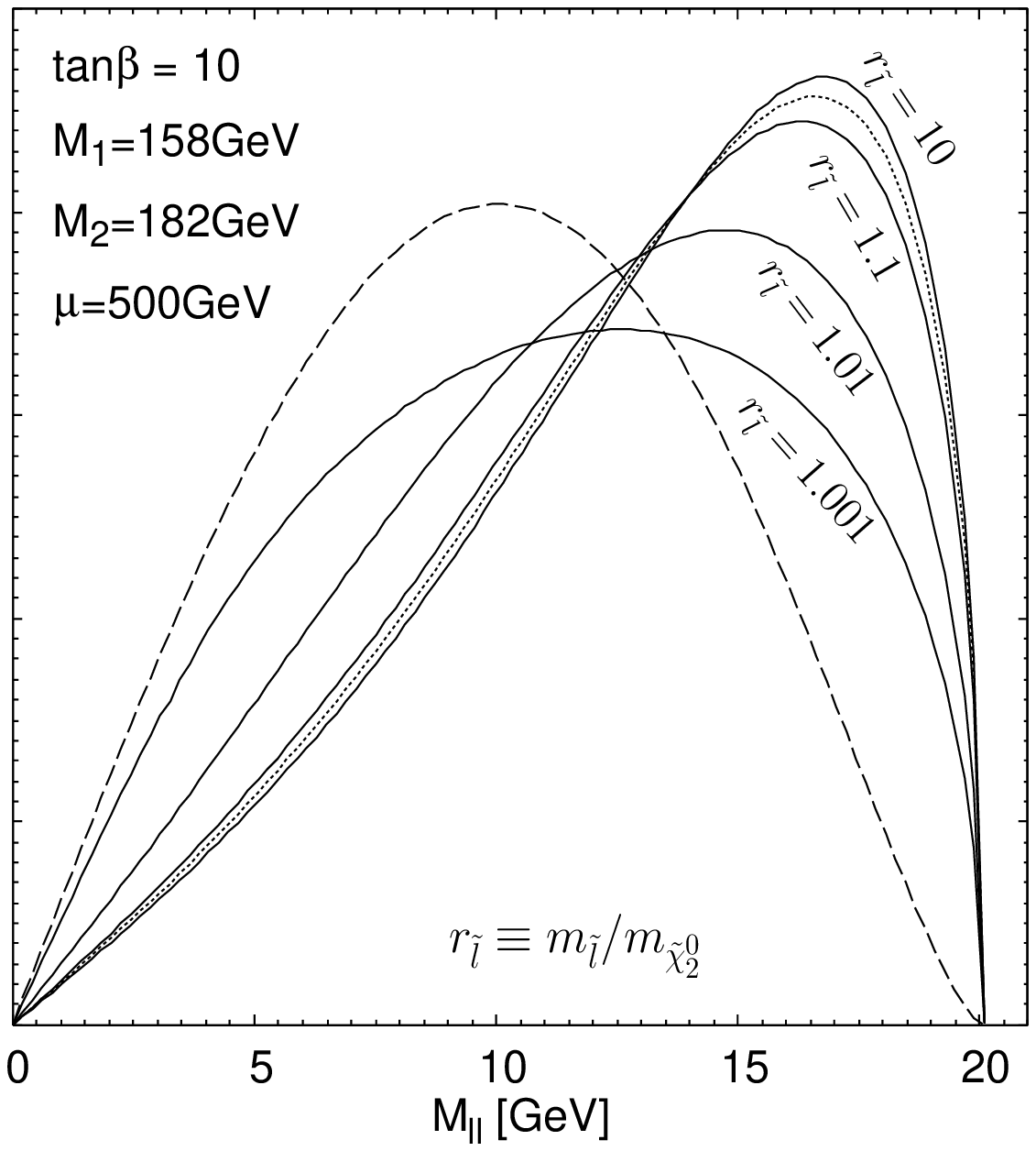}
\end{center}
\caption{Dilepton invariant mass distribution from the $\tilde{\chi}^0_2 
 \rightarrow \tilde{\chi}^0_1\, l^+ l^-$ decay.  Solid lines represent 
 the distributions for the Higgsino LSP case (left panel) and the 
 case of gaugino-like $\tilde{\chi}^0_1$ and $\tilde{\chi}^0_2$ with 
 ${\rm sgn}(M_1) = {\rm sgn}(M_2)$ (right panel).  The curves with dashed 
 and dotted lines represent the ones obtained from Eq.~(\ref{eq:Mll-distr-2}) 
 with $\eta_\chi = -1$ and $+1$, respectively (in both panels).}
\label{fig:Mll}
\end{figure}

In Fig.~\ref{fig:Mll} we plot $M_{ll}$ distributions calculated for 
several different choices for the soft supersymmetry breaking parameters. 
In the left panel, we plot the $M_{ll}$ distribution for the Higgsino 
LSP case by a solid curve.  The curve is drawn using the complete 
expression with $\tan\beta = 10$, $M_1 = M_2 = 480~{\rm GeV}$ 
and $\mu = 170~{\rm GeV}$, which gives $m_{\tilde{\chi}^0_2} 
- m_{\tilde{\chi}^0_1} \simeq 20~{\rm GeV}$.  We have varied 
the slepton masses $m_{\tilde{l}} \equiv m_{\tilde{l}_L} 
= m_{\tilde{l}_R}$ in the range $1.001 \leq r_{\tilde{l}} \equiv 
m_{\tilde{l}}/m_{\tilde{\chi}^0_2} \leq 10$, but it does not lead to any 
visible change of the curve.  In the figure we have also drawn curves 
obtained using the approximate expression of Eq.~(\ref{eq:Mll-distr-2}) 
for $\eta_\chi = -1$ (dashed line) and $\eta_\chi = +1$ (dotted 
line). We find that the expression of Eq.~(\ref{eq:Mll-distr-2}) 
with $\eta_\chi = -1$ well approximates the full result. 
The small discrepancy arises from corrections higher order in 
$(\varDelta m/m_Z)^2$. The right panel shows the $M_{ll}$ distribution 
in the case of gaugino-like $\tilde{\chi}^0_1$ and $\tilde{\chi}^0_2$ 
with ${\rm sgn}(M_1) = {\rm sgn}(M_2)$ (solid lines).  The parameters 
are chosen to be $\tan\beta = 10$ and $\mu = 500~{\rm GeV}$, and $M_1$ 
and $M_2$ are chosen such that the same values of $m_{\tilde{\chi}^0_1}$ 
and $m_{\tilde{\chi}^0_2}$ as in the left panel are obtained: 
$M_1 = 158~{\rm GeV}$ and $M_2 = 182~{\rm GeV}$.  The slepton masses 
are varied as $r_{\tilde{l}} \equiv m_{\tilde{l}}/m_{\tilde{\chi}^0_2} 
= 10, 1.1, 1.01, 1.001$, and we find that the $M_{ll}$ distribution 
in this case depends on $m_{\tilde{l}}$ but only when it is very close 
to the $\tilde{\chi}^0_2$ mass.  As in the left panel, we also draw 
curves obtained from Eq.~(\ref{eq:Mll-distr-2}) with $\eta_\chi = -1$ 
(dashed line) and $\eta_\chi = +1$ (dotted line).  We find that the 
approximate curve with $\eta_\chi = +1$ well reproduces the full 
result with $r_{\tilde{l}} \simgt 1.1$.  

The plots in Fig.~\ref{fig:Mll} clearly show that the Higgsino 
LSP case can be discriminated from the case with $\eta_\chi \equiv 
{\rm sgn}(M_{\tilde{\chi}^0_1})\, {\rm sgn}(M_{\tilde{\chi}^0_2}) 
= +1$, even with the same mass difference $m_{\tilde{\chi}^0_2} 
- m_{\tilde{\chi}^0_1}$, regardless of the values for the 
other supersymmetry breaking parameters such as the slepton 
masses.  In particular, we find that the behavior of the $M_{ll}$ 
distribution near the endpoint is completely different between the 
two cases of $\eta_\chi = -1$ and $+1$.  This can be understood in 
terms of the selection rule for the orbital angular momentum due to 
the $CP$ properties of the two neutralinos $\tilde{\chi}^0_1$ and 
$\tilde{\chi}^0_2$~\cite{Choi:2005gt}.  For $\eta_\chi = +1$ ($-1$), 
the $M_{ll}$ distribution near the kinematical endpoint, which 
corresponds to the limit of slow moving $\tilde{\chi}^0_1$ in the 
$\tilde{\chi}^0_2$ rest frame, should give an $S$-wave ($P$-wave) 
behavior, leading to $M_{ll} \propto (\varDelta m - M_{ll})^{1/2}$ 
($M_{ll} \propto (\varDelta m - M_{ll})^{3/2}$) near the endpoint. 
While gaugino-like $\tilde{\chi}^0_1$ and $\tilde{\chi}^0_2$ could 
potentially give a similar distribution if ${\rm sgn}(M_1) = 
-{\rm sgn}(M_2)$, the shape of the $M_{ll}$ distribution together 
with the smallness of the endpoint can provide a powerful tool to 
test the Higgsino LSP scenario considered here, which necessarily 
leads to $\varDelta m \ll m_Z$ and $\eta_\chi = -1$.%
\footnote{It is interesting to point out that the signatures 
discussed here arise only from the fact that the neutral Higgsinos 
are pseudo-Dirac fermions.  The same technique, therefore, can also 
be used to test the idea of pseudo-Dirac gauginos, depending on the 
spectrum for the other superparticles (for examples of theories 
giving pseudo-Dirac gauginos, see~\cite{Hall:1990hq}).  For instance, 
if the LSP is one of the pseudo-Dirac bino or wino states, similar 
signatures may arise in the dilepton invariant mass distribution, 
depending on the existence of other sources of leptons and/or 
patterns of cascade decays for the superparticles.}

To demonstrate that the signatures discussed above are really useful 
at the LHC, we must check that there are no other leptons from the 
superparticle cascade decays which bury the signatures.  We must 
also show that the shape of the $M_{ll}$ distribution from the 
$\tilde{\chi}^0_2 \rightarrow \tilde{\chi}^0_1\, l^+ l^-$ decay 
is preserved under selection cuts in the analysis at a level that 
different shapes for $\eta_\chi = +1$ and $-1$ can be discriminated 
in a realistic detector.  We address the first issue in the next 
subsection, in the context of a class of theories discussed in 
subsections~\ref{subsec:low-med} and \ref{subsec:spectra}.  The 
second issue will be addressed in section~\ref{sec:model}, where 
we explicitly demonstrate that the $M_{ll}$ distribution can indeed 
be used to test the Higgsino LSP scenario, as well as to extract 
the information on the neutralino masses, by performing Monte Carlo 
simulations using a specific theory.

\subsection{Higgsino LSP with quasi-degenerate gauginos and sfermions}
\label{subsec:cascade}

Let us consider the pattern of the superparticle masses depicted 
in Fig.~\ref{fig:spectra}(b), which can naturally arise in a theory 
where moduli-type, or boundary condition, supersymmetry breaking is 
employed with small $M_{\rm mess}$.  Specifically, we consider the 
following spectrum.  The three gauginos are almost degenerate at 
the weak scale with the masses denoted by $m_{\tilde{g}}$.  The squarks 
and sleptons are also nearly degenerate with the masses denoted by 
$m_{\tilde{q}}$, although the two top squarks, $\tilde{t}_1$ and 
$\tilde{t}_2$, can have substantially different masses because of 
a large mass splitting due to large $A_t$.  The mass splittings among 
the gauginos and among different squarks and sleptons arise at higher 
order, but they are expected to be small and of $O(10\%)$.  The $\mu$ 
parameter is smaller than about $200~{\rm GeV}$, so that it is 
smaller than both $m_{\tilde{g}}$ and $m_{\tilde{q}}$.

We first consider the case with $m_{\tilde{q}} < m_{\tilde{g}}$. 
In this case, a squark cannot decay into a gluino, so that it decays 
as $\tilde{q} \rightarrow \tilde{\chi}^+_1 q'$, $\tilde{\chi}^0_1 q$ 
or $\tilde{\chi}^0_2 q$.  Here, $q$ and $q'$ represent quarks having 
the same and different flavors with $\tilde{q}$, and the lightest 
chargino, $\tilde{\chi}^+_1$, and the lightest two neutralinos, 
$\tilde{\chi}^0_{1,2}$, are the charged and neutral Higgsinos, 
respectively, with small mixings with the gaugino states.  On the 
other hand, the gluino, once produced, decays into a quark and 
a squark $g \rightarrow q \tilde{q}$, followed by squark decay 
discussed above.  In these decay chains, leptons arise only from 
decays of $\tilde{\chi}^+_1$ and $\tilde{\chi}^0_2$: $\tilde{\chi}^+_1 
\rightarrow \tilde{\chi}^0_1\, l^+ \nu$ and $\tilde{\chi}^0_2 \rightarrow 
\tilde{\chi}^0_1\, l^+ l^-$.  In the $M_{ll}$ distribution analysis, 
we select events having two and only two leptons with the same 
flavor and opposite charge.  The number of dileptons arising from 
$\tilde{\chi}^+_1$ decays is then small over the relevant energy 
region of $M_{ll} = O(10~{\rm GeV})$, compared with that from 
$\tilde{\chi}^0_2$ decay.  (The background from $\tilde{\chi}^+_1$ 
decays can actually be estimated using opposite-sign opposite-flavor 
leptons and thus subtracted using the combination $e^+e^- + 
\mu^+\mu^- - e^+\mu^- - \mu^+e^-$.)  The only remaining issue 
then is the squark branching ratio into $\tilde{\chi}^0_2$ and the 
$\tilde{\chi}^0_2$ branching ratio into leptons.  In the parameter 
region we consider, the former is typically of $O(10\%)$ and 
the latter is $\approx 3\%$ for both $e^+e^-$ and $\mu^+\mu^-$ 
modes, which are large enough to produce an appreciable number 
of dilepton events.%
\footnote{Strictly speaking, opposite-sign same-flavor dileptons 
may also arise from the three-body decay of the gluino $\tilde{g} 
\rightarrow \tilde{W} q q$ followed by the wino decay $\tilde{W} 
\rightarrow l \tilde{l} \rightarrow l l \tilde{\chi}^0_1$, if the 
gluino is slightly heavier than the wino, e.g. by about $O(10\%)$, 
due to higher order effects.  The branching ratio of this mode, however, 
is extremely small because of a large phase space suppression, so 
that the resulting dileptons are completely negligible compared with 
the ones arising from $\tilde{\chi}^0_2$ decay.  A similar comment 
also applies to $\tilde{q} \rightarrow q l \tilde{l}$ followed by 
$\tilde{l} \rightarrow l \tilde{\chi}^0_1$.}
We therefore conclude that the signatures discussed in the previous 
section can be used at the LHC for $m_{\tilde{q}} < m_{\tilde{g}}$. 
This claim will be confirmed in the next section, where we perform 
an explicit study using Monte Carlo simulations. 

In the case of $m_{\tilde{q}} > m_{\tilde{g}}$, a squark decays 
mainly into a quark and a gluino, although it may also decay 
into a wino or bino by a small amount.  The gluino then undergoes 
three-body decays: $\tilde{g} \rightarrow \tilde{\chi}^+_1 q \bar{q}'$, 
$\tilde{\chi}^0_1 q \bar{q}$ or $\tilde{\chi}^0_2 q \bar{q}$. 
The branching ratios for $\tilde{g} \rightarrow \tilde{W} q 
\bar{q}^{(\prime)}$, $\tilde{B} q \bar{q}$ may also be non-zero if 
the gluino is slightly heavier than the wino and/or bino, but these 
modes are highly suppressed by smallness of the phase space.  The 
gluino branching ratio into $\tilde{\chi}^0_2$ is of $O(10\%)$, and 
the $\tilde{\chi}^0_2$ branching ratio into leptons (including both 
$e$ and $\mu$) is $\approx 6\%$, implying that an appriciable number 
of dilepton events can be obtained from $\tilde{\chi}^0_2$ decay. 
The background from $\tilde{\chi}^+_1$ decay is, again, not important. 
The only sources of leptons that could potentially destroy the 
signatures are decays of $\tilde{W}$ and $\tilde{B}$, produced by 
squark decay. The dominant decay modes of $\tilde{W}$ and $\tilde{B}$, 
however, will be into a Higgsino (either one of $\tilde{\chi}^+_1$, 
$\tilde{\chi}^0_1$ and $\tilde{\chi}^0_2$) and a $W$, $Z$ or Higgs 
boson, so that the dangerous modes giving $l^+ l^-$ directly are 
suppressed by the three-body phase space and a small gaugino-Higgsino 
mixing.  The number of dangerous dileptons from $\tilde{W}$ and 
$\tilde{B}$ decays, therefore, is at most of the same order as the 
ones from $\tilde{\chi}^0_2$ decay.  Since the $M_{ll}$ endpoint 
for these dileptons is about $m_{\tilde{g}} - |\mu|$ and the 
distribution is suppressed for $M_{ll}$ much smaller than this 
values, it is unlikely that these leptons destroy the signatures 
from $\tilde{\chi}^0_2$ decay.  (We select events having two and only 
two leptons when doing the $M_{ll}$ analysis.)  We thus find that 
the Higgsino LSP signatures discussed in the previous subsection are 
also useful in the case of $m_{\tilde{q}} > m_{\tilde{g}}$ at the LHC.

We finally consider the case where $m_{\tilde{q}} = m_{\tilde{g}}$ 
at the leading order (at tree level).  In this case we expect that the 
masses of the gluino and squarks, $m_{\tilde{g}}$ and $m_{\tilde{q}}$, 
are slightly ($O(10\%)$) larger than those of the wino, bino and 
sleptons, $m_{\tilde{W}}$, $m_{\tilde{B}}$ and $m_{\tilde{l}}$, due 
to higher order (radiative) effects.  The orderings among $m_{\tilde{g}}$ 
and $m_{\tilde{q}}$ and among $m_{\tilde{W}}$, $m_{\tilde{B}}$ and 
$m_{\tilde{l}}$ are model-dependent.  With these spectra, gluinos and 
squarks once produced decay mostly into $\tilde{W}$ or $\tilde{B}$ plus 
a few jets, although a small fraction decays directly into light Higgsino 
states, $\tilde{\chi}^+_1$, $\tilde{\chi}^0_1$ and $\tilde{\chi}^0_2$. 
Decays of the electroweak gauginos differ depending on the ordering of 
$m_{\tilde{W}}$, $m_{\tilde{B}}$ and $m_{\tilde{l}}$.  If there is a 
slepton with the mass smaller than that of a gaugino, e.g. $m_{\tilde{l}} 
< m_{\tilde{W}}$, there is a sizable branching ratio for the gaugino 
decaying into the slepton, $\tilde{W} \rightarrow l\tilde{l}$.  This 
gives a large amount of opposite-sign same-flavor dileptons through 
the slepton decay $\tilde{l} \rightarrow l \tilde{\chi}^0_1$, which 
can potentially destroy the signatures from $\tilde{\chi}^0_2$ decay. 
On the other hand, if all the sleptons are heavier than $\tilde{W}$ 
and $\tilde{B}$, these gauginos decay mainly into a Higgsino (one of 
$\tilde{\chi}^+_1$, $\tilde{\chi}^0_1$ and $\tilde{\chi}^0_2$) and 
a $W$, $Z$ or Higgs boson.  The branching ratio into $\tilde{\chi}^0_2$ 
is typically of $O(10\%)$, and the desired signatures are obtained from 
$\tilde{\chi}^0_2$ decay.  There are other dileptons from three-body 
decays of $\tilde{W}$ and $\tilde{B}$, such as $\tilde{W} \rightarrow 
\tilde{\chi}^+_1 ll$, but the number of these dileptons is sufficiently 
small.  We thus expect that the Higgsino LSP signatures are useful for 
$m_{\tilde{q}} = m_{\tilde{g}}$ as long as all the sleptons are heavier 
than the electroweak gauginos.

We have seen that the Higgsino LSP signatures discussed in 
subsection~\ref{subsec:dilepton} are useful, i.e. not buried by 
dileptons from other superparticle decays, in a large class of 
theories motivated by solving the supersymmetric fine-tuning problem. 
In the next section, we explicitly demonstrate that these dilepton 
signatures can indeed be used in realistic analyses by performing 
Monte Carlo simulations in a theory with $m_{\tilde{g}} > m_{\tilde{q}}$. 
We also present a technique which can essentially determine all the 
superparticle masses in a class of theories discussed here, up to 
small mass splittings of $O(10\%)$ among different squarks and sleptons 
and a smaller splitting between the electroweak gaugino masses.

\section{Natural Supersymmetry at the LHC}
\label{sec:model}

In this section, we perform a Monte Carlo study for a class of theories 
discussed in the previous section and in subsections~\ref{subsec:low-med} 
and \ref{subsec:spectra}, which naturally leads to the correct scale 
for electroweak symmetry breaking.  We explicitly demonstrate that 
the signatures discussed in the previous section can be used to 
test the Higgsino LSP and extract the small mass difference between 
the two neutral Higgsinos at the LHC.  We also devise a series of 
cuts that allows us to determine all the relevant superparticle masses 
in theories with $m_{\tilde{g}} > m_{\tilde{q}}$: $m_{\tilde{\chi}^0_1}$, 
$m_{\tilde{q}}$, $\varDelta m$ and $m_{\tilde{g}}$.  In particular, we 
perform the analysis in the context of the model based on mixed moduli 
and anomaly mediated supersymmetry breaking, discussed at the end of 
subsection~\ref{subsec:low-med}, and show that the model can be tested 
at the LHC, up to theoretical uncertainties of $\approx 15\%$ on various 
superparticle masses.

\subsection{Framework}
\label{subsec:setup}

The basic setup for our analysis is the same as that in 
subsection~\ref{subsec:cascade} (and subsection~\ref{subsec:low-med}). 
The three gauginos are almost degenerate at the weak scale, 
$m_{\tilde{g}} \equiv M_1 \simeq M_2 \simeq M_3$, and the squarks 
and sleptons are also nearly degenerate, $m_{\tilde{q}} \equiv 
(m_Q^2)^{1/2} \simeq (m_U^2)^{1/2} \simeq (m_D^2)^{1/2} \simeq 
(m_L^2)^{1/2} \simeq (m_E^2)^{1/2}$.  The $A$ parameters are nearly 
universal at the weak scale, $A \equiv A_u \simeq A_d \simeq A_e$, 
with $A$ satisfying $|A/m_{\tilde{q}}| \simgt O(1)$.  The (top) 
squark masses, $m_{\tilde{q}}$, should not be very large, and the 
ratio of the electroweak VEVs should satisfy $\tan\beta \simgt 5$. 
In the analysis in this section, we only consider the case 
$m_{\tilde{g}} > m_{\tilde{q}}$. 

To perform an explicit Monte Carlo study, we must choose particular 
parameter points.  For this purpose, we take the model based on mixed 
moduli and anomaly mediated supersymmetry breaking, discussed at the 
end of subsection~\ref{subsec:low-med}, and choose the parameters 
within the region satisfying the condition $\Delta^{-1} \geq 20\%$. 
In particular, we take~\cite{Kitano:2005wc,Choi:2005hd}
\begin{equation}
  M_{1,2,3} = M_0,
\qquad
  m_{Q,U,D,L,E}^2 = \frac{M_0^2}{2},
\qquad
  A_{u,d,e} = -M_0,
\qquad
  m_{H_u,H_d}^2 = O\left(\frac{M_0^2}{8\pi^2}\right),
\label{eq:moduli-anomaly}
\end{equation}
at some scale $M_{\rm mess} = O(100~{\rm GeV}\!\sim\!{\rm TeV})$, where 
\begin{equation}
  450~{\rm GeV} \simlt M_0 \simlt 900~{\rm GeV},
\label{eq:M0-range}
\end{equation}
for $\tan\beta \simgt 20$.  For smaller $\tan\beta$, the lower bound 
in Eq.~(\ref{eq:M0-range}) increases; for $\tan\beta = 10$ ($5$) 
the lower bound becomes $\approx 550~{\rm GeV}$ ($900~{\rm GeV}$). 
In our analysis, we take $\mu$, $m_A$ and $\tan\beta$ to be free 
parameters in the Higgs sector, which are left undetermined after the 
electroweak symmetry breaking condition is imposed on $\mu$, $\mu B$, 
$m_{H_u}^2$ and $m_{H_d}^2$.  As shown in Ref.~\cite{Kitano:2005ew} 
the constraint from the $b \rightarrow s \gamma$ decay chooses the 
sign of $\mu$ to be positive.  We thus have
\begin{equation}
  {\rm sgn}(\mu) = +1,
\qquad
  |\mu| \simlt 190~{\rm GeV},
\qquad
  m_A \simlt 300~{\rm GeV},
\qquad
  \tan\beta \simgt 5.
\label{eq:mu-tanbeta-range}
\end{equation}
The last three conditions come from the naturalness criterion, 
$\Delta^{-1} \geq 20\%$.  

In our Monte Carlo study, we choose two parameter points given in 
Table~\ref{table:points}, satisfying Eqs.~(\ref{eq:moduli-anomaly},%
~\ref{eq:M0-range},~\ref{eq:mu-tanbeta-range}). 
\begin{table}[t]
\begin{center}
 \begin{tabular}[t]{|c|c|c|} \hline
   & point~I & point~II \\ \hline
  $M_0$ [GeV]    &   600 &   900 \\
  $\mu$ [GeV]    &   170 &   170 \\
  $m_A$ [GeV]    &   250 &   250 \\
  $\tan \beta$   &    15 &    15 \\
  $M_{\rm mess}$ & 1 TeV & 1 TeV \\ \hline
 \end{tabular}
\end{center}
\caption{Two representative parameter points of the model used 
 in Monte Carlo simulations.}
\label{table:points}
\end{table}
The point~I is representative for the case with a relatively low 
superparticle mass scale, while the point~II for the case with 
a high superparticle mass scale.  These points satisfy the experimental 
constraints such as the ones coming from the Higgs boson mass and the 
$b \rightarrow s \gamma$ decay, within theoretical uncertainties.  We 
note, however, that the constraints from the Higgs boson mass and 
$b \rightarrow s \gamma$ are not very important in our present context, 
because they are sensitive to the parameters in the Higgs sector, 
such as $m_A$ and $\tan\beta$, whose precise values are not relevant 
in our LHC study below.

The physical masses for the superparticles are obtained from the 
inputs in Table~\ref{table:points} as follows.  We interpret the input 
masses as the running masses in the $\overline{\rm DR}'$ scheme at the 
scale $M_{\rm mess}$, and evolve them down to the superparticle mass 
scale using renormalization group equations.  We then add the effects 
of electroweak symmetry breaking, such as the $D$-term contributions 
to the scalar masses and the gaugino-Higgsino mixings, and convert 
the running masses to the pole masses by including finite threshold 
corrections, using the code {\tt SuSpect\,2.3}~\cite{Djouadi:2002ze}. 
The resulting superparticle masses are given in Table~\ref{table:masses} 
for the two parameter points in Table~\ref{table:points}. 
\begin{table}[t]
\begin{center}
 \begin{tabular}[t]{|c|c|c|} \hline
  & point~I & point~II \\ \hline
 $\tilde{g}$             & 623 & 917 \\ \hline
 $\tilde{\chi}^+_1$      & 167 & 170 \\
 $\tilde{\chi}^+_2$      & 600 & 893 \\ \hline
 $\tilde{\chi}^0_1$      & 161 & 166 \\
 $\tilde{\chi}^0_2$      & 177 & 176 \\
 $\tilde{\chi}^0_3$      & 584 & 882 \\
 $\tilde{\chi}^0_4$      & 603 & 894 \\ \hline
 $\tilde{u}_L$           & 473 & 686 \\
 $\tilde{u}_R$           & 471 & 684 \\
 $\tilde{d}_L$           & 480 & 691 \\
 $\tilde{d}_R$           & 472 & 685 \\
 $\tilde{e}_L$           & 433 & 643 \\
 $\tilde{e}_R$           & 429 & 640 \\
 $\tilde{\nu}_{eL}$      & 425 & 638 \\ \hline
 $\tilde{t}_1$           & 365 & 571 \\
 $\tilde{t}_2$           & 576 & 783 \\
 $\tilde{b}_1$           & 463 & 678 \\
 $\tilde{b}_2$           & 481 & 691 \\
 $\tilde{\tau}_1$        & 424 & 636 \\
 $\tilde{\tau}_2$        & 437 & 646 \\
 $\tilde{\nu}_{\tau L}$  & 425 & 638 \\ \hline
 \end{tabular}
\end{center}
\caption{Superparticle masses in GeV for points~I and II in 
 Table~\ref{table:points}.  The masses for the second generation 
 squarks and sleptons are not listed because they are nearly degenerate 
 with the corresponding first generation squarks and sleptons.}
\label{table:masses}
\end{table}
Strictly speaking, this procedure is not quite meaningful in the 
context of the model under study, because we generically expect 
unknown higher order corrections of $O(1/8\pi^2)$ in the expression 
of Eq.~(\ref{eq:moduli-anomaly}), which can be comparable to some of 
the low energy corrections included here.  Nevertheless, this procedure 
allows us to incorporate the fact that the colored particles are 
systematically heavier than the non-colored ones by $O(10\%)$, which 
we expect to hold in realistic situations.  We thus perform our 
Monte Carlo study using the masses given in Table~\ref{table:masses}, 
although one should remember that there are intrinsic theoretical 
uncertainties of $O(10\%)$ for the superparticle masses in the model.

To perform the analysis, we generate both supersymmetric and standard 
model events using \textsc{Pythia}~6.324~\cite{Sjostrand:2000wi}. 
We generate supersymmetric events for the two parameter points 
in Table~\ref{table:points}.  The number of events generated for each 
point is equivalent to the integrated luminosity of $30~{\rm fb}^{-1}$, 
which corresponds to the three-year running of the LHC at low 
luminosity.  The superparticle decays are calculated using the 
code {\tt SDECAY\,1.1a}~\cite{Muhlleitner:2003vg}, with the results 
transferred to \textsc{Pythia} and used in the event generation. 
For the estimation of standard model background, we have generated 
0.5M QCD 2$\to$2 events for each bin of the transverse momentum: 
$100~{\rm GeV} < p_T < 200~{\rm GeV}$, $200~{\rm GeV} < p_T < 
400~{\rm GeV}$, $400~{\rm GeV} < p_T < 800~{\rm GeV}$, and $p_T 
> 800~{\rm GeV}$. We have also generated the $W$+jets events with 
$W \rightarrow e \nu, \mu \nu, \tau \nu$ (0.5M events for $50~{\rm GeV} 
< p_T < 200~{\rm GeV}$ and 0.2M events for $p_T > 200~{\rm GeV}$), 
the $Z$+jets events with $Z \rightarrow \nu \bar{\nu}, \tau^+ \tau^-$ 
(0.5M events for $50~{\rm GeV} < p_T < 150~{\rm GeV}$ and 0.2M 
events for $p_T > 150~{\rm GeV}$), 1M events for the $t\bar{t}$ 
production, and 0.2M events for each $ZZ$, $ZW$ and $WW$ production. 
These standard model events are simply scaled to $30~{\rm fb}^{-1}$ 
when estimating standard model backgrounds.  While our background 
estimations are correct probably only to a factor of a few due to 
inherent uncertainties associated with the QCD effects, we expect 
that our analysis is not much affected by this because the standard 
model background can be pretty much reduced by our cut selections, 
as we will see later.  Some of the analysis, e.g. the effective mass 
analysis in subsections~\ref{subsec:gluino} and \ref{subsec:M0=900}, 
may be affected by these uncertainties, but then we can always raise 
the cut on $E_T^{\rm miss}$ and recover the usefulness of the analysis.

For the detector simulation, we use {\tt AcerDET\,1.0}~%
\cite{Richter-Was:2002ch}, a generic fast detector simulation and 
reconstruction package for the LHC, which has a similar principle 
of operation to the official fast simulation package of the ATLAS 
detector, {\tt ATLFAST}~\cite{ATLFAST}.  The package performs 
identification and isolation of leptons, photons and jets in terms 
of detector coordinates: pseudorapidity $\eta$, azimuthal angle $\phi$, 
and cone size $\varDelta R = \sqrt{(\varDelta \phi)^2 + (\varDelta 
\eta)^2}$.  Lepton, photon and jet four-momenta are smeared, and the 
cluster selections are made based on $p_T$ and $|\eta|$.  Isolation 
criteria are applied to leptons and photons in terms of the distance 
from other clusters, $\varDelta R > 0.4$, and of maximum transverse 
energy deposited in cells in a cone $\varDelta R = 0.2$ around the 
cluster.  The calibration of jet four-momenta is also performed, 
and each jet is labeled either as a light jet, $b$-jet, $c$-jet or 
$\tau$-jet, using information from event generators.  (We use default 
parameters for these selection, isolation, calibration and labeling 
processes.)  For the $b$-jet identification, we further implement 
$b$-tagging efficiency of $60\%$ per a $b$-labeled jet, with 
mistagging probability of $10\%$ for a $c$-labeled jet and $1\%$ 
for a light jet.  For the $\tau$-jets, we use efficiency of $50\%$ 
per a $\tau$-labeled jet, with the mistagging probability of 
$10\%$ for other jets.

For each event, we apply the following trigger selections~\cite{ATLFAST}: 
one isolated electron with $p_T > 20~{\rm GeV}$, one isolated photon 
with $p_T > 40~{\rm GeV}$, two isolated electrons/photons with 
$p_T > 15~{\rm GeV}$, one muon with $p_T > 20~{\rm GeV}$, two muons 
with $p_T > 6~{\rm GeV}$, one isolated electron with $p_T > 15~{\rm GeV}$ 
and one isolated muon with $p_T > 6~{\rm GeV}$, one jet with $p_T > 
180~{\rm GeV}$, three jets with $p_T > 75~{\rm GeV}$, and four jets 
with $p_T > 55~{\rm GeV}$, where isolated electrons/photons, isolated 
muons and jets must be in the central regions of pseudorapidity $|\eta| 
< 2.5$, $2.4$, and $3.2$, respectively.  In our analysis, we consider 
only events passing one of these criteria.  

In our study, we ignore possible systematic errors caused by cuts 
and choices of fitting functions and regions, and take only into 
account the statistical errors.  Based on good agreements between 
the input values and the fit results obtained in subsections~%
\ref{subsec:squark-Higgsino}, \ref{subsec:gluino} and \ref{subsec:M0=900}, 
however, we expect that these neglected errors are not much larger than 
the statistical errors included in the analysis.

\subsection{Characteristic features: short cascades and fewer leptons}
\label{subsec:features}

We begin by identifying characteristic features relevant to the LHC 
arising for the superparticle spectra under consideration.  At the 
LHC, most of the superparticle productions comes for $m_{\tilde{q}}, 
m_{\tilde{g}} \simlt 1~{\rm TeV}$ from squark and gluino productions. 
For point~I (point~II) in Table~\ref{table:points}, the cross sections 
are $\simeq 3.8~{\rm pb}$, $25.3~{\rm pb}$ and $23.0~{\rm pb}$ 
($0.27~{\rm pb}$, $2.98~{\rm pb}$ and $3.71~{\rm pb}$) for the $\tilde{g} 
\tilde{g}$, $\tilde{g} \tilde{q}$ and $\tilde{q} \tilde{q}$ productions, 
respectively, where the total superparticle production cross section 
is about $55.2~{\rm pb}$ ($9.22~{\rm pb}$).  We find that the 
squark-gluino pair production and the squark pair production dominate 
in our parameter space.  After the production, a squark decays mostly 
into $\tilde{\chi}^+_1$ or $\tilde{\chi}^0_1$ and a quark, $\tilde{q} 
\rightarrow \tilde{\chi}^+_1 q', \tilde{\chi}^0_1 q$, but it also decays 
into $\tilde{\chi}^0_2$, $\tilde{q} \rightarrow \tilde{\chi}^0_2 q$, 
with a small branching ratio of $O(10\%)$.  The $\tilde{\chi}^+_1$ and 
$\tilde{\chi}^0_2$ produced then decay into $\tilde{\chi}^0_1$ and 
quarks/leptons through three-body decays, giving leptons with branching 
fractions of $O(10\%)$.  For the gluino, it decays into a quark and 
a squark with the branching ratio of $\approx 100\%$, which is followed 
by squark decay.

An important feature of these decays is that the decay chains are 
relatively short.  Compared with the case where the wino decay 
into a slepton is open, for example, decay chains with the present 
spectra are shorter in average.  Another important feature is that 
the number of leptons arising in the cascades is significantly smaller 
than in the case where colored and non-colored superparticles have 
a large mass hierarchy, e.g. as in a typical mSUGRA parameter point. 
We can thus use (\#~of~1~lepton~events)/(\#~of~0~lepton~events) 
to make a first guess that the superparticles may have a spectrum 
like the one considered here.  In fact, the features described here 
allow us to test certain generic aspects of the spectra, such as 
the nature of the LSP, and to determine the basic mass parameters 
in a simple manner.  These information can then be used to test 
or discriminate between possible models, as will be discussed 
in subsection~\ref{subsec:model-test}.

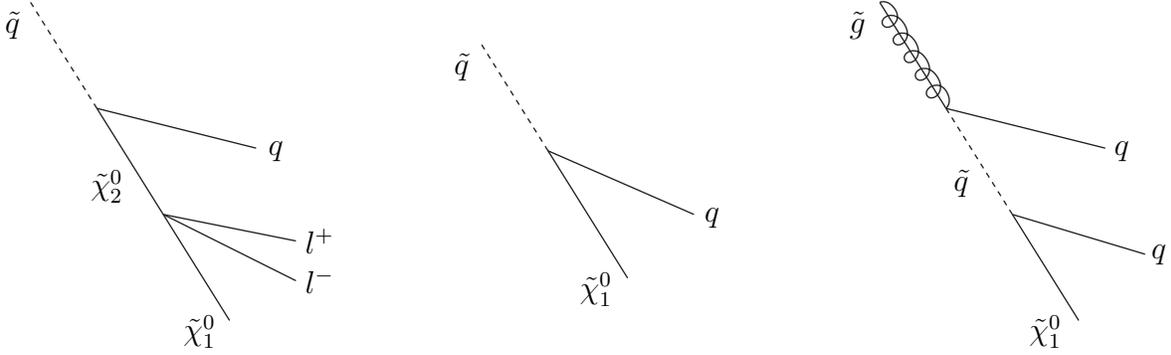
\begin{figure}[t]
\begin{center}
\begin{picture}(450,130)(0,-15)
  \DashLine(10,120)(35,80){2} \Text(6,117)[tr]{$\tilde{q}$}
  \Line(35,80)(60,40)  \Text(45,57)[tr]{$\tilde{\chi}^0_2$}
  \Line(60,40)(85,0)   \Text(80,2)[tr]{$\tilde{\chi}^0_1$}
  \Line(35,80)(95,65)  \Text(100,64)[l]{$q$}
  \Line(60,40)(110,30) \Text(114,30)[l]{$l^+$}
  \Line(60,40)(110,15) \Text(114,15)[l]{$l^-$}
  \DashLine(180,104)(205,64){2} \Text(176,101)[tr]{$\tilde{q}$}
  \Line(205,64)(235,16) \Text(230,18)[tr]{$\tilde{\chi}^0_1$}
  \Line(205,64)(260,40) \Text(265,39)[l]{$q$}
  \Line(330,120)(355,80) \Gluon(330,120)(355,80){3.5}{5}
  \Text(326,117)[tr]{$\tilde{g}$}
  \DashLine(355,80)(380,40){2}  \Text(365,57)[tr]{$\tilde{q}$}
  \Line(380,40)(405,0)  \Text(400,2)[tr]{$\tilde{\chi}^0_1$}
  \Line(355,80)(415,65) \Text(420,64)[l]{$q$}
  \Line(380,40)(430,25) \Text(434,25)[l]{$q$}
\end{picture}
\caption{Decay cascades used in the analysis.}
\label{fig:cascades}
\end{center}
\end{figure}

In our analysis, the following decay cascades are used: $\tilde{q} 
\rightarrow \tilde{\chi}^0_2 q \rightarrow \tilde{\chi}^0_1 l^+ l^- q$, 
$\tilde{q} \rightarrow \tilde{\chi}^0_1 q$, and $\tilde{g} \rightarrow 
\tilde{q} q \rightarrow \tilde{\chi}^0_1 q q$, which are depicted in 
Fig.~\ref{fig:cascades}. 
Here, we have not discriminated between quarks and antiquarks. 
Using the kinematics of these cascades, we can determine the 
masses of the gluino, squarks and neutralinos, as well as 
the small mass difference between $\tilde{\chi}^0_1$ and 
$\tilde{\chi}^0_2$, model independently.  This will be shown in 
subsections~\ref{subsec:squark-Higgsino} and \ref{subsec:gluino} 
for the case of point~I and in subsection~\ref{subsec:M0=900} 
for the case of point~II.  Various kinematical endpoints, such 
as the ones for dileptons, two leptons plus a jet, and a combination 
of two jets, will be used.  Precisions of order a few to ten 
percent are achieved, as will be shown later below.

\subsection{Determination of the squark and neutral Higgsino masses}
\label{subsec:squark-Higgsino}

In this subsection we show that using the kinematics of the 
cascade decay $\tilde{q} \rightarrow \tilde{\chi}^0_2 q \rightarrow 
\tilde{\chi}^0_1 l^+ l^- q$ and those for the squark pair production 
with $\tilde{q} \rightarrow \tilde{\chi}^0_1 q$, we can determine 
$m_{\tilde{q}}$, $m_{\tilde{\chi}^0_1}$ and $m_{\tilde{\chi}^0_2}$ 
without using input from particular models.  This information 
can thus be used for nontrivial tests of the model predictions, 
as will be discussed in subsection~\ref{subsec:model-test}.  The 
analysis also demonstrates that the $M_{ll}$ distribution discussed 
in section~\ref{sec:higgsino} is indeed useful in testing the Higgsino 
nature of the lightest neutralinos.  In this and the next subsections, 
we use point~I in Table~\ref{table:points} ($M_0 = 600~{\rm GeV}$) 
for the analysis.  The same analysis will be repeated for point~II 
($M_0 = 900~{\rm GeV}$) in subsection~\ref{subsec:M0=900}.

We first look at the $M_{ll}$ distribution from the three-body decay 
$\tilde{\chi}^0_2 \rightarrow \tilde{\chi}^0_1 l^+ l^-$.  As discussed 
in section~\ref{sec:higgsino} the endpoint and the shape of the 
distribution measure the mass difference of the two neutralinos, 
$\varDelta m$, as well as the relative $CP$ property of the two 
neutralinos, $\eta_\chi$.

\begin{figure}[t]
\begin{center}
  \includegraphics[height=6.5cm]{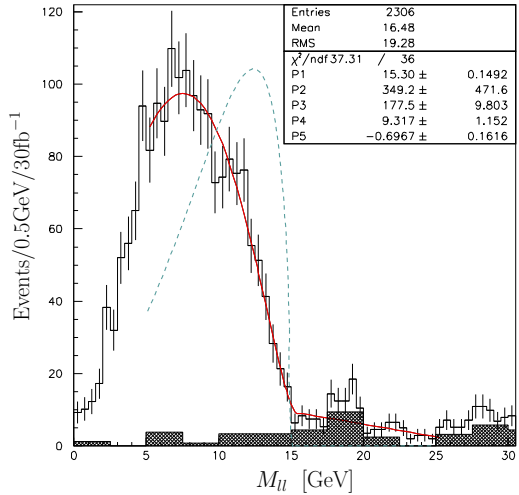}
\end{center}
\caption{The distribution of the dilepton invariant mass, $M_{ll}$, for 
 point~I.  Hatched histogram represents the standard model background, 
 which is smeared over five bins in order not to magnify the statistical 
 uncertainty due to the scaling of the events.  The solid line is the 
 best fit function for the signal plus background, obtained using the 
 theoretical curve with $\eta_\chi = -1$.  It can be clearly distinguished 
 from the $\eta_\chi = +1$ case, drawn by the dashed line.  The endpoint 
 is extracted to be $15.30 \pm 0.15~{\rm GeV}$.}
\label{fig:mll.600}
\end{figure}

We select the dilepton events with the following cuts:
\begin{itemize}
  \item $E_T^{\rm miss} > 300~{\rm GeV}$
  \item At least two jets with $p_T > 50~{\rm GeV}$
  \item Two and only two leptons with the same flavor and opposite charge
  \item Veto $b$-jets
\end{itemize}
In Fig.~\ref{fig:mll.600}, we show the $M_{ll}$ distribution obtained 
with the cuts described above.
The standard model backgrounds are effectively reduced by the cuts 
(hatched histogram).  We can clearly see the endpoint of $M_{ll}$ 
around $15~{\rm GeV}$, which can be the first test for the Higgsino 
LSP scenario.  Note that while $M_{ll}$ is small, $p_T$'s of the 
leptons are not so small because of the parent $\tilde{\chi}^0_2$'s 
transverse momenta, so that these leptons are not much affected by 
the trigger selections of subsection~\ref{subsec:setup}.

We have performed a fit of the distribution by using the theoretical 
curve including the effect of the finite $Z$-boson mass, with a linear
background distribution.  In this fitting process, the standard model 
background events are smeared in order not to artificially magnify 
the statistical uncertainty due to the scaling of the generated 
events to $30~{\rm fb}^{-1}$.  A reasonable fit is possible only 
for the $\eta_\chi = -1$ case (solid line), and we obtain the 
endpoint value
\begin{equation}
  M_{ll}^{\rm max} = 15.30 \pm 0.15~{\rm GeV},
\label{eq:fit-Mll}
\end{equation}
which is consistent with the input value of the mass difference 
$\varDelta m = 15.40~{\rm GeV}$ (see Eq.~(\ref{eq:endpoint})).  The 
theoretical curve with the $\eta_\chi = +1$ case is superimposed 
in the plot (dashed line).  We can clearly see the deviation from 
the simulated distribution, especially near the endpoint.  We thus 
conclude that the smallness of the endpoint, $M_{ll}^{\rm max}$, 
together with the shape of the $M_{ll}$ distribution characteristic 
of $\eta_\chi = -1$, provide an extremely powerful test for the 
Higgsino LSP scenario.  If these signals are actually observed, they 
strongly suggest that the LSP is one of the nearly degenerate neutral 
Higgsinos.

At this stage, we only have information on the neutralino mass 
difference, but further information can be obtained by stepping up 
the cascade, i.e. by combining dileptons with the quark jet from 
the squark decay.  We can construct two independent Lorentz-invariant 
quantities $M_{llq}$ and $M_{lq}$, whose endpoint values are given by 
\begin{equation}
  M_{llq}^{\rm max} 
  = M_{l q}^{\rm max} = m_{\tilde{q}} 
    \Biggl( 1 - \frac{m_{\tilde{\chi}^0_2}^2}{m_{\tilde{q}}^2} \Biggr)^{1/2}
    \Biggl( 1 - \frac{m_{\tilde{\chi}^0_1}^2}{m_{\tilde{\chi}^0_2}^2} 
    \Biggr)^{1/2}.
\label{eq:formula_mllq}
\end{equation}
The two endpoints coincide because the final state leptons come from 
the three-body decay.  The measurement of the two endpoints, therefore, 
can give only one additional information.

For the event selection for the $M_{llq}$ and $M_{lq}$ measurements, 
we have imposed a cut on $M_{ll}$
\begin{itemize}
  \item $M_{ll} < 15~{\rm GeV}$
\end{itemize}
in addition to the cuts for the $M_{ll}$ measurement, in order to 
reject incorrect lepton pairs.  The jet which is combined with the 
lepton(s) is selected from the two largest $p_T$ jets.  We choose 
the one that gives the smaller $M_{llq}$ to see the endpoint of 
the distribution.

\begin{figure}[t]
\begin{center}
  \includegraphics[height=6.5cm]{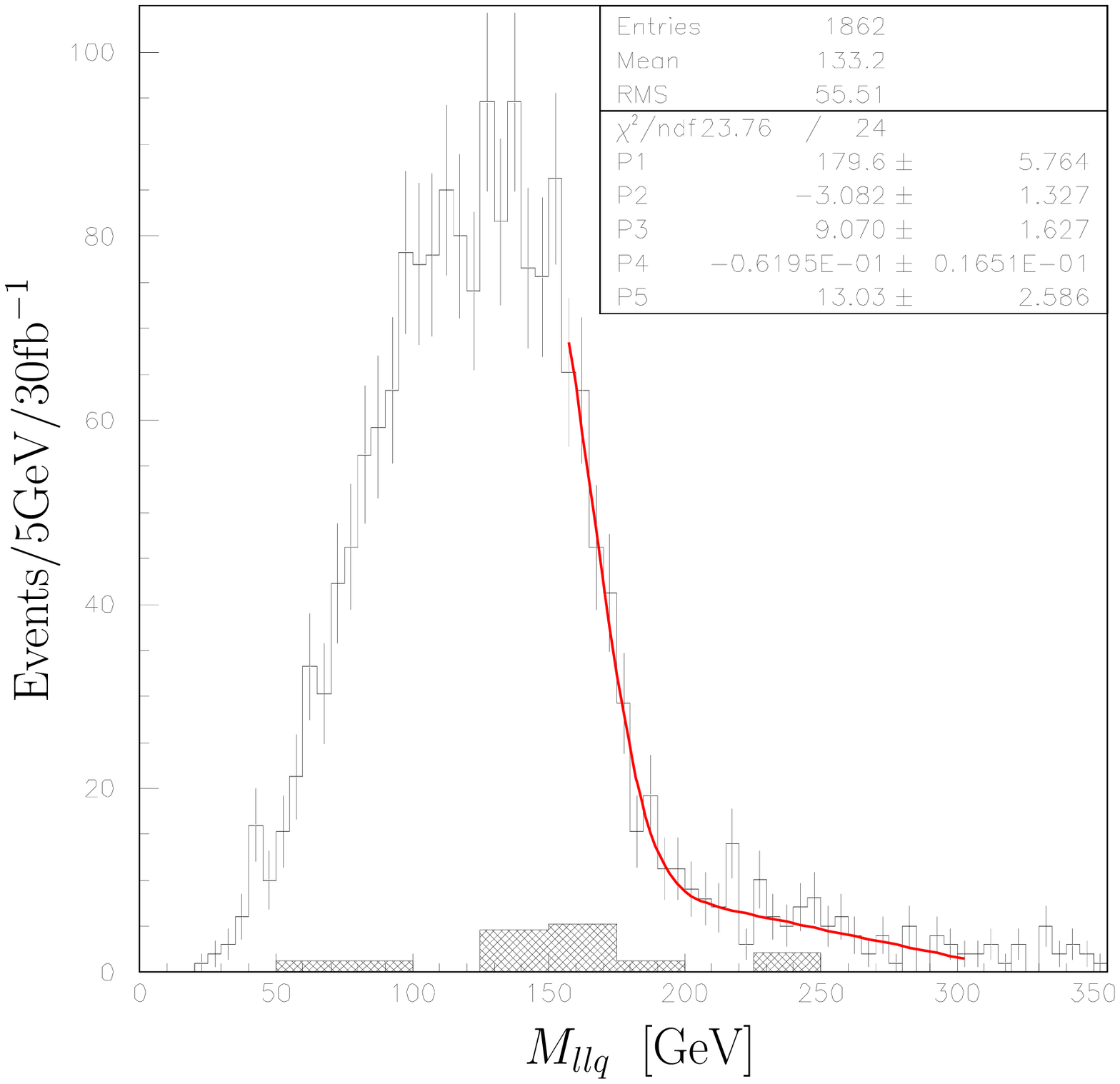}
  \hspace*{.5cm}
  \includegraphics[height=6.5cm]{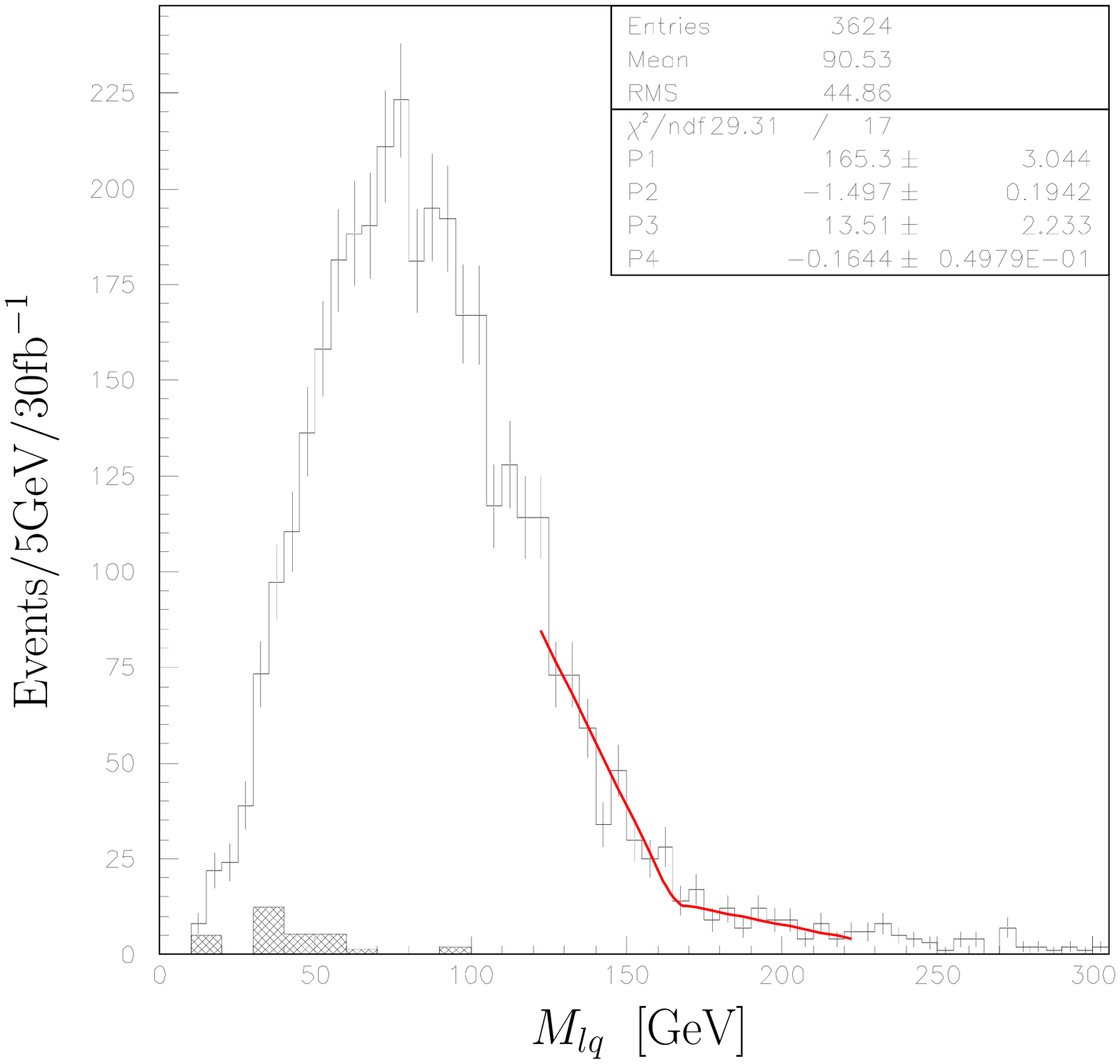}
\end{center}
\caption{The distributions of $M_{llq}$ (left) and $M_{lq}$ (right) 
 for point~I.  Hatched histogram represents the standard model background, 
 which for $M_{llq}$ is smeared over five bins in order not to magnify 
 the statistical uncertainty due to the scaling of the events. The solid 
 line is the best fit function for the signal plus background near the 
 endpoint, obtained by using a linear function with Gaussian smearing 
 for the signal and a linear function for the background.  The endpoint 
 is extracted to be $179.6 \pm 5.8~{\rm GeV}$ for $M_{llq}$.  The $M_{lq}$ 
 endpoint cannot be extracted clearly using a simple linear function.}
\label{fig:mllq.600}
\end{figure}

In Fig.~\ref{fig:mllq.600}, we show the distributions of $M_{llq}$ 
(left) and $M_{lq}$ (right).
The endpoint is clearer in $M_{llq}$ than in $M_{lq}$.  We thus use 
the $M_{llq}$ endpoint for the mass determination.  We have performed 
a fit of the $M_{llq}$ distribution near the endpoint, with a linear 
function with Gaussian smearing.  A linear function is also assumed 
for the background distribution.  The best fit function is drawn in 
the figure, and we obtain
\begin{equation}
  M_{llq}^{\rm max} = 179.6 \pm 5.8~{\rm GeV}.
\label{eq:fit-Mllq}
\end{equation}
This is consistent with the expected endpoint obtained using the 
formula of Eq.~(\ref{eq:formula_mllq}), which is $\simeq 180~{\rm GeV}$. 
We have also tried to estimate the endpoint in the $M_{lq}$ distribution. 
A reasonable fit, however, cannot be obtained with linear functions. 
Better functions are needed if one wants to extract the endpoint 
from the $M_{lq}$ distribution.  From the $M_{ll}$ and $M_{llq}$ 
endpoint measurements, we now have two relations among three mass 
parameters, $m_{\tilde{\chi}^0_1}$, $m_{\tilde{\chi}^0_2}$ and 
$m_{\tilde{q}}$.  We still need one more independent quantity to 
determine all the three masses.

In principle, the threshold value of $M_{llq}$ with the cut 
$M_{ll} > \xi M_{ll}^{\rm max}$ $(0<\xi<1)$ could provide the 
required additional information:
\begin{eqnarray}
  M_{llq}^{\rm min}|_{M_{ll} > \xi M_{ll}^{\rm max}} &&
\nonumber\\
  && \hspace*{-3cm}
  = \frac{M_{llq}^{\rm max}}{\sqrt{2}} 
  \left[
  1 - \frac{\Bigl( \Bigl( 1+\xi^2+(1-\xi^2) 
    \frac{m_{\tilde{\chi}^0_1}}{m_{\tilde{\chi}^0_2}} \Bigr)^2
    - 4 \xi^2 \Bigr)^{1/2}}
    {1 + \frac{m_{\tilde{\chi}^0_1}}{m_{\tilde{\chi}^0_2}}}
  + \xi^2 \frac{1+\frac{m_{\tilde{\chi}^0_2}^2}{m_{\tilde{q}}^2}}
    {1-\frac{m_{\tilde{\chi}^0_2}^2}{m_{\tilde{q}}^2}}
    \frac{1-\frac{m_{\tilde{\chi}^0_1}}{m_{\tilde{\chi}^0_2}}}
    {1+\frac{m_{\tilde{\chi}^0_1}}{m_{\tilde{\chi}^0_2}}}
  \right]^{1/2}.
\end{eqnarray}
However, with the limited statistics and the narrow physical $M_{ll}$ 
region of the Higgsino LSP scenario, the threshold of $M_{llq}$ is not 
quite useful for the mass determination.  With fixed $M_{ll}^{\rm max}$ 
and $M_{llq}^{\rm max}$, $M_{llq}^{\rm min}$ has a very little 
sensitivity to the mass parameters.

We therefore have to look for another quantity to determine the three 
masses.  Such a quantity can be obtained by analyzing the squark pair 
production process followed by the two squarks decaying into two jets 
and two $\tilde{\chi}^0_1$'s.  Although we cannot reconstruct the 
squark four-momenta due to two escaping invisible neutralinos by 
the event by event analysis, we can extract a relation between 
$m_{\tilde{q}}$ and $m_{\tilde{\chi}^0_1}$ by the endpoint analysis 
of the $M_{T2}$ variable defined in Ref.~\cite{Lester:1999tx}.  This 
variable is designed to take the maximal value at the squark mass 
when we input the correct $m_{\tilde{\chi}^0_1}$ in the calculation. 
The definition is given by
\begin{equation}
  M_{T2}^2 = \min_{{\bf p}_{T1}^{\rm miss}+{\bf p}_{T2}^{\rm miss} 
    = {\bf p}_T^{\rm miss}}
  \left[ \max \{ m_T^2({\bf p}_T^{j1}, {\bf p}_{T1}^{\rm miss}),
    m_T^2({\bf p}_T^{j2}, {\bf p}_{T2}^{\rm miss}) \} \right],
\end{equation}
where ${\bf p}_T^{j1}$ and ${\bf p}_T^{j2}$ are the transverse momenta 
of the jets from the squark decays, and ${\bf p}_T^{\rm miss}$ is the 
missing transverse momentum.  The transverse mass, $m_T^2$, is defined 
by
\begin{equation}
  m_T^2 ({\bf p}_T^a, {\bf p}_T^b)
  = m_a^2 + m_b^2 + 2 (E_T^a E_T^b - {\bf p}_T^a \cdot {\bf p}_T^b).
\end{equation}
By identifying the endpoints of $M_{T2}$ for various input 
values of $m_{\tilde{\chi}^0_1}$, we can obtain a relation between 
$m_{\tilde{q}}$ and $m_{\tilde{\chi}^0_1}$, which can provide the 
last information to determine the three masses, $m_{\tilde{\chi}^0_1}$, 
$m_{\tilde{\chi}^0_2}$ and $m_{\tilde{q}}$.

To select the squark pair production events, we use the following cuts:
\begin{itemize}
  \item $E_T^{\rm miss} > 300~{\rm GeV}$
  \item Veto leptons, $b$-jets, $\tau$-jets
  \item Two and only two jets with $p_T > 50~{\rm GeV}$
\end{itemize}
For our assumptions on the $b$-tagging and $\tau$-tagging efficiencies, 
see subsection~\ref{subsec:setup}.

\begin{figure}[t]
\begin{center}
  \includegraphics[height=6.5cm]{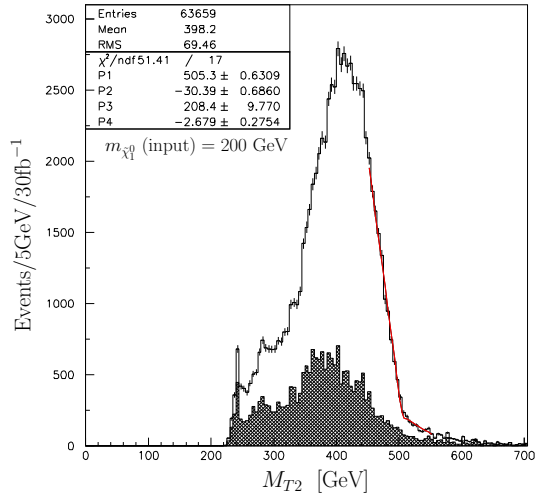}
\end{center}
\caption{The $M_{T2}$ (defined in the text) distribution for the input 
 value of the neutralino mass $m_{\tilde{\chi}^0_1} = 200~{\rm GeV}$ 
 for point~I.  Hatched histogram is the standard model background. 
 The endpoint is extracted by fitting the signal plus background 
 histogram with a linear function, and the background near the 
 endpoint by a linear function. The endpoint is obtained to be 
 $505.3 \pm 0.6~{\rm GeV}$.}
\label{fig:mT2.600}
\end{figure}

In Fig.~\ref{fig:mT2.600}, we show the $M_{T2}$ distribution 
for the input value of $m_{\tilde{\chi}^0_1} = 200~{\rm GeV}$ 
as an example.
We can see a clear edge in the distribution around $500~{\rm GeV}$. 
Fitting with a linear function with a linear background, we obtain 
the endpoint $505.3 \pm 0.6~{\rm GeV}$.  With the rich statistics 
(63,859 events survive the cuts) and the sharp edge, we can measure 
the endpoint quite accurately.  Note that this is not the squark 
mass itself --- it is the value of some quantity that would become 
the squark mass if our hypothetical input neutralino mass is in 
fact the true neutralino mass.

\begin{figure}[t]
\begin{center}
  \includegraphics[height=6.5cm]{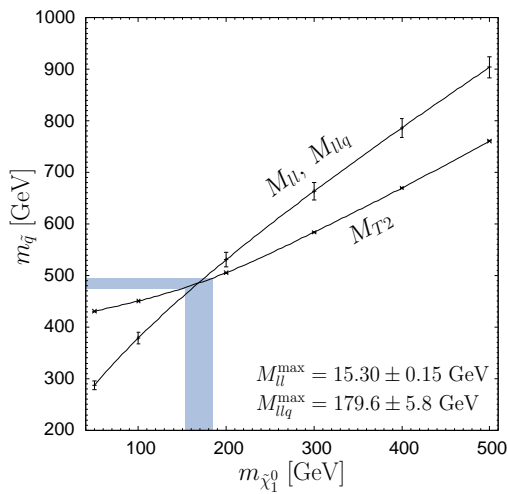}
\end{center}
\caption{Two curves on the $m_{\tilde{\chi}^0_1}$--$m_{\tilde{q}}$ 
 plane deduced from the cascade decay analysis, $M_{ll}^{\rm max}$ 
 and $M_{llq}^{\rm max}$, and the squark pair production analysis, 
 $M_{T2}^{\rm max}$, for point~I.  Both curves are obtained by 
 inputting hypothetical values of $m_{\tilde{\chi}^0_1}$, which 
 is taken as the horizontal axis.  The intersection determines 
 the real values of $m_{\tilde{\chi}^0_1}$ and $m_{\tilde{q}}$. 
 The obtained masses with the $1 \sigma$ statistical errors are 
 shown by shaded bands.}
\label{fig:mchiVSmsq.600}
\end{figure}

Combining the $M_{ll}^{\rm max}$ and $M_{llq}^{\rm max}$ measurements 
with the $M_{T2}$ analysis, we can now determine all the three masses, 
$m_{\tilde{q}}$, $m_{\tilde{\chi}^0_1}$ and $m_{\tilde{\chi}^0_2}$, 
as follows.  First, the endpoint analysis of the cascade decay, 
$M_{ll}^{\rm max}$ and $M_{llq}^{\rm max}$, gives two constraints 
on the three mass parameters, leaving one parameter unfixed.  If 
we take this parameter to be $m_{\tilde{\chi}^0_1}$, we can draw 
a curve on the $m_{\tilde{\chi}^0_1}$--$m_{\tilde{q}}$ plane, using 
the constraints from $M_{ll}^{\rm max}$ and $M_{llq}^{\rm max}$. 
On the other hand, the $M_{T2}$ analysis of the squark pair 
production gives another relation between $m_{\tilde{\chi}^0_1}$ 
and $m_{\tilde{q}}$, giving an independent curve on the same 
plane. The intersection of the two curves will then give the 
real values of ($m_{\tilde{\chi}^0_1}$, $m_{\tilde{q}}$).  In 
Fig.~\ref{fig:mchiVSmsq.600} we show the two curves explained 
above with the statistical errors.
The $M_{T2}$ curve is obtained by performing the $M_{T2}$ endpoint 
measurements for six different input values of $m_{\tilde{\chi}^0_1}$ 
and then interpolating them with a smooth curve.  The measured values 
of $m_{\tilde{\chi}^0_1}$ and $m_{\tilde{q}}$ by this combined analysis 
are indicated by shaded bands with the $1 \sigma$ statistical errors.

\begin{figure}[t]
\begin{center}
  \includegraphics[height=5.7cm]{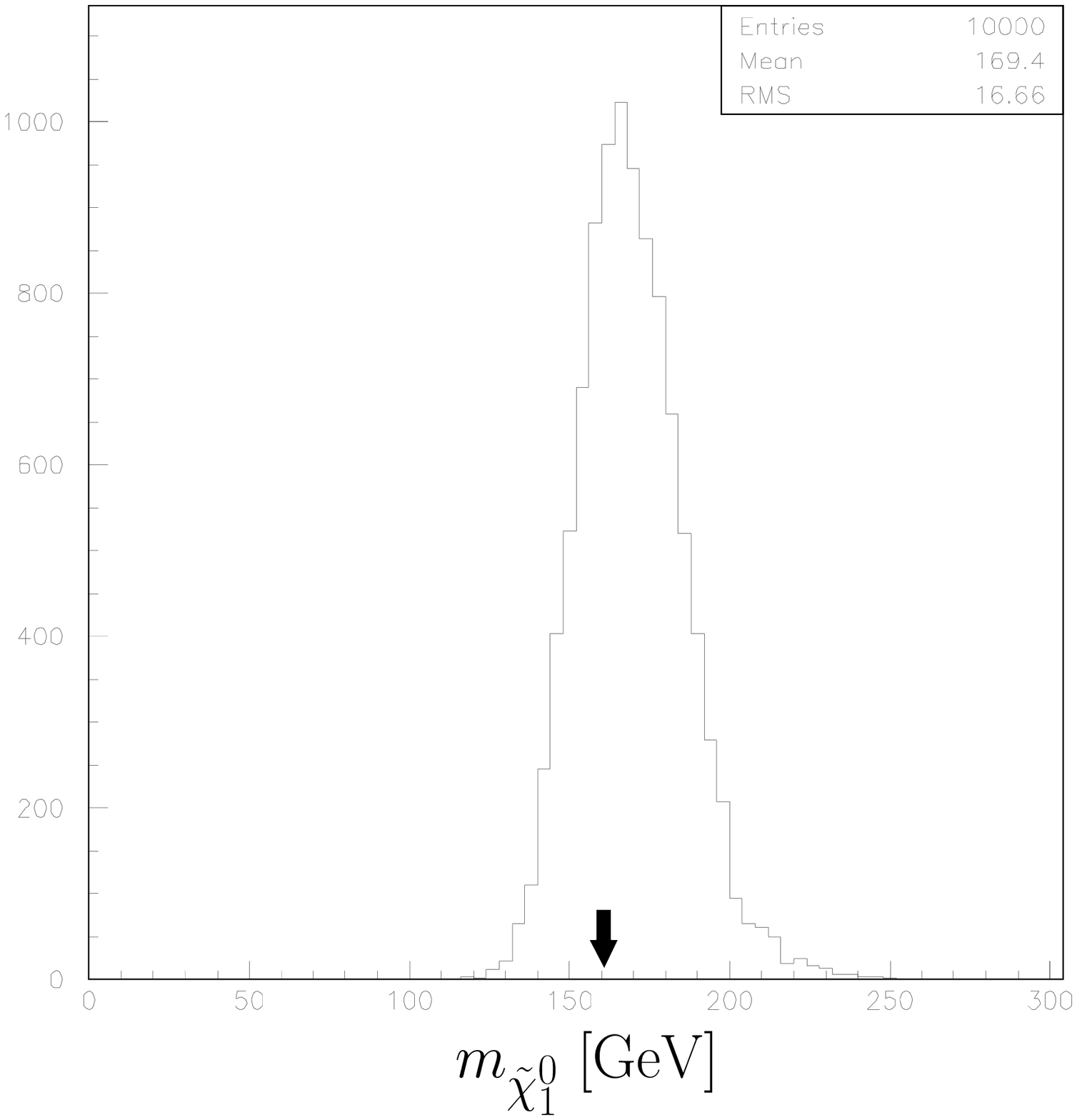}
  \includegraphics[height=5.7cm]{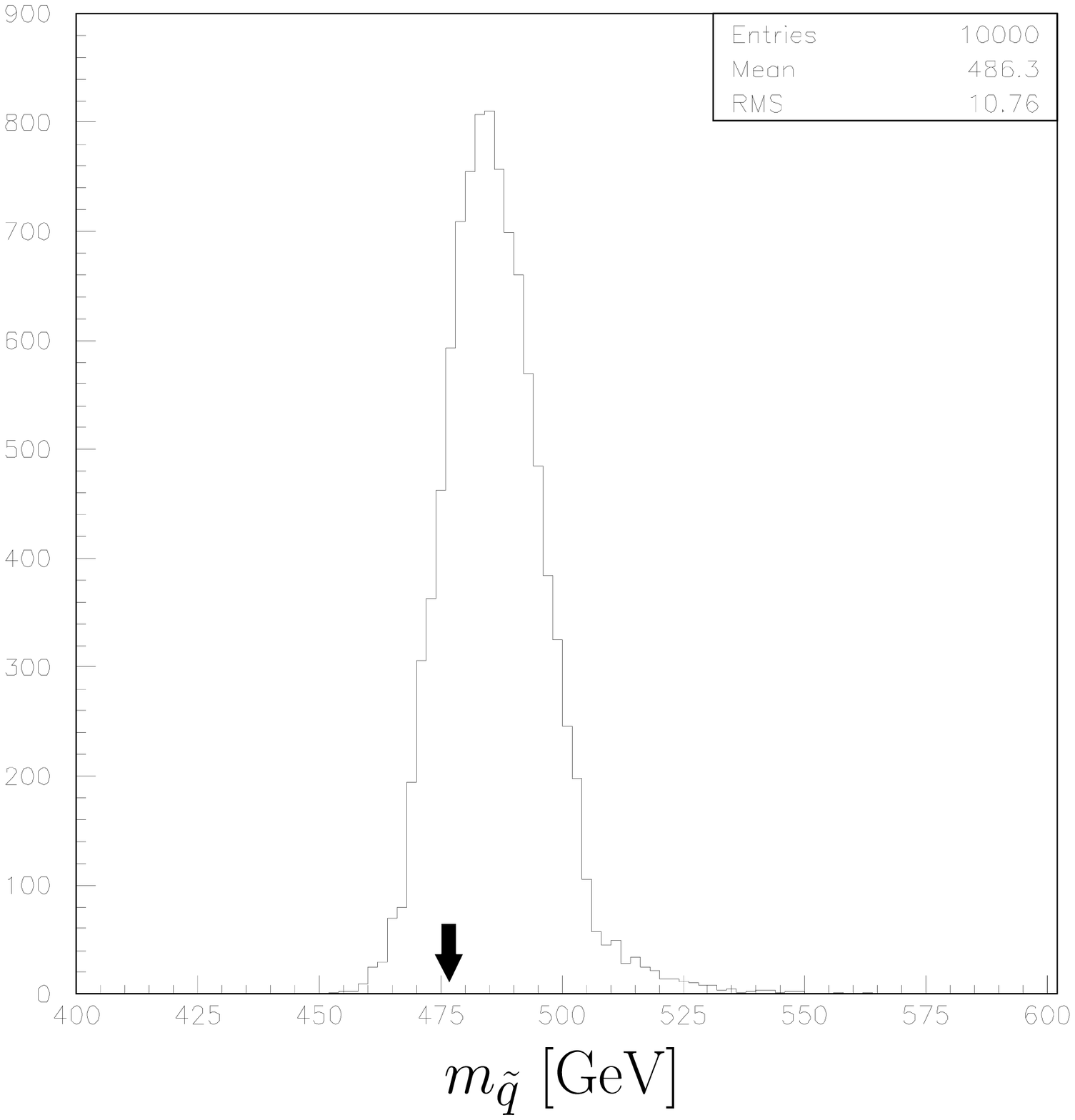}
  \includegraphics[height=5.7cm]{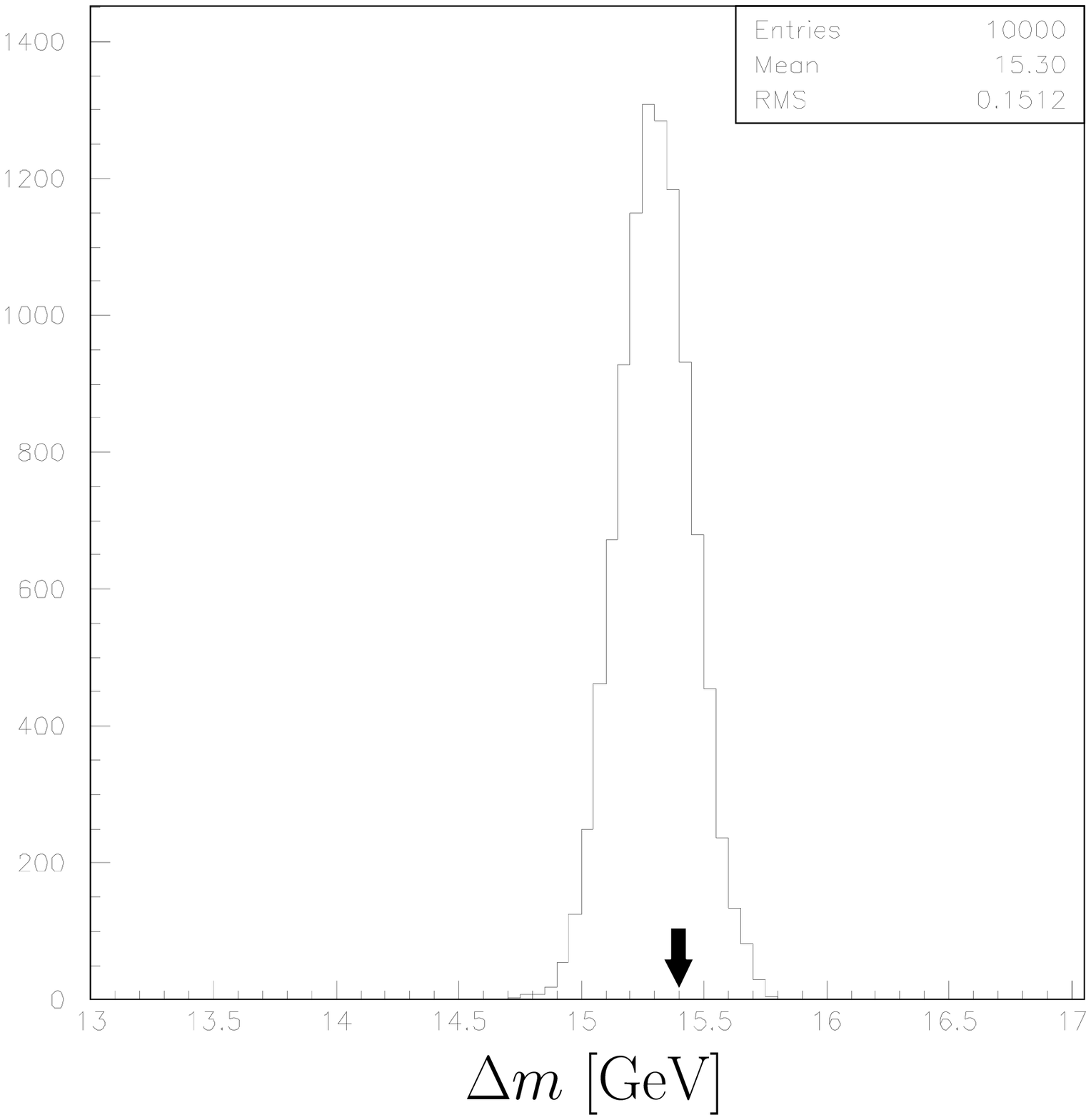}
\end{center}
\caption{The statistical errors of the measured mass parameters 
 $m_{\tilde{\chi}^0_1}$, $m_{\tilde{q}}$ and $\varDelta m$ for 
 point~I.  Arrows indicate the input values used in the Monte Carlo 
 event generation.  The accuracy at the level of $10\%$, $2\%$ and 
 $1\%$ is obtained for $m_{\tilde{\chi}^0_1}$, $m_{\tilde{q}}$ and 
 $\varDelta m$, respectively.}
\label{fig:error.600}
\end{figure}

The effectiveness and accuracy of the method are demonstrated 
in Fig.~\ref{fig:error.600}.
To draw the figure, we have generated 10,000 ``experiments'' and 
considered that in these experiments the values of $M_{ll}^{\rm max}$, 
$M_{llq}^{\rm max}$ and $M_{T2}^{\rm max}$ are determined according 
to the Gaussian distributions with the statistical errors given in 
Eqs.~(\ref{eq:fit-Mll},~\ref{eq:fit-Mllq}) and by the $M_{T2}$ fit. 
We have then calculated the three mass parameters using the method 
described above and have plotted their distributions.  These plots 
show that $m_{\tilde{\chi}^0_1}$ and $m_{\tilde{q}}$ have larger tails 
in large mass regions.  This represents the fact that the two curves 
in Fig.~\ref{fig:mchiVSmsq.600} are more similar in a larger 
$m_{\tilde{\chi}^0_1}$ region than in a lower region.  The input 
values of the mass parameters are indicated in Fig.~\ref{fig:error.600} 
by arrows, and we find that the correct values are obtained within 
reasonable statistical uncertainties.  By fitting the histograms 
with the Gaussian distribution, we obtain
\begin{equation}
  m_{\tilde{\chi}^0_1} = 169 \pm 17~{\rm GeV},
\quad
  m_{\tilde{q}} = 486 \pm 11~{\rm GeV},
\quad 
  \varDelta m = 15.30 \pm 0.15~{\rm GeV}.
\end{equation}
This demonstrates that the neutralino and squark masses can be 
measured with $10\%$ and $2\%$ level accuracy, respectively, at 
the LHC in the Higgsino LSP scenario.  The mass difference between 
the two neutral Higgsinos can be measured at $1\%$ accuracy.  The 
information on these masses are very useful to test particular 
models, as will be discussed in subsection~\ref{subsec:model-test}.

\subsection{Determination of the gluino mass}
\label{subsec:gluino}

With the knowledge of the squark and neutralino masses, we 
can determine the gluino mass using the kinematics of the 
$\tilde{g} \rightarrow \tilde{q}q \rightarrow \tilde{\chi}^0_1 q q$ 
cascade decay.  The invariant mass of the two final jets have the 
maximal value at
\begin{equation}
  M_{jj}^{\rm max} = m_{\tilde{g}} 
    \Biggl( 1 - \frac{m_{\tilde{q}}^2}{m_{\tilde{g}}^2} \Biggr)^{1/2}
    \Biggl( 1 - \frac{m_{\tilde{\chi}^0_1}^2}{m_{\tilde{q}}^2} 
  \Biggr)^{1/2}.
\label{eq:mjj-endpoint}
\end{equation}
Therefore, if the endpoint of the $M_{jj}$ distribution arising from 
this cascade is measured, we can determine the gluino mass.

In supersymmetric models, the gluino production is mostly from 
the $\tilde{g}+\tilde{q}$ or $\tilde{g}+\tilde{g}$ pair production, 
which necessarily gives additional jets from the other side of the 
squark/gluino decay.  Those additional jets cause an uncertainty 
for the selection of the correct jet pair.  In order to reduce 
this combinatorial background, we select the events with three hard 
jets, which is typical in the $\tilde{g}+\tilde{q}$ production, and 
choose a pair of jets which gives the smallest $M_{jj}$ among three 
combinations such that the calculated $M_{jj}$ would not exceed 
the endpoint in Eq.~(\ref{eq:mjj-endpoint}).  (Note that this does 
not mean that the selected pair is necessarily the correct one, 
but it guarantees that the endpoint of the $M_{jj}$ distribution 
is given by the right formula, Eq.~(\ref{eq:mjj-endpoint}).)

\begin{figure}[t]
\begin{center}
  \includegraphics[height=6.5cm]{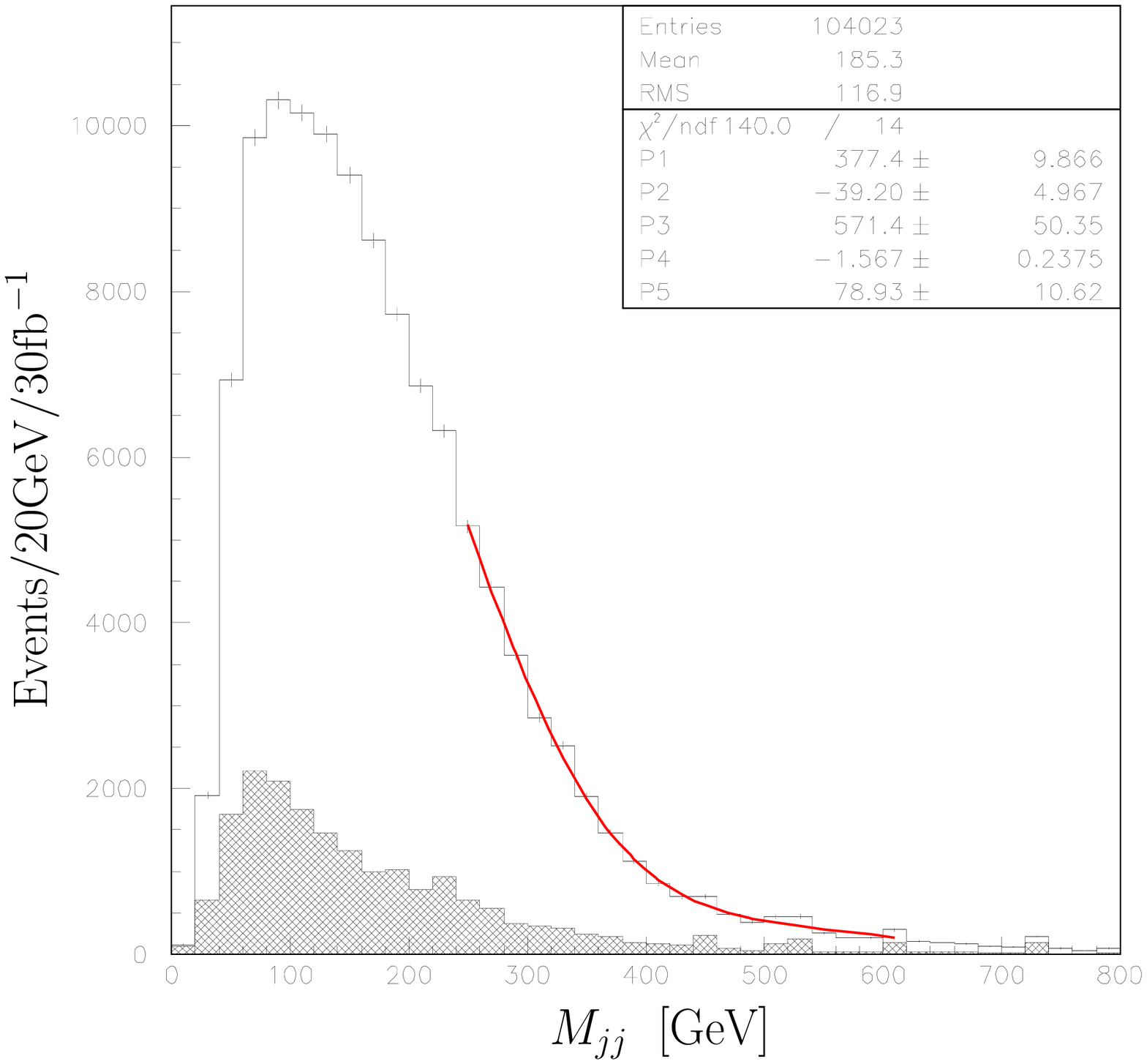}
  \hspace*{.5cm}
  \includegraphics[height=6.5cm]{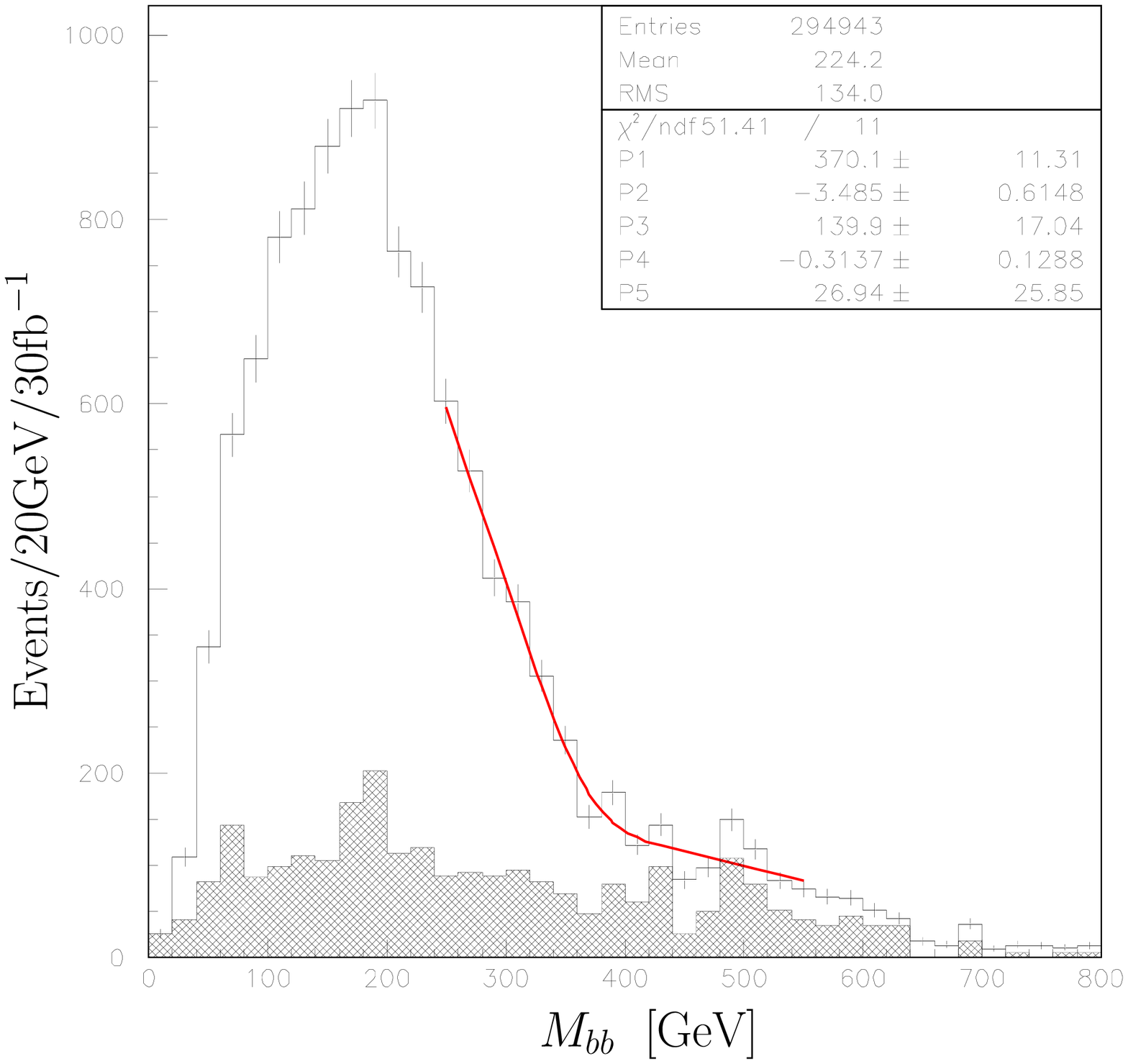}
\end{center}
\caption{The $M_{jj}$ (left) and $M_{bb}$ (right) invariant mass 
 distributions for point~I.  In the $M_{jj}$ analysis, we select 
 events with three hard jets and take a combination that gives the 
 smallest $M_{jj}$, out of the three combinations, in order to see 
 the endpoint.  A fit is performed with a linear function with Gaussian 
 smearing together with a linear function for the background events. 
 The endpoint is obtained as $377 \pm 10~{\rm GeV}$.  In the $M_{bb}$ 
 analysis, we select events with three hard jets including two hard 
 $b$-jets.  A similar endpoint, $370 \pm 11~{\rm GeV}$, is obtained 
 using the same fitting function.}
\label{fig:mjj.600}
\end{figure}

The selection cuts we use are the following:
\begin{itemize}
  \item $E_T^{\rm miss} > 300~{\rm GeV}$
  \item Veto leptons, $b$-jets, $\tau$-jets
  \item Three and only three jets with $p_T > 50~{\rm GeV}$
\end{itemize}
With these cuts, we obtain the $M_{jj}$ distribution shown in the 
left panel of Fig.~\ref{fig:mjj.600}.
The endpoint structure is visible around $400~{\rm GeV}$. By fitting 
the histogram near the endpoint by a linear function with Gaussian 
smearing and a linear background shape, we obtain
\begin{equation}
  M_{jj}^{\rm max} = 377 \pm 10~{\rm GeV}.
\end{equation}
The large value of $\chi^2$ shown in the plot is caused by the 
artificially magnified statistical uncertainty of the standard model 
background due to the scaling of the events.  We have checked that 
the reasonable value of $\chi^2$ is obtained without the standard 
model background.

We can perform the same analysis for the $\tilde{g} \rightarrow 
\tilde{b}b \rightarrow \tilde{\chi}^0_1 b b$ decay by requiring 
two $b$-jets.  We show the resulting $M_{bb}$ distribution in the 
right panel of Fig.~\ref{fig:mjj.600}.  In this case, we do not suffer 
from the combinatorial background, although the statistics is reduced. 
If we assume $m_{\tilde{b}} \simeq m_{\tilde{q}}$, which is indeed the 
case here, we can use this endpoint for the gluino mass reconstruction. 
We, however, do not use this analysis in the following because it will 
make our analysis more model dependent.  For the purpose of testing 
the particular model, however, the $M_{bb}$ analysis can be used as 
a consistency check (or to extract some information on the value of 
$\tan\beta$).

\begin{figure}[t]
\begin{center}
  \includegraphics[height=6.5cm]{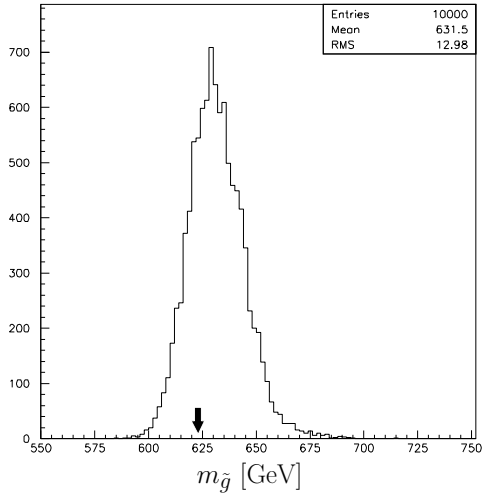}
\end{center}
\caption{The statistical error of the gluino mass for point~I. 
 The arrow indicates the input value used in the Monte Carlo event 
 generation.  The accuracy at the level of $2\%$ is obtained.}
\label{fig:error.mgl.600}
\end{figure}

Combining the information of $M_{jj}^{\rm max}$ with the analysis 
in the last subsection, we can determine the gluino mass.  The 
reconstructed gluino mass is shown in Fig.~\ref{fig:error.mgl.600}.
The input value of $m_{\tilde{g}} = 623~{\rm GeV}$ is indicated by 
the arrow.  We see that the reconstruction is successful within the 
statistical uncertainty.  With the approximation of the statistical 
fluctuation to take the Gaussian form, we obtain
\begin{equation}
  m_{\tilde{g}} = 632 \pm 13~{\rm GeV}.
\label{eq:mgl_mjj}
\end{equation}
We find that a quite accurate ($\approx 2\%$) measurement of the 
gluino mass is possible by this method. 

\begin{figure}[t]
\begin{center}
  \includegraphics[height=6.5cm]{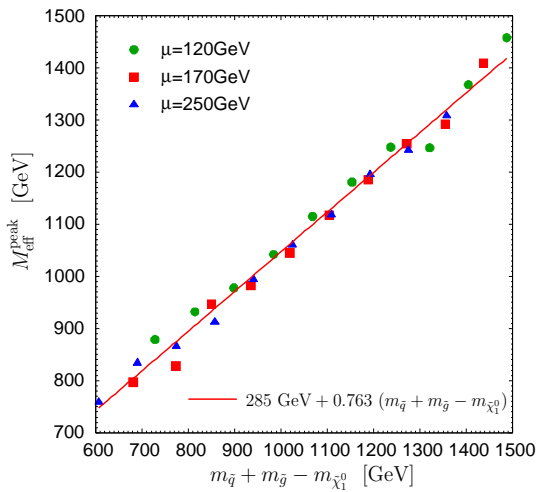}
\end{center}
\caption{An empirical relation between the peak location of $M_{\rm eff}$ 
 and a combination $m_{\tilde{q}} + m_{\tilde{g}} - m_{\tilde{\chi}^0_1}$. 
 We have generated 50,000 events for each 30 sample points in the mixed 
 moduli-anomaly mediation model.  A very good linear relation is obtained. 
 Note that this relation will be modified if different selection cuts 
 are used.  The standard model background has not been included in 
 searching the peak location.}
\label{fig:meff_relation}
\end{figure}

There is another method of extracting the gluino mass, which can be 
used in any model within the class considered here.  This is to use 
the effective mass $M_{\rm eff}$ defined by
\begin{equation}
  M_{\rm eff} = E_T^{\rm miss} + \sum_i p_T^i,
\end{equation}
where the sum runs over all the jets.  The peak location of this 
variable is known to have a correlation with the gluino and squark 
masses~\cite{Baer:1995nq,Hinchliffe:1996iu}.  In particular, as we 
will see below, we have a definite relation between the peak location 
of $M_{\rm eff}$ and the superparticle masses within the model used 
here.  To perform this analysis, we use the cut criteria listed in 
Ref.~\cite{Tovey:2000wk} to select the events, except for the lepton,
$b$-jet and $\tau$-jet vetoes:
\begin{itemize}
  \item $\geq$ 4 jets with $p_T \geq 50~{\rm GeV}$
  \item $\geq$ 2 jets with $p_T \geq 100~{\rm GeV}$
  \item $E_T^{\rm miss} \geq 
        \max\bigl\{ 100~{\rm GeV},\ 0.25 \sum_i p_T^i \bigr\}$
  \item Transverse sphericity $S_T \geq 0.2$
  \item $\varDelta \phi_{({\bf p}_T^1,{\bf p}_T^2)} \leq 170^\circ$
  \item $\varDelta \phi_{({\bf p}_T^1+{\bf p}_T^2,{\bf p}_T^{\rm miss})}
        \leq 90^\circ$
  \item Veto leptons, $b$-jets, $\tau$-jets
\end{itemize}
By generating supersymmetric events for various parameter points in the 
model, we find an excellent linear relation between $M_{\rm eff}^{\rm 
peak}$ and $m_{\tilde{q}} + m_{\tilde{g}} - m_{\tilde{\chi}^0_1}$ as 
shown in Fig.~\ref{fig:meff_relation}.
The relation is given by
\begin{equation}
  M_{\rm eff}^{\rm peak} = 285~{\rm GeV} 
    + 0.763\, (m_{\tilde{q}}+m_{\tilde{g}}-m_{\tilde{\chi}^0_1}).
\label{eq:meff_relation}
\end{equation}
Note that this should be regarded as a sort of theoretical prediction 
as we have not included the standard model background in producing 
the plot.  By using this empirical fact, we can extract the combination 
$m_{\tilde{q}} + m_{\tilde{g}} - m_{\tilde{\chi}^0_1}$ from the 
$M_{\rm eff}$ peak analysis.

\begin{figure}[t]
\begin{center}
  \includegraphics[height=6.5cm]{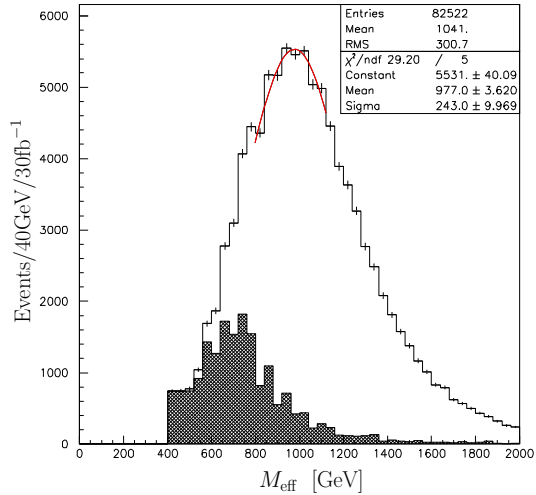}
\end{center}
\caption{The distribution of $M_{\rm eff}$ for point~I. The QCD 
 background is smeared in order not to artificially magnify the 
 statistical uncertainty due to the scaling of the events.  The 
 location of the peak is determined by fitting with a Gaussian 
 function near the peak.  It is given by $M_{\rm eff}^{\rm peak} 
 = 977~{\rm GeV}$.}
\label{fig:meff.600}
\end{figure}

In Fig.~\ref{fig:meff.600} we show the distribution of $M_{\rm eff}$ 
described above.
We obtain the peak location $M_{\rm eff}^{\rm peak} = 977~{\rm GeV}$ by 
fitting the histogram near the peak with a Gaussian function.  With the 
assumption that the theoretical relation in Eq.~(\ref{eq:meff_relation}) 
is accurate at a 5\% level, we obtain the gluino mass
\begin{equation}
  m_{\tilde{g}} = 590 \pm 62~{\rm GeV},
\end{equation}
by combining the $M_{\rm eff}^{\rm peak}$ analysis here with the 
analysis in the last subsection.  We find that the error amounts to 
$O(10\%)$.  We thus conclude that the $M_{jj}$ endpoint analysis is 
more useful than the $M_{\rm eff}^{\rm peak}$ analysis to determine 
the gluino mass not only because it is more model independent but 
also because it has better accuracy.

\subsection{The case with large superparticle masses}
\label{subsec:M0=900}

In this subsection we repeat the analyses in 
subsections~\ref{subsec:squark-Higgsino} and \ref{subsec:gluino} 
for the case of large superparticle masses (point~II in 
Table~\ref{table:points}) to see if accurate measurements are 
still possible despite the smaller statistics due to smaller 
superparticle production cross sections.  The analyses below 
show that essentially the same method can be used to determine the 
mass parameters in good accuracy.  In fact, similar or even better 
accuracy is obtained for the $m_{\tilde{q}}$ and $m_{\tilde{g}}$ 
determination compared to the case with low superparticle masses, 
as we will see below.

\begin{figure}[t]
\begin{center}
  \includegraphics[height=6.5cm]{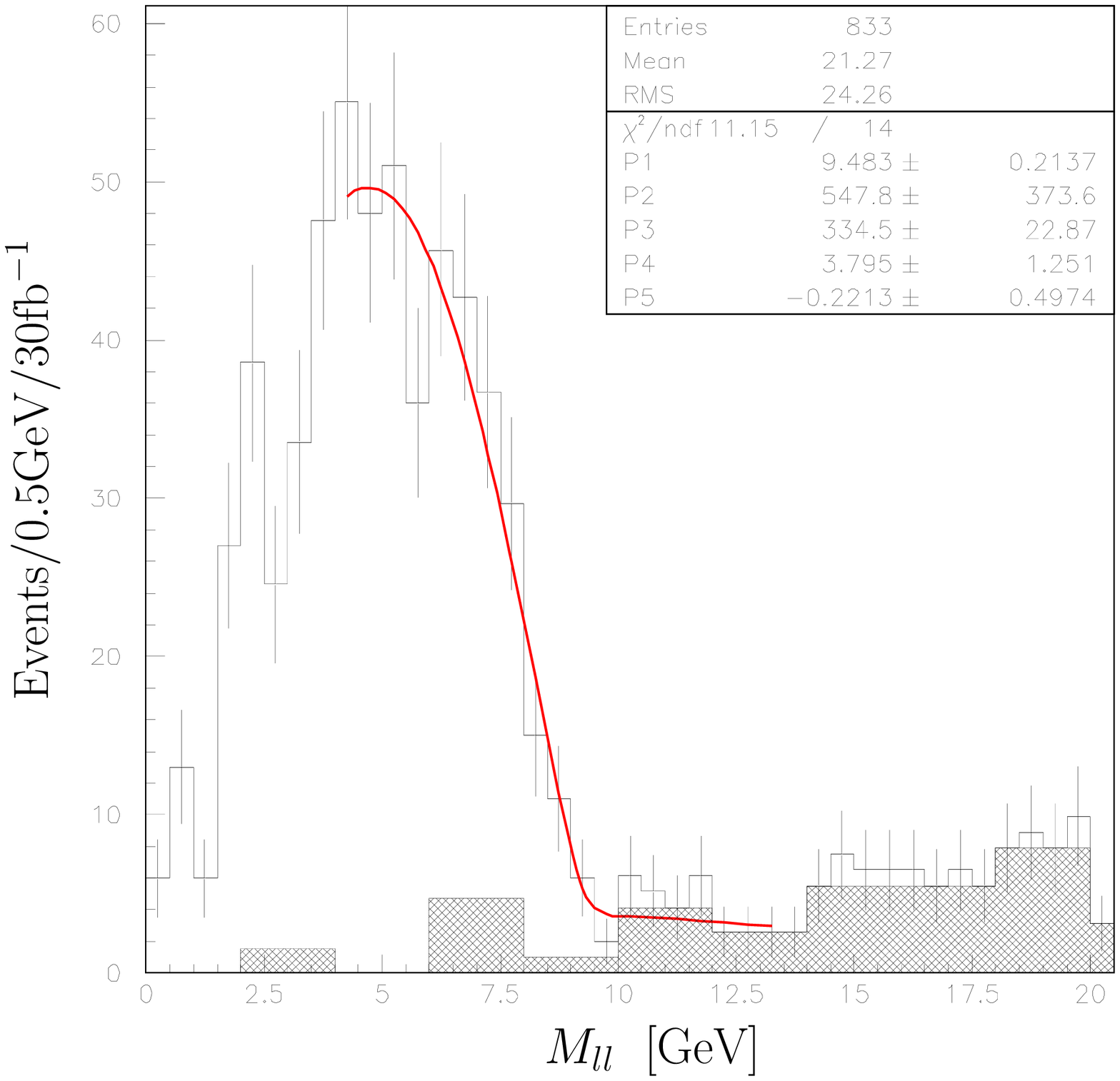}
  \hspace*{.5cm}
  \includegraphics[height=6.5cm]{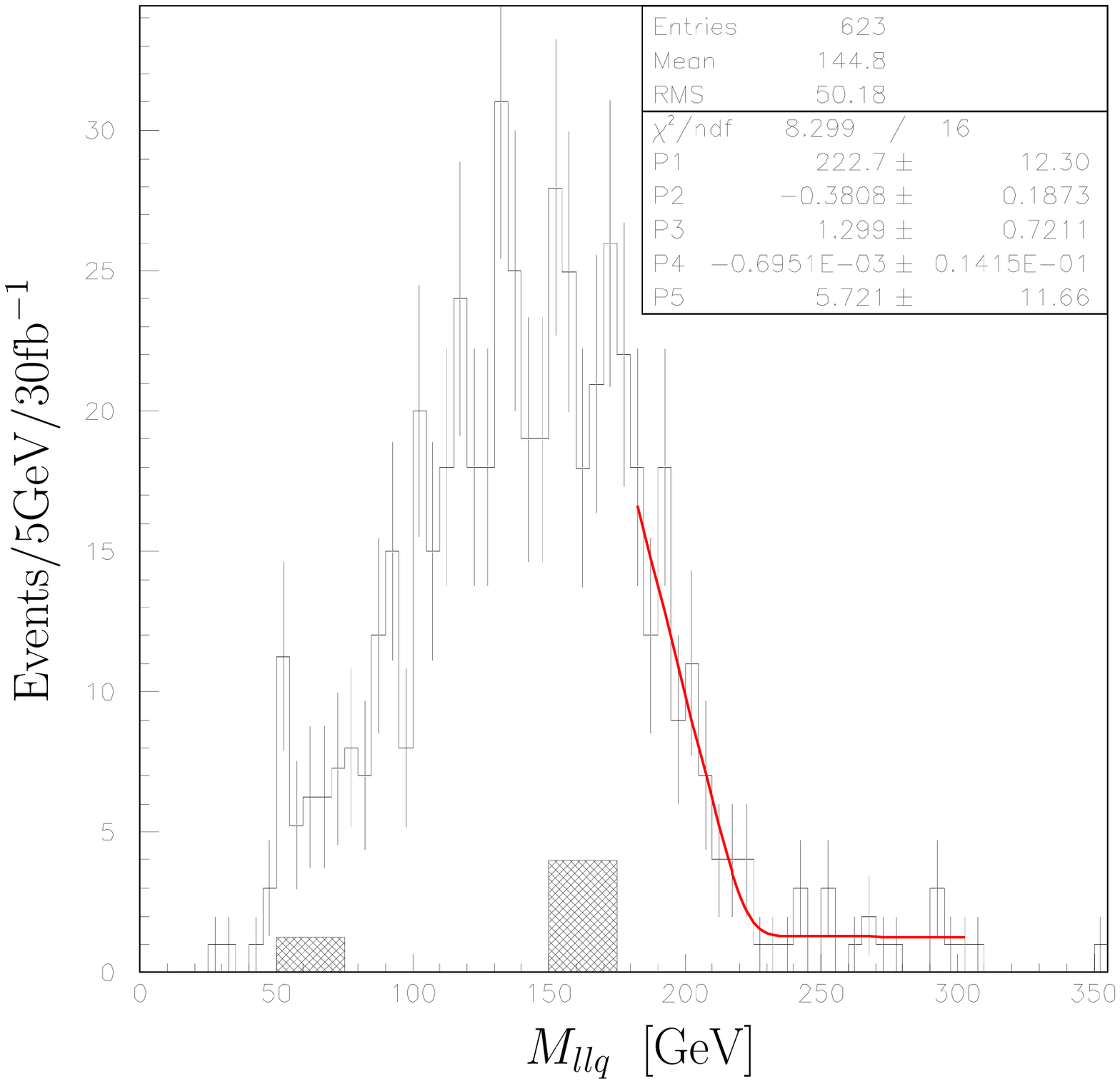}
\end{center}
\caption{The distributions of $M_{ll}$ (left) and $M_{llq}$ (right) 
 for point~II.  Hatched histogram represents the standard model 
 background, which is smeared over five bins in order not to magnify 
 the statistical uncertainty due to the scaling of the events.  For 
 $M_{ll}$, we have searched the endpoint by fitting with the theoretical 
 curve assuming $\eta_\chi = -1$.  The endpoint is obtained to be 
 $9.48 \pm 0.21~{\rm GeV}$.  For $M_{llq}$, we have used a linear 
 function with Gaussian smearing for the signal and a linear 
 function for the background.  The endpoint is obtained to be 
 $223 \pm 12~{\rm GeV}$.}
\label{fig:mll.900}
\end{figure}

The $M_{ll}$ and $M_{llq}$ distributions are shown in 
Fig.~\ref{fig:mll.900}.
We have used the same cuts with those in 
subsection~\ref{subsec:squark-Higgsino} for the event selection. 
We can see the clear endpoints in both distributions.  The $M_{ll}$ 
distribution is fitted with the theoretical curve with $\eta_\chi 
= -1$.  A good fit is obtained only for $\eta_\chi = -1$, allowing a 
successful measurement of $\eta_\chi$.  The endpoints are obtained as:
\begin{equation}
  M_{ll}^{\rm max} = 9.48 \pm 0.21~{\rm GeV},
\quad
  M_{llq}^{\rm max} = 223 \pm 12~{\rm GeV}.
\end{equation}
where the $M_{llq}$ endpoint is determined by a fit using a linear 
function with Gaussian smearing together with a linear background shape.

\begin{figure}[t]
\begin{center}
  \includegraphics[height=6.5cm]{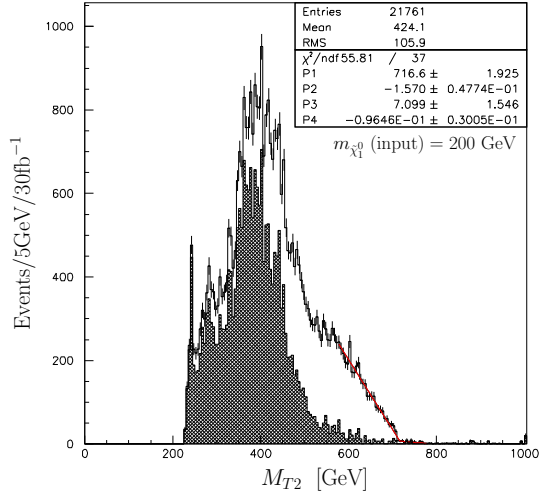}
\end{center}
\caption{The $M_{T2}$ distribution for the input value of the neutralino 
 mass $m_{\tilde{\chi}^0_1} = 200~{\rm GeV}$ for point~II.  Hatched 
 histogram is the standard model background.  The endpoint is extracted 
 by fitting the signal plus background histogram with a linear function, 
 and the background near the endpoint by a linear function.  The endpoint 
 is obtained to be $716.6 \pm 1.9~{\rm GeV}$.}
\label{fig:mT2.900}
\end{figure}

The $M_{T2}$ distribution with an input $m_{\tilde{\chi}^0_1} = 
200~{\rm GeV}$ is shown in Fig.~\ref{fig:mT2.900}.
The signal to background ratio is not so large, but there is no 
significant background near the endpoint because of the large squark 
mass.  The endpoint measurement, therefore, does not suffer seriously 
from the standard model background.  We obtain $716.6 \pm 1.9~{\rm GeV}$ 
for this value of the input neutralino mass.  We then repeat the 
analysis for six different input neutralino masses and obtain a 
curve on the $m_{\tilde{\chi}^0_1}$--$m_{\tilde{q}}$ plane by 
interpolating these points.

\begin{figure}[!ht]
\vspace{0.5cm}
\begin{center}
  \includegraphics[height=6.5cm]{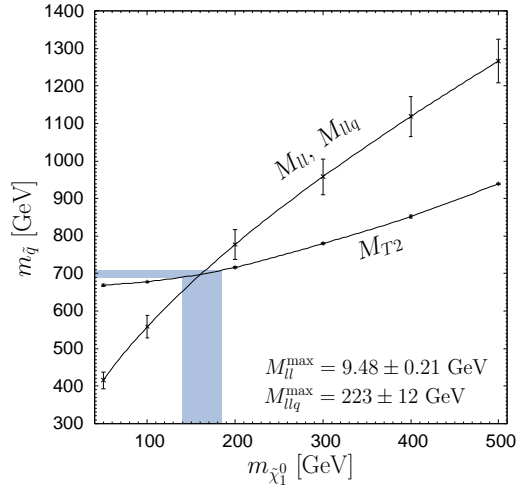}
\end{center}
\caption{Two curves on the $m_{\tilde{\chi}^0_1}$--$m_{\tilde{q}}$ 
 plane deduced from the cascade decay analysis, $M_{ll}^{\rm max}$ 
 and $M_{llq}^{\rm max}$, and the squark pair production analysis, 
 $M_{T2}^{\rm max}$, for point~II. Both curves are obtained by 
 inputting hypothetical values of $m_{\tilde{\chi}^0_1}$.  The 
 intersection determines the real values of $m_{\tilde{\chi}^0_1}$ 
 and $m_{\tilde{q}}$.  The obtained masses with the $1 \sigma$ 
 statistical errors are shown by shaded bands.}
\label{fig:mchiVSmsq.900}
\end{figure}

The two curves obtained from the $M_{ll}$ and $M_{llq}$ endpoints 
and the $M_{T2}$ endpoint are shown in Fig.~\ref{fig:mchiVSmsq.900}.
We find that the crossing angle of the curves is larger compared to 
the case in Fig.~\ref{fig:mchiVSmsq.600}.  This property makes it 
possible to measure the squark mass in this point with similar accuracy 
to the case of low superparticle masses, even though we have larger 
statistical uncertainty.

\begin{figure}[t]
\begin{center}
  \includegraphics[height=6.5cm]{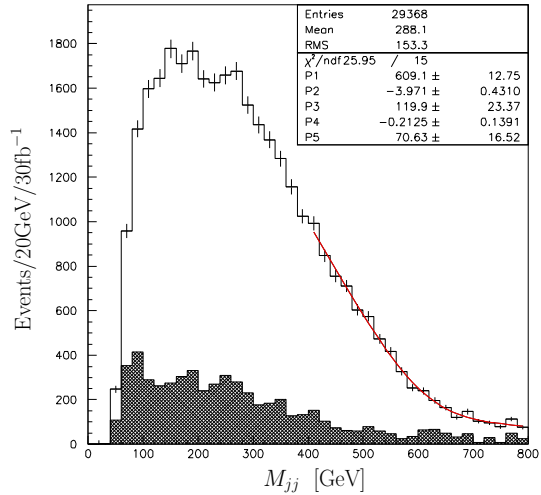}
\end{center}
\caption{The $M_{jj}$ invariant mass distribution for point~II. 
 We select events with three hard jets and take a combination that 
 gives the smallest $M_{jj}$, out of the three combinations, in 
 order to see the endpoint.  A fit is performed with a linear 
 function with Gaussian smearing together with a linear function 
 for the background events.  The endpoint is obtained as 
 $609 \pm 13~{\rm GeV}$.}
\label{fig:mjj.900}
\end{figure}

For the $M_{jj}$ endpoint measurement, we use a different cut for 
the jet $p_T$.  We take
\begin{itemize}
  \item Three and only three jets with $p_T > 100~{\rm GeV}$
\end{itemize}
instead of $p_T > 50~{\rm GeV}$, used in the analysis of the low 
superparticle mass case, because a clear endpoint is not obtained 
with the $50~{\rm GeV}$ cut.  The resulting $M_{jj}$ distribution 
is shown in Fig.~\ref{fig:mjj.900}.
By fitting with a linear function with Gaussian smearing and 
a linear background, we obtain
\begin{equation}
  M_{jj}^{\rm max} = 609 \pm 13 ~{\rm GeV}.
\end{equation}

\begin{figure}[!ht]
\begin{center}
  \includegraphics[height=5.cm]{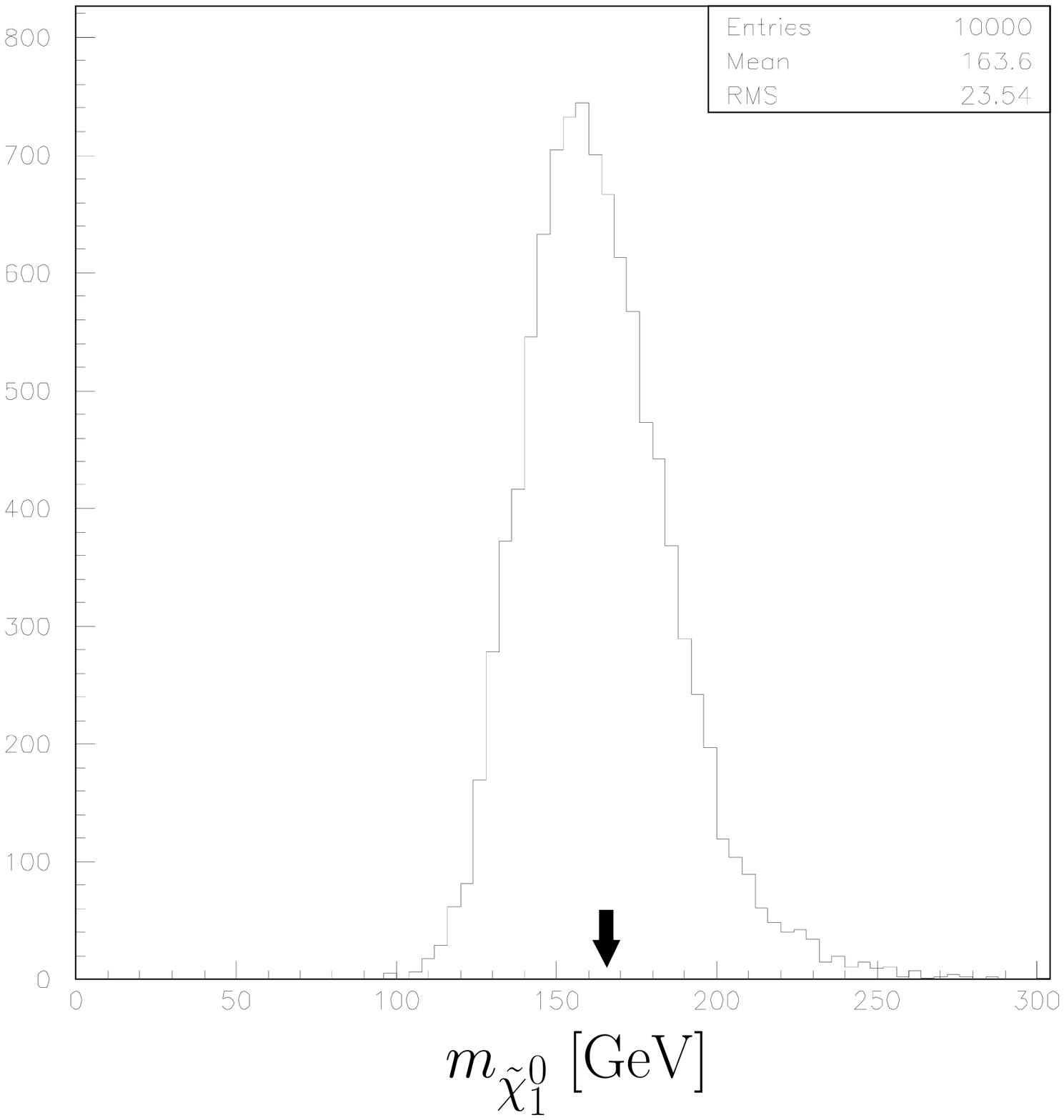}
\hspace*{.5cm}
  \includegraphics[height=5.cm]{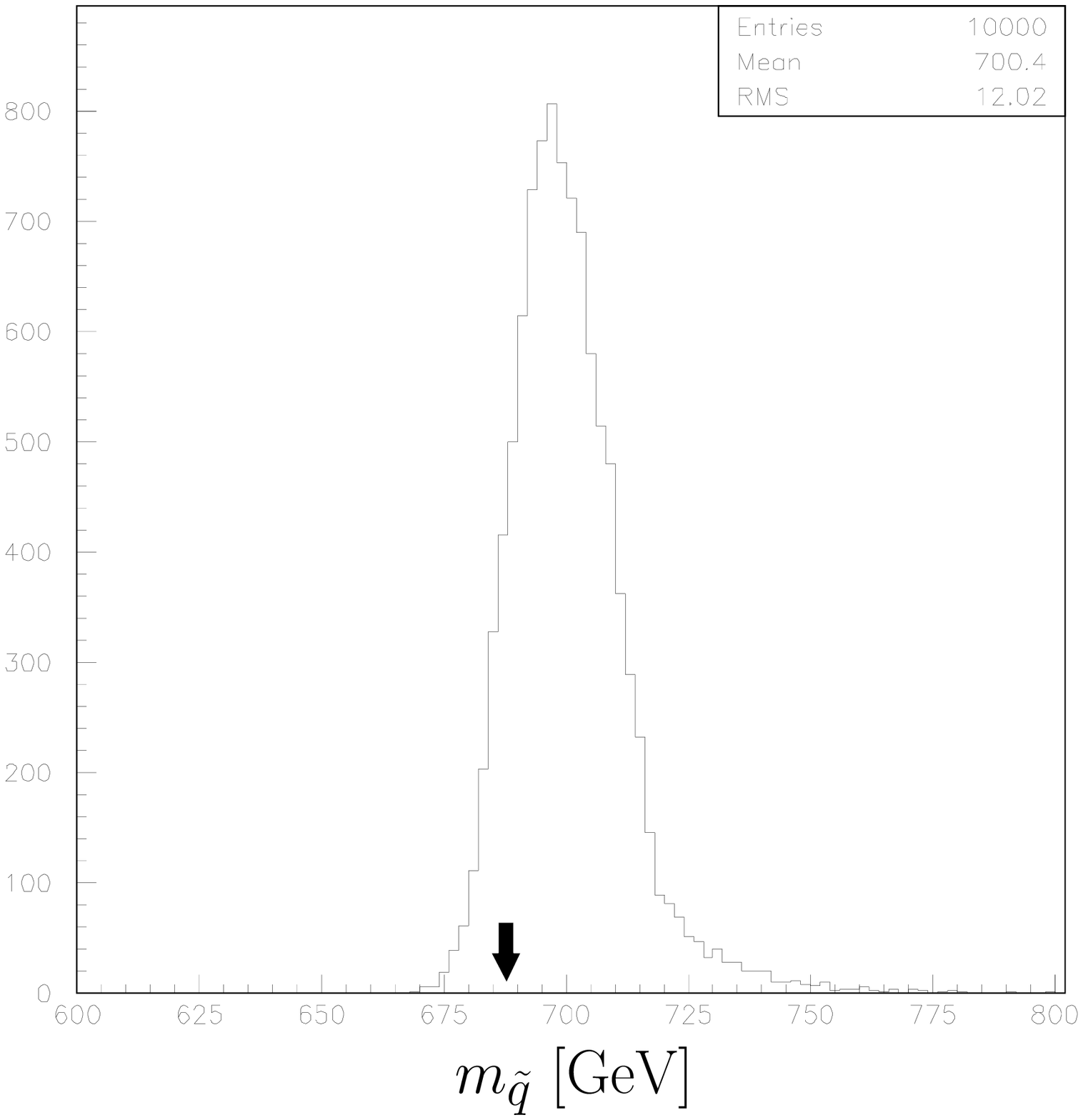}
\\ \vspace*{.5cm}
  \includegraphics[height=5.cm]{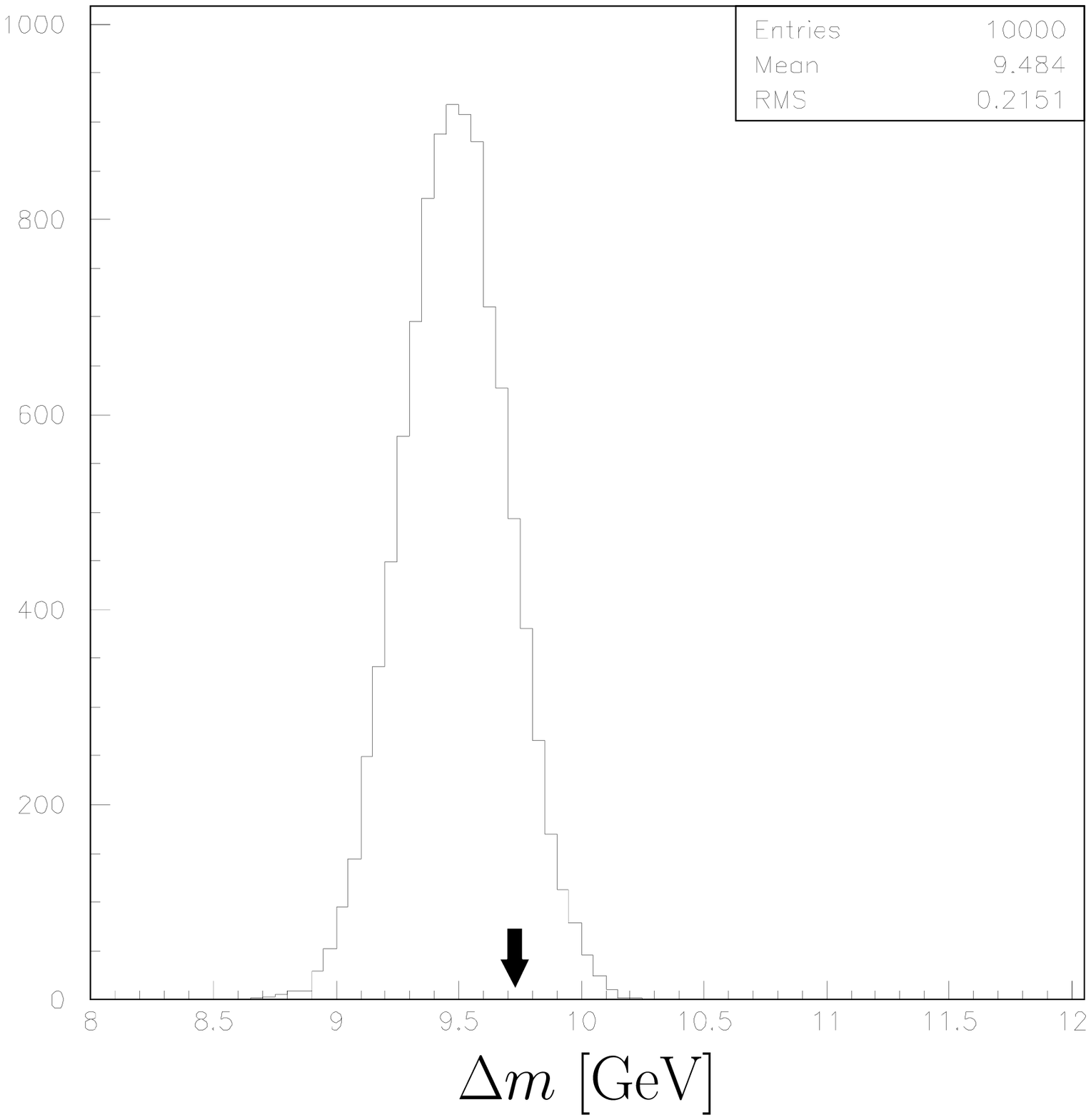}
\hspace*{.5cm}
  \includegraphics[height=5.cm]{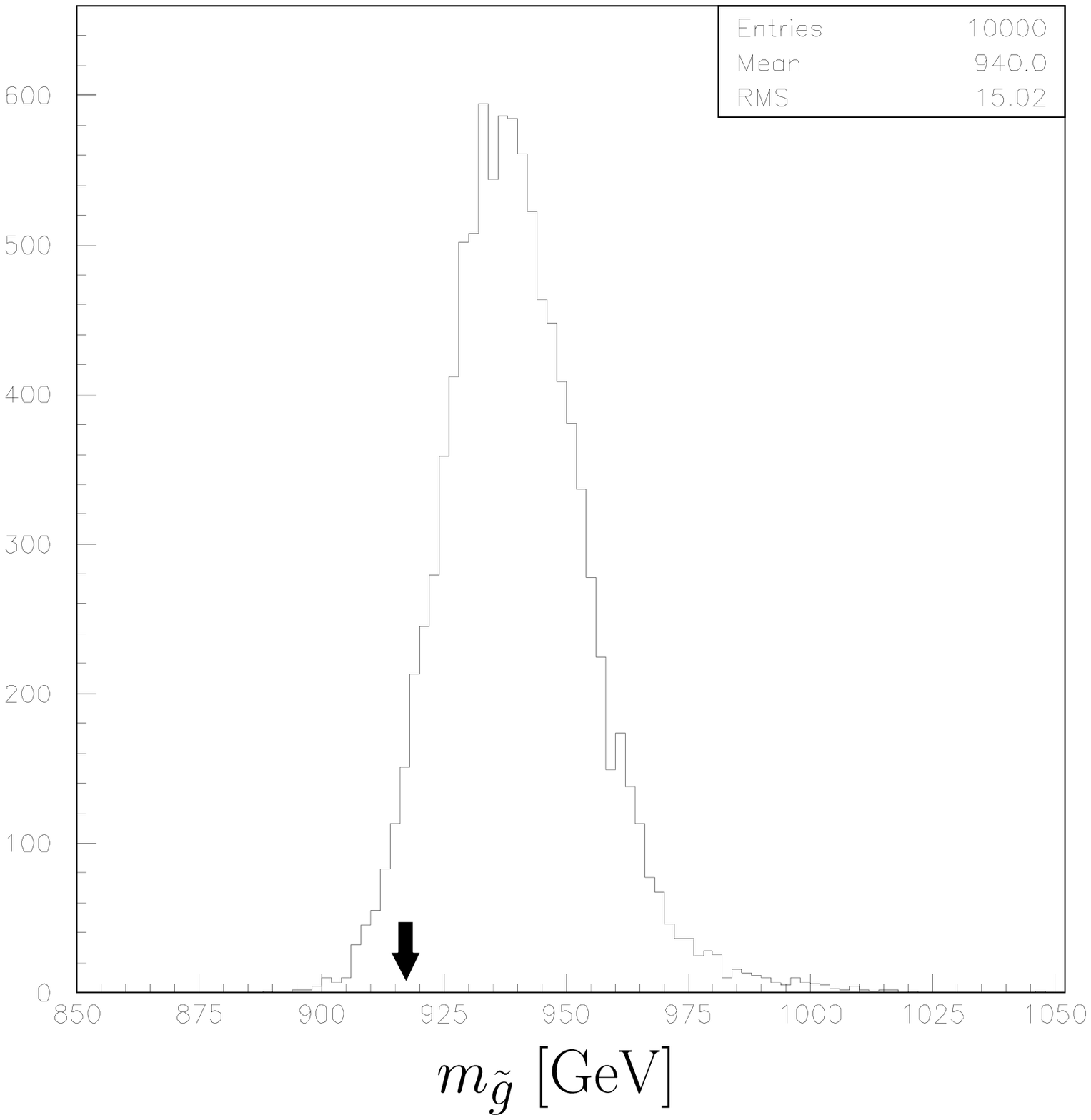}
\end{center}
\caption{The statistical errors of the measured mass parameters 
 $m_{\tilde{\chi}^0_1}$, $m_{\tilde{q}}$, $\varDelta m$ and 
 $m_{\tilde{g}}$ for point~II.  Arrows indicate the input values 
 used in the Monte Carlo event generation.  The accuracy at 
 the level of $15\%$, $2\%$, $2\%$ and $2\%$ is obtained for 
 $m_{\tilde{\chi}^0_1}$, $m_{\tilde{q}}$, $\varDelta m$ and 
 $m_{\tilde{g}}$, respectively.}
\label{fig:error.900}
\end{figure}

Combining all the results, we can determine the four mass parameters, 
$m_{\tilde{\chi}^0_1}$, $m_{\tilde{q}}$, $\varDelta m$ and 
$m_{\tilde{g}}$.  The estimation of the statistical uncertainties 
is given in Fig.~\ref{fig:error.900}.
The input values are indicated with the arrows, which are all within 
reasonable statistical uncertainties.  The gluino mass is obtained 
with slightly larger values.  This is caused by the systematics that 
the $M_{jj}$ endpoint tends to give larger values than the one obtained 
in Eq.~(\ref{eq:mjj-endpoint}) when we use the $p_T > 100~{\rm GeV}$ 
cut for jets.  Therefore, the estimation of the systematic error will 
be important in this analysis.  By fitting the histograms with Gaussian 
functions, we obtain
\begin{equation}
  m_{\tilde{\chi}^0_1} = 164 \pm 24~{\rm GeV},
\quad
  m_{\tilde{q}} = 700 \pm 12~{\rm GeV},
\quad
  \varDelta m = 9.5 \pm 0.2~{\rm GeV},
\end{equation}
\begin{equation}
  m_{\tilde{g}} = 940 \pm 15 ~{\rm GeV}.
\label{eq:mgl.900}
\end{equation}
The neutralino, squark and gluino masses can be measured at $15\%$, 
$2\%$ and $2\%$ level accuracy, respectively.  Since this point 
represents the case of the highest superparticle masses from the 
naturalness requirement (see Eq.~(\ref{eq:M0-range})), the above 
analysis shows that the method of mass determination developed here 
covers the entire region of the parameter space.

\begin{figure}[t]
\begin{center}
  \includegraphics[height=6.5cm]{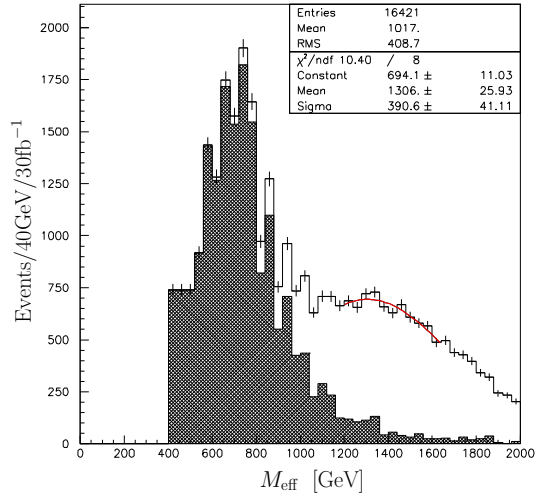}
\end{center}
\caption{The $M_{\rm eff}$ distribution for point~II.  The QCD 
 background is smeared in order not to artificially magnify the 
 statistical uncertainty due to the scaling of the events.  The peak 
 location is determined by fitting with a Gaussian function near the 
 peak.  It is found to be $M_{\rm eff}^{\rm peak} = 1306~{\rm GeV}$.}
\label{fig:meff.900}
\end{figure}

For completeness, we show in Fig.~\ref{fig:meff.900} the $M_{\rm eff}$ 
distribution.
The standard model background is huge, although it is not so 
significant around the peak location.  By using the relation in 
Eq.~(\ref{eq:meff_relation}), we obtain the gluino mass
\begin{equation}
  m_{\tilde{g}} = 801 \pm 85~{\rm GeV},
\end{equation}
which is slightly deviated from the input value of $917~{\rm GeV}$ 
(by $1.4\, \sigma$).  This little discrepancy is mainly caused 
by the shift of $M_{\rm eff}$ in the lower direction due to the 
standard model background.  For a realistic use of the $M_{\rm eff}$ 
analysis in the high superparticle mass region, one needs to develop 
a better understanding of the shape of the standard model background 
and/or to devise a better cut (especially on $E_T^{\rm miss}$) to 
reduce the standard model background.  Note that the relation in 
Eq.~(\ref{eq:meff_relation}) will be modified if different cuts 
are used.

\subsection{Testing the model with mixed moduli-anomaly mediation}
\label{subsec:model-test}

One of the most important features of the LHC experiment is its 
potential of ruling out models.  With the limited precisions of 
various measurements, it is extremely important to develop methods 
of testing model predictions rather than just measuring parameters 
under the assumption of a particular model.  We here take the 
model with mixed moduli-anomaly mediation as an example and discuss 
possible ways of testing the model.

We have already seen that one of the characteristic features, the 
Higgsino LSP, can be tested using the distribution of the dilepton 
invariant mass from $\tilde{\chi}^0_2$ decay.  Other interesting 
features of the model include approximate universality of the gaugino 
masses and the definite ratio between the sfermion and gaugino 
masses, given in Eq.~(\ref{eq:moduli-anomaly}).  We can test these 
features by using the mass parameters obtained in the previous 
analyses.  Specifically, we can calculate the parameter $M_0$ in 
three different ways and compare them with each other.  If the 
relations predicted in the model hold, the three values must 
coincide within (mostly theoretical) uncertainties/corrections.

The first way of calculating $M_0$ is to use the measured neutralino 
mass difference, $\varDelta m$.  By using Eq.~(\ref{eq:chi0_2-chi0_1}) 
with $M_1 \simeq M_2 \simeq M_0$, we can extract $M_0$ as
\begin{equation}
  M_0 \simeq \frac{m_Z^2}{ \varDelta m}.
\label{eq:M0-1}
\end{equation}
This must give the correct value of $M_0$ up to corrections of 
$\approx 15\%$, which come mainly from $O(|\mu|\sin 2\beta/M_0)$ 
corrections in diagonalizing the neutralino mass matrix and from 
the effect of running between the effective messenger scale, 
$M_{\rm mess}$, and the gaugino mass scale.  The other two ways 
use direct measurements of the squark and gluino masses:
\begin{equation}
  M_0 \simeq \sqrt{2} m_{\tilde{q}},
\qquad
  M_0 \simeq m_{\tilde{g}},
\label{eq:M0-2}
\end{equation}
which must also give the correct value of $M_0$ up to corrections. 
The corrections come, for example, from running between $M_{\rm mess}$ 
and the superparticle mass scale as well as from finite supersymmetric 
QCD corrections.  For the squark masses, the $SU(2)_L$ and $U(1)_Y$ 
$D$-terms also give corrections of $O(m_Z^2/M_0^2)$.  All these corrections, 
again, amount to $\approx 15\%$.  We thus conclude that the three values 
of $M_0$ calculated using Eqs.~(\ref{eq:M0-1},~\ref{eq:M0-2}) must 
coincide within $15\%$, if the model is actually realized.  This can 
provide a rather nontrivial test for the model.

\begin{figure}[t]
\begin{center}
  \includegraphics[height=6.5cm]{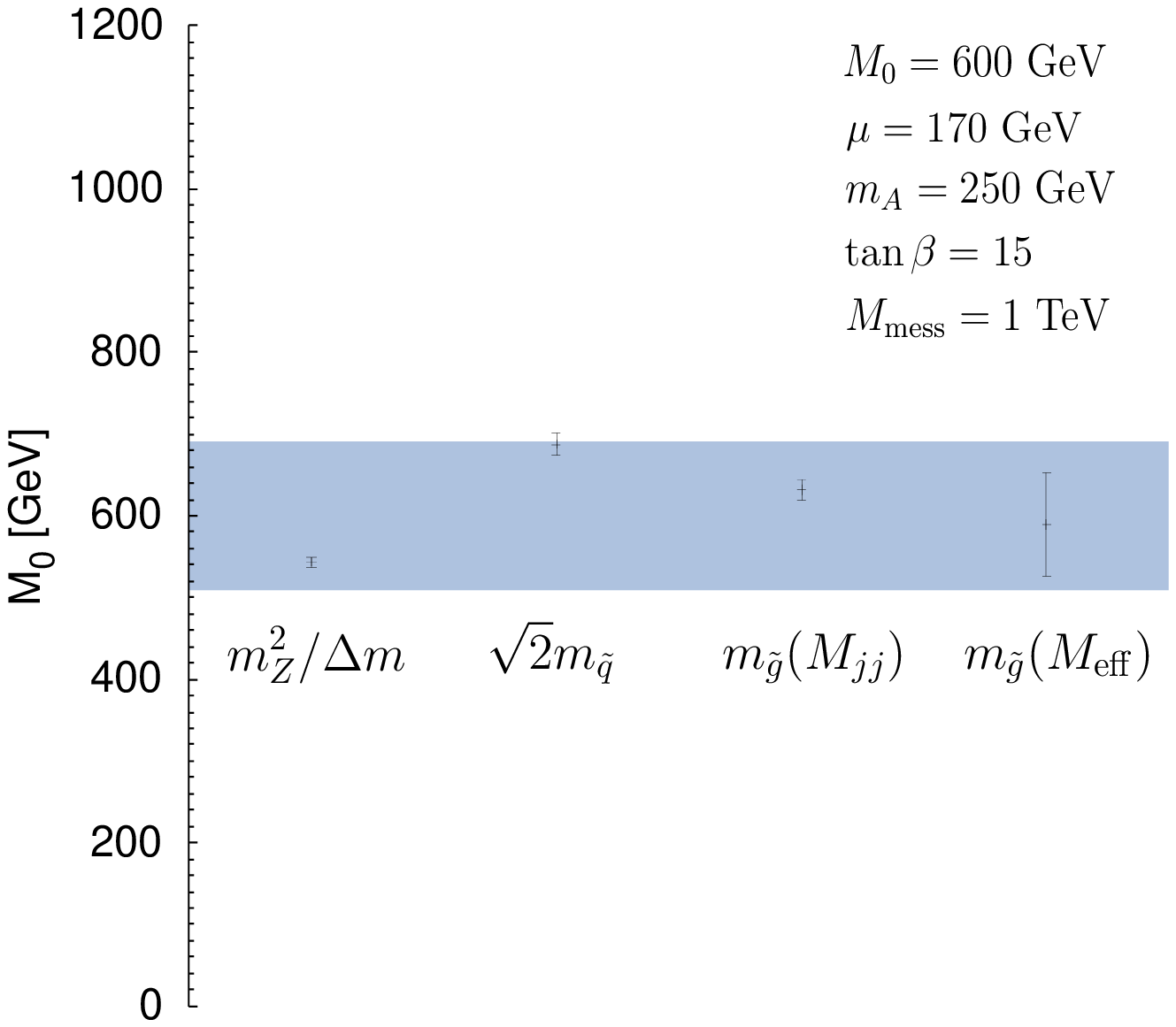}
\hspace*{0.5cm}
  \includegraphics[height=6.5cm]{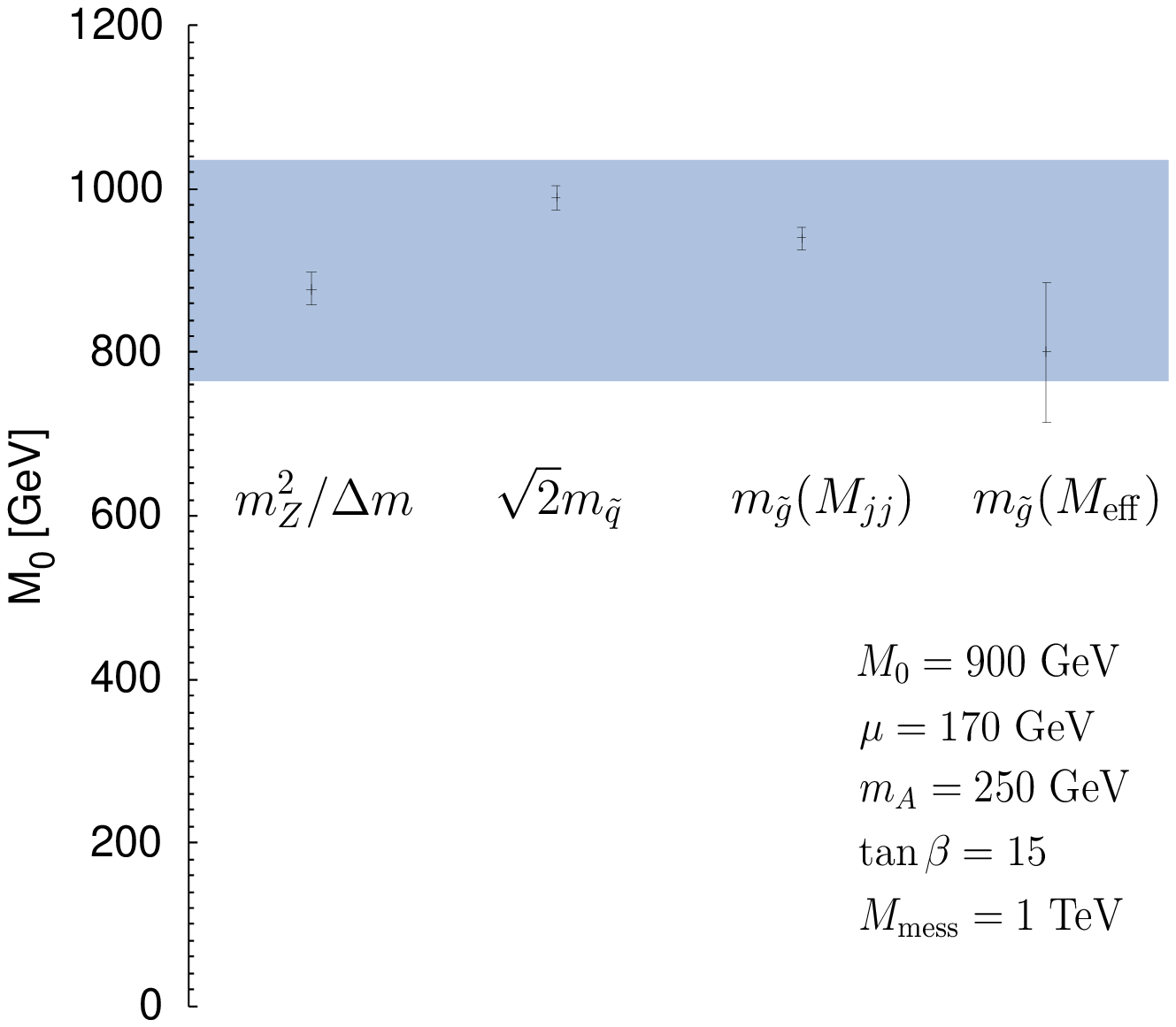}
\end{center}
\caption{A nontrivial test for the mixed moduli-anomaly mediation 
 model in the case of $M_0 = 600~{\rm GeV}$ (left) and $ M_0 = 
 900~{\rm GeV}$ (right).  Three ways of calculating $M_0$, i.e. 
 $m_Z^2/\varDelta m$, $\sqrt{2} m_{\tilde{q}}$ and $m_{\tilde{g}}$, give 
 the same value within $\approx 15\%$ theoretical uncertainties.  For 
 $m_{\tilde{g}}$, we have plotted the values obtained in two different 
 ways, i.e. from $M_{jj}^{\rm max}$ and $M_{\rm eff}^{\rm peak}$.}
\label{fig:model}
\end{figure}

We show in Fig.~\ref{fig:model} the values of $M_0$ calculated in 
three different ways for point~I ($M_0 = 600~{\rm GeV}$; left panel) 
and for point~II ($M_0 = 900~{\rm GeV}$; right panel).
The shaded regions indicate the $\pm 15\%$ range around the true values 
of $M_0$ ($= 600~{\rm GeV}$ and $900~{\rm GeV}$).  For $m_{\tilde{g}}$, 
we have plotted both values obtained from the $M_{jj}$ endpoint and 
the $M_{\rm eff}$ peak analyses.  As we can see, all measurements 
agree with each other within $15\%$ undertainties, both for the cases 
of point~I and point~II.  Moreover, we can even understand a nature of 
the dispersions in the plot.  We find that the $O(|\mu|\sin 2\beta/M_0)$ 
correction to $m_Z^2/\varDelta m$ is negative whereas the QCD corrections 
to the squark and gluino masses are positive, as expected from theory. 
Since experimental errors on determining these quantities are rather 
small, the structure of these dispersions might be useful for deducing 
further detailed structures of the underlying theory, such as the 
structure of higher order corrections to the superparticle masses.

\section{Discussion and Conclusions}
\label{sec:concl}

Weak scale supersymmetry is often said to be fine-tuned, especially 
if the matter content is minimal.  Is it true?  If the LEP~II 
bound on the Higgs boson mass pushed the top squark masses above 
a TeV, as is sometimes stated in literature, this would be 
a true statement.  The size of the top-stop loop contribution to 
the Higgs boson mass-squared parameter, $m_{H_u}^2$, would then be 
larger than about $(250~{\rm GeV})^2$ even for a unit logarithm, 
$\ln(M_{\rm mess}/m_{\tilde{t}}) \approx 1$, leading to fine-tuning 
of order $10\%$ {\it independently of} an underlying mechanism of 
supersymmetry breaking (see Eqs.~(\ref{eq:higgs-mass},~\ref{eq:mh2})). 
In fact, in most supersymmetry breaking models, the logarithm 
is (much) larger than 1, leading to (much) severer fine-tuning. 
For example, fine-tuning is already as bad as $2\%$ for $M_{\rm mess} 
\approx 100~{\rm TeV}$.  Such heavy top squarks, therefore, would 
not allow natural electroweak symmetry breaking in the context 
of minimal supersymmetry.

The situation, however, is not as described above if there is 
a large $A_t$ term at the weak scale.  For $|A_t/m_{\tilde{t}}| 
\approx (1.5\!\sim\!2.5)$, the top squark masses can be as small 
as $m_{\tilde{t}} \approx (250\!\sim\!400)~{\rm GeV}$ to evade 
the LEP~II constraint of $M_{\rm Higgs} \simgt 114.4~{\rm GeV}$, 
so that the ``model-independent'' contribution to $m_{H_u}^2$ from 
top-stop loop is only about $(100~{\rm GeV})^2$, even including the 
contribution from the $A_t$ term.  (The lower bound on $m_{\tilde{t}}$ 
here, $\approx 250~{\rm GeV}$, arises in fact from the direct search 
bound of the top squark, $m_{\tilde{t}_1} \simgt 100~{\rm GeV}$, 
and not from the Higgs boson mass bound.)  Such light top squarks 
allow ``fine-tuning'' to be only of $O(20\%)$ for $M_{\rm mess} 
= O(10~{\rm TeV})$ and allow completely natural electroweak symmetry 
breaking for smaller values of $M_{\rm mess}$.  With these low 
energy spectra, the amount of fine-tuning, $\Delta^{-1}$, can 
also be small for theories with large logarithms.  We find that 
$\Delta^{-1} \approx 10\%$ is possible even in theories with 
$M_{\rm mess} \sim M_{\rm unif}$.  Therefore, under the current 
experimental constraints, the supersymmetric fine-tuning problem 
should {\it not} be regarded as the problem of minimal supersymmetry 
itself but as the problem of specific supersymmetry breaking mechanisms.

Characteristic features required to have (relatively) natural electroweak 
symmetry breaking in minimal supersymmetric theories are (i) a large $A$ 
term for the top squarks, $|A_t/m_{\tilde{t}}| \approx (1.5\!\sim\!2.5)$ 
(ii) light top squarks (iii) a moderately large ratio of the electroweak 
VEVs, $\tan\beta \simgt 5$, and (iv) a small $\mu$ term, $|\mu| \simlt 
190~{\rm GeV}$ ($270~{\rm GeV}$) for $\Delta^{-1} \geq 20\%$ ($10\%$). 
A generic implication of these low energy spectra is a relatively 
light Higgs boson, $M_{\rm Higgs} \simlt 120~{\rm GeV}$.  There will 
be classes of theories leading to these features/spectra, and we 
have identified two representative ones (see Fig.~\ref{fig:spectra}). 
Among them, a class giving nearly universal gaugino and sfermion masses 
at low energies (see Fig.~\ref{fig:spectra}(b)) can make electroweak 
symmetry breaking most natural.  Examples for these theories are obtained 
by employing moduli-type, or boundary condition, supersymmetry breaking 
(effectively) at a low scale.  A consistency with the desert can be 
explicitly recovered if the setup is realized through mixed moduli 
and anomaly mediated supersymmetry breaking.

An important consequence of the class of theories described above is 
that the Higgsinos are the lightest among the superpartners of the 
standard model particles.  Assuming that the gravitino is not much lighter 
than the Higgsinos, which is actually the case in the model with mixed 
moduli-anomaly mediation, the existence of the nearly degenerate Higgsino 
states ($\tilde{\chi}^0_1$, $\tilde{\chi}^0_2$ and $\tilde{\chi}^+_1$) 
can give interesting signals at the LHC.  The signals arise in the 
invariant mass distribution of dileptons arising from $\tilde{\chi}^0_2$ 
decay: a smallness of the endpoint and a particular shape determined by 
the relative $CP$ property of the two neutralinos, $\tilde{\chi}^0_1$ 
and $\tilde{\chi}^0_2$.  We have argued that these signals are indeed 
useful in a wide variety of circumstances within the class of theories 
considered here.

We have explicitly demonstrated the usefulness of the signals in 
realistic analyses by performing Monte Carlo simulations, including 
detector simulations and background estimations.  We have also presented 
a method that allows the determination of all the relevant superparticle 
masses, $m_{\tilde{\chi}^0_1}$, $m_{\tilde{q}}$, $m_{\tilde{g}}$ 
and $\varDelta m = m_{\tilde{\chi}^0_2} - m_{\tilde{\chi}^0_1}$, 
independently of the details of the model.  This allows us to determine 
all the superparticle masses within the class of models considered, up 
to {\it theoretical} uncertainties of $\approx 15\%$.  Note that some 
of the existing techniques, e.g. the $M_{llq}$ threshold analysis, is 
not very useful here because of the near degeneracy of the Higgsino 
states.  We can, nevertheless, determine the four mass parameters 
with the precisions of order a few to ten percent, by combining 
various endpoint analyses.  This is extremely important because it 
provides ways to test various possible models, which generically give 
nontrivial relations among these parameters.  We have demonstrated 
this in the case of the model with mixed moduli-anomaly mediation, 
and shown that the model can indeed be tested (and thus can be 
discriminated from models that give significantly different relations 
among the four parameters).

It will be possible to perform further tests for the class of models 
discussed here.  An important issue is to measure the $A$ parameters, 
especially that of the top squarks.  This may be done, for example, 
along the lines presented in~\cite{Hisano:2002xq}.  There is also 
an important interplay between collider physics and cosmology.  As 
discussed in Ref.~\cite{Kitano:2005ew}, the present class of models 
has a large discovery potential in ongoing and future direct dark 
matter detection experiments, such as CDMS~II, if the Higgsino LSP 
composes the dark matter of the universe.  Now suppose that the mass 
and the detection cross section for the dark matter are measured in 
(one of) these experiments.  The results from the LHC can then be 
used to perform a consistency check on the LSP mass and to provide 
a constraint on the Higgs sector parameters, $m_A$ and $\tan\beta$, 
because the detection cross section depends strongly on these parameters. 
Together with the other data from the LHC, such as the one for the 
Higgs boson mass, we will be able to determine all the parameters 
of the model with certain accuracy. 

In performing all these analyses, inputs from a particular model(s) 
will be very important/useful, especially if one wants to pin 
down the parameter point of minimal weak scale supersymmetry.  The 
model(s) assumed is then better to be a ``likely'' one, compared with 
other possible models that can also ``accommodate'' the same set of 
measurements.  Naturalness of electroweak symmetry breaking, together 
with the simplicity of a model, will then keep playing an important 
guiding role in these ``model selection'' processes, which will 
most likely take the form of processes of ``slowly convincing 
ourselves.''

\section*{Acknowledgments}

The work of R.K. was supported by the U.S. Department of Energy under 
contract number DE-AC02-76SF00515.  The work of Y.N. was supported 
in part by the Director, Office of Science, Office of High Energy 
and Nuclear Physics, of the US Department of Energy under Contract 
DE-AC02-05CH11231, by the National Science Foundation under grant 
PHY-0403380, by a DOE Outstanding Junior Investigator award, and 
by an Alfred P. Sloan Research Fellowship.

\appendix
\section{Renormalization Group Properties of Moduli Mediation Models}

We here study interesting renormalization group (RG) properties 
of moduli mediated supersymmetry breaking models.  It is found in 
Ref.~\cite{Choi:2005uz} that mixed moduli-anomaly mediation models 
have an interesting RG property --- the contributions from anomaly 
mediation can be canceled by the actual one-loop running effect at 
a certain energy scale, mimicking a pure moduli mediated model at 
that energy scale.  Since a clear derivation of the result is not 
available in the paper, we present it here.  We show that the result 
can be understood as a rather straightforward consequence of special 
properties of moduli mediated models.  We use the method of 
``analytic continuation into superspace''~\cite{Arkani-Hamed:1998kj}, 
which provides a powerful way of analyzing RG equations in softly 
broken supersymmetric theories.  We explain the basic mechanism of 
this interesting property and the origin of the conditions for the 
cancellation to happen.  We also point out that the effects of mass 
thresholds are under a good theoretical control.

We consider the case where the effective Lagrangian is given at 
a scale $\Lambda$ as follows:
\begin{eqnarray}
  {\cal L} &=& \int\! d^4\theta\, (T+T^\dagger)^{r_i} 
    Q_i^\dagger e^{-2V} Q_i
  + \left( \int\! d^2 \theta\, \frac{\lambda_{ijk}}{6} 
    Q_i Q_j Q_k + {\rm h.c.} \right)
\nonumber\\
  && + \left( \int\! d^2 \theta\, T\, {\cal W}^\alpha 
    {\cal W}_\alpha + {\rm h.c.} \right),
\label{app:lagrangian}
\end{eqnarray}
where $T$ is a spurion field with a non-vanishing $F$-component. 
This is the form of the Lagrangian obtained in moduli (radion) mediated 
models.  The lowest component of $T$ represents the volume of the 
extra dimensions in which the gauge fields propagate.  The rational 
number $r_i$ represents how much fraction of the extra dimensions 
the matter $Q_i$ propagates, compared with the gauge fields.  For 
example, if the gauge fields and the matter field $Q_i$ propagate 
in six and five dimensional spacetime, respectively, $r_i$ is given 
by $(5-4)/(6-4) = 1/2$.

We now exploit the following property of moduli mediated models to 
show certain special RG properties of these models.  We first note 
that at tree level there are following simple scaling relations 
associated with the rescaling of the moduli field $T$:
\begin{equation}
  S \rightarrow a S,
\quad
  Z_i \rightarrow a^{r_i} Z_i
\quad
  {\mbox{for}}\quad  T \rightarrow a T,
\label{app:transf}
\end{equation}
where $S$ and $Z_i$ are the gauge kinetic function and the wavefunction 
factor defined by ${\cal L} \ni [S {\cal W}^\alpha {\cal W}_\alpha]_F$ 
and ${\cal L} \ni [Z_i Q_i^\dagger Q_i]_D$, respectively.  We then 
find that these scaling relations can be extended to the one-loop level 
if the moduli rescaling, $T \rightarrow a T$, is supplemented by the 
following rescalings of the RG scale $\mu_R$ and the Yukawa couplings:
\begin{equation}
  \ln \frac{\mu_R}{\Lambda} \rightarrow a \ln \frac{\mu_R}{\Lambda},
\quad
  \lambda_{ijk} \rightarrow a^{(r_i+r_j+r_k-1)/2} \lambda_{ijk},
\label{app:transf2}
\end{equation}
where $\lambda_{ijk}$ are the superpotential Yukawa couplings appearing 
in Eq.~(\ref{app:lagrangian}) (the ``physical'' Yukawa couplings are 
given by $y_{ijk} \equiv \lambda_{ijk}/(Z_i Z_j Z_k)^{1/2}$).  Once 
this property is proved (see later), we can use these scaling relations 
to show that the gauge kinetic function, $S$, and the wavefunction 
factor, $Z_i$, take the following form:
\begin{equation}
  S = T \cdot \hat{S}
    \biggl( \frac{\ln(\mu_R/\Lambda)}{T} \biggr),
\label{app:s}
\end{equation}
\begin{equation}
  Z_i = (T + T^\dagger)^{r_i} \cdot \hat{Z}_i 
    \biggl( \frac{|\lambda_{ijk}|^2}{(T+T^\dagger)^{r_i+r_j+r_k-1}},\,
    \frac{1}{T+T^\dagger} \ln\frac{\mu_R^2}{|\Lambda|^2} \biggr),
\label{app:z}
\end{equation}
at an arbitrary scale $\mu_R$.  Here, $\hat{S}$ and $\hat{Z}_i$ are 
some functions, and we have used the fact that $Z_i$ can depend only 
on $T$ through the combination $T+T^\dagger$ because of the invariance 
of the Lagrangian under the transformation $T \rightarrow T + i \beta$.

Now, suppose that the condition $r_i + r_j + r_k = 1$ is satisfied 
for the fields having the Yukawa interaction $\lambda_{ijk}$.  We then 
find that the $T$ dependencies in $\hat{S}$ and $\hat{Z}_i$ appear 
only with the renormalization scale $\mu_R$.  In this case, the gaugino 
masses, $A$ terms, and soft scalar squared masses are simply given by:
\begin{equation}
  m_\lambda = \frac{1}{2} [\ln S]_F 
  = M_0 \biggl[ 1 + \frac{2bg^2}{(4\pi)^2} \ln\frac{\mu_R}{\Lambda} \biggr],
\label{app:gaugino_mass}
\end{equation}
\begin{equation}
  A_{ijk} = - ( [\ln Z_i]_F + [\ln Z_j]_F + [\ln Z_k]_F )
  = - M_0 \biggl[ 1 + 2 (\gamma_i+\gamma_j+\gamma_k) 
    \ln\frac{\mu_R}{\Lambda} \biggr],
\label{app:A_term}
\end{equation}
\begin{equation}
  m_{i}^2 = - [\ln Z_i]_D
  = M_0^2 \biggl[ r_i + 4 \gamma_i \ln\frac{\mu_R}{\Lambda}
    + 2 \dot{\gamma}_i \Bigl( \ln\frac{\mu_R}{\Lambda} \Bigr)^2 \biggr],
\label{eq:soft_mass}
\end{equation}
at an arbitrary renormalization scale $\mu_R$.  Here, $g$ and $b$ 
represent the gauge couplings and the beta function coefficients, 
$d\ln(1/g^2)/d\ln\mu_R = -2b/(4\pi)^2$, respectively, and $\gamma_i$ 
and $\dot{\gamma}_i$ are the anomalous dimensions, $d\ln Z_i/d \ln\mu_R 
= -2\gamma_i$, and their derivatives with respect to the scale $\mu_R$, 
$\dot{\gamma}_i = d\gamma_i/d\ln\mu_R$.  The overall supersymmetry 
breaking parameter $M_0$ is defined by
\begin{equation}
  M_0 = \frac{[T]_F}{[T+T^\dagger]_A},
\label{app:M0}
\end{equation}
where the subscript $A$ denotes the lowest component.  Note that the 
soft supersymmetry breaking parameters in Eqs.~(\ref{app:gaugino_mass}~%
--~\ref{eq:soft_mass}) are given by simple functions of the quantities 
at the scale $\mu_R$, the gauge couplings, beta functions and anomalous 
dimensions, as well as $\ln(\mu_R/\Lambda)$.

It is rather simple to prove the scaling relations in 
Eqs.~(\ref{app:transf},~\ref{app:transf2}).  The RG equations for the 
gauge couplings (the gauge kinetic functions $S = 1/2g^2 + \cdots$) 
are given by
\begin{equation}
  \frac{d}{dt} S(T,t) = - \frac{b}{(4\pi)^2},
\end{equation}
where we have defined $t \equiv \ln (\mu_R/\Lambda)$.  This obviously 
leads to
\begin{equation}
  \frac{d}{dt} S(T,t) = \frac{d}{d(at)} a S(T,t) 
    = - \frac{b}{(4\pi)^2}.
\end{equation}
On the other hand, the RG equations for $S(aT,at)$ are given by
\begin{equation}
  \frac{d}{d(at)} S (aT,at) = - \frac{b}{(4\pi)^2}.
\end{equation}
We thus find that $aS(T,t)$ satisfies the same RG equation as 
$S(aT,at)$.  With the initial condition $S(T,0)=T$, we can 
determine the integration constant:
\begin{equation}
  S(aT,at) = a S(T,t).
\label{app:s-weight}
\end{equation}
This implies that $S$ scales as $S \rightarrow a S$ for $(T,t) 
\rightarrow (aT,at)$, which is the relation we wanted to prove. 
In fact, the result here is not a special property of moduli mediated 
models.  The effective Lagrangian in softly broken supersymmetric 
theories can always be recast in the form of Eq.~(\ref{app:lagrangian}) 
as far as the gauge sector is concerned, so that the expression for 
the gaugino masses in Eq.~(\ref{app:gaugino_mass}) is a (well-known) 
general result.

We can prove the scaling properties of $Z_i$'s along the same lines. 
The RG equations for $Z_i$'s are
\begin{equation}
  \frac{d}{dt} \ln Z_i = -2 \gamma_i 
  = \frac{1}{2} \sum_{j,k} \frac{|\lambda_{ijk}|^2}{Z_i Z_j Z_k} 
    - 2 (S + S^\dagger)^{-1} C_2^{(i)}(R),
\label{app:homo}
\end{equation}
at one loop, where $C_2^{(i)}(R)$ is the quadratic Casimir operator 
for the superfield $Q_i$.%
\footnote{Precisely speaking, $S+S^\dagger$ in Eq.~(\ref{app:homo}) 
should be the real gauge coupling superfield $R$ defined by 
$R - (T_G/8\pi^2) \ln R = S + S^\dagger - \sum_i (T_i/8\pi^2) 
\ln Z_i$ (in the NSVZ scheme).  The difference, however, is 
irrelevant at the one-loop level.}
We can again transform this to
\begin{eqnarray}
  \frac{d}{d(at)} \ln (a^{r_i} Z_i) 
  &=& \frac{1}{2} \sum_{j,k} \frac{a^{-1}|\lambda_{ijk}|^2}{Z_i Z_j Z_k}
    -2 a^{-1} \Bigl( S(T,t) + S(T,t)^\dagger \Bigr)^{-1} C_2^{(i)}(R)
\nonumber\\
  &=& \frac{1}{2} \sum_{j,k}
    \frac{a^{r_i+r_j+r_k-1}|\lambda_{ijk}|^2}
    {(a^{r_i} Z_i) (a^{r_j} Z_j) (a^{r_k} Z_k)}
    -2 \Bigl( S(aT,at) + S(aT,at)^\dagger \Bigr)^{-1} C_2^{(i)}(R),
\end{eqnarray}
where we have used Eq.~(\ref{app:s-weight}) in the second equation. 
This equation shows that $a^{r_i} Z_i(T,\lambda_{ijk},t)$ satisfies 
the same RG equation as $Z_i(aT,a^{(r_i+r_j+r_k-1)/2}\lambda_{ijk},at)$. 
With the initial condition of $Z_i(T,\lambda_{ijk},0)=(T+T^\dagger)^{r_i}$, 
the integration constant is determined and we obtain
\begin{equation}
  Z_i(aT,a^{(r_i+r_j+r_k-1)/2}\lambda_{ijk},at)
  = a^{r_i} Z_i(T,\lambda_{ijk},t).
\end{equation}
This is the scaling relation for $Z_i$ given in 
Eqs.~(\ref{app:transf},~\ref{app:transf2}).

There is a subtlety if the gauge group contains a $U(1)$ factor. 
In this case the Fayet-Iliopoulos term
\begin{equation}
  \int\! d^4\theta\, \xi V_Y,
\end{equation}
may be induced at one loop, which contributes to the soft scalar 
squared masses as
\begin{equation}
  m_i^2 = - [\ln Z_i]_D + g_Y^2 \xi Y_i,
\end{equation}
where $g_Y$ is the $U(1)$ gauge coupling and $Y_i$ is the $U(1)$ 
charge of the superfield $Q_i$.  The RG equation for $\xi$ is given 
by~\cite{Jack:1999zs}
\begin{equation}
  \frac{d}{dt} \xi
  = \frac{-2}{(4\pi)^2} \biggl[ \sum_i Y_i \ln Z_i \biggr]_D
    + \frac{2 g_Y^2}{(4\pi)^2}\, \xi \sum_i Y_i^2.
\end{equation}
Since the combination of $\sum_i Y_i \ln Z_i$ is RG invariant, i.e. 
$\sum_i Y_i \gamma_i=0$, $\xi$ is never generated if $\sum_i Y_i 
\ln Z_i = 0$ (and $\xi = 0$) at the classical level.  Therefore, 
we can neglect the contributions to $m_i^2$ from the Fayet-Iliopoulos 
term if the condition $\sum_i Y_i r_i = 0$ is satisfied.

In the case where there is a mass threshold $M$, we find that the 
scaling properties of Eqs.~(\ref{app:transf},~\ref{app:transf2}) 
are maintained if the rescalings of $T$, $\mu_R$ and $\lambda_{ijk}$ 
are supplemented by
\begin{equation}
  \ln\frac{M}{\Lambda} \rightarrow a \ln\frac{M}{\Lambda}.
\end{equation}
The functions $\hat{S}$ and $\hat{Z}_i$, in this case, can depend 
on $\ln(M/\Lambda)/T$ and $\ln(|M|^2/|\Lambda|^2)/(T+T^\dagger)$, 
respectively.  The soft supersymmetry breaking terms, therefore, 
obtain additional contributions:
\begin{equation}
  \varDelta m_\lambda = \frac{2 \varDelta b\, g^2}{(4\pi)^2} 
    M_0 \ln\frac{M}{\Lambda},
\label{app:delta-mhalf}
\end{equation}
\begin{equation}
  \varDelta A_{ijk} = - 2 (\varDelta\gamma_i 
    + \varDelta\gamma_j + \varDelta\gamma_k)
    M_0 \ln\frac{M}{\Lambda},
\label{app:delta-A}
\end{equation}
\begin{equation}
  \varDelta m_i^2 = M_0^2 
    \left[ 4 \varDelta\gamma_i \ln\frac{M}{\Lambda}
    + 2 \varDelta\dot{\gamma}_i \left(\ln\frac{M}{\Lambda}\right)^2
    \right],
\label{app:delta-msq}
\end{equation}
where $\varDelta b$, $\varDelta \gamma_i$ and $\varDelta \dot{\gamma}_i$ 
are the changes of $b$, $\gamma_i$ and $\dot{\gamma}_i$ at the scale $M$ 
($({\rm high\,\, scale\,\, value}) - ({\rm low\,\, scale\,\, value})$). 
The gauge coupling $g$ in Eq.~(\ref{app:delta-mhalf}) is the one at 
the scale $\mu_R$.

The derivation here should make it clear the origins of the special 
properties of Eqs.~(\ref{app:gaugino_mass}~--~\ref{eq:soft_mass}) and 
the required condition $r_i + r_j + r_k = 1$.  Since the RG equations 
for the gauge and Yukawa couplings take the form of $dg/dt \sim g^3$ 
and $dy/dt \sim y^3 + y g^2$ at one loop, and $g^2 \propto 1/T$ and 
$y^2 \propto (Z_i Z_j Z_k)^{-1} \propto (T+T^\dagger)^{-(r_i+r_j+r_k)}$ 
in moduli mediated models, it is clear that one-loop RG equations 
are invariant under the rescaling $(T,t) \rightarrow (aT, at)$ 
if $r_i + r_j + r_k = 1$ is chosen.  This scaling property then 
guarantees the forms of Eqs.~(\ref{app:s},~\ref{app:z}), leading 
to Eqs.~(\ref{app:gaugino_mass}~--~\ref{eq:soft_mass}).  (This also 
makes it clear that these properties persist under the existence 
of arbitrary generational mixings.)  This simple scaling property 
clearly cannot persist at higher loop orders, so that the properties 
of Eqs.~(\ref{app:gaugino_mass}~--~\ref{eq:soft_mass}) are that of 
one-loop RG equations.

Inclusion of anomaly mediation is straightforward at this point. 
We should simply replace $\Lambda$ in Eqs.~(\ref{app:s},~\ref{app:z}) 
by $\Lambda \Phi$, where $\Phi\, (= 1 + m_{3/2} \theta^2)$ is the chiral 
compensator field.  (Note that the compensator field $\Phi$ does not 
couple to $T$ as $T$ is a dimensionless chiral superfield.)  A curious
similarity between moduli and anomaly mediations is manifest here. 
In particular, at the scale
\begin{equation}
 \ln \frac{\Lambda}{\mu_R} = \frac{m_{3/2}}{2M_0},
\end{equation}
the $F$-component of $\ln(\mu_R/\Lambda \Phi)/T$ as well as $F$- and 
$D$-components of ${\ln(\mu_R^2/|\Lambda \Phi|^2)}/{(T+T^\dagger)}$ 
vanish.  Therefore, either $\hat{S}$ or $\hat{Z}$ does not have $F$- 
or $D$-components if the condition $r_i + r_j + r_k = 1$ is satisfied. 
The solutions of the RG equations at this scale are remarkably simple:
\begin{equation}
  m_\lambda = M_0,
\qquad
  A_{ijk} = - M_0,
\qquad
  m_i^2 = r_i M_0^2.
\end{equation}
If there is a mass threshold, the solutions are 
obtained by simply adding the contributions in 
Eqs.~(\ref{app:delta-mhalf}~--~\ref{app:delta-msq}), because of 
the ultraviolet insensitivity of anomaly mediated contributions.

\end{document}